\documentclass[pra,showpacs,twocolumn,eqsecnum]{revtex4}
\usepackage{dcolumn}
\usepackage{bm}
\usepackage{amssymb}
\usepackage{amsmath}
\usepackage{amsfonts}

\usepackage[]{graphicx}
\begin{document}
\bibliographystyle{prsty}
\title{Quantizing polaritons in inhomogeneous dissipative systems}
\author{ Aur\'elien Drezet $^{1}$}
\address{(1) Univ.~Grenoble Alpes, CNRS, Institut N\'{e}el, F-38000 Grenoble, France}
\begin{abstract}
In this article we provide a general analysis of canonical quantization for polaritons in dispersive and dissipative electromagnetic inhomogeneous media. We compare several approaches based either on the Huttner Barnett model [B. Huttner, S. M. Barnett, Phys. Rev. A \textbf{46}, 4306 (1992)] or the Green function, Langevin noise method [T. Gruner, D.-G. Welsch, Phys. Rev. A \textbf{53}, 1818 (1996)] which includes only material oscillators as fundamental variables. We show that in order to preserve unitarity, causality and time symmetry  one must necessarily include with an equal footing both  electromagnetic modes and material fluctuations in the evolution equations. This becomes particularly relevant for all  nanophotonics and plasmonics problems involving spatially localized antennas or devices.  
\end{abstract}

 \maketitle
\section{Introduction}
Tremendous progress has been realized in the last decades concerning the theoretical foundation of quantum optics in dielectric media. While the historical approach proposed by Jauch and Watson~\cite{Jauch1948} was already based on the standard canonical quantization formalism for fields,  it neglected dispersion and dissipation which are intrinsic properties of any causal dielectric media satisfying Kramers Kr\"{o}nig relations.  Since then several important studies were devoted to the extension of the method to inhomogeneous and artificially structured media which are central issues in modern micro and nano photonics~\cite{Knoll1987,Glauber1991,Wubs2003}. Furthermore, theoretical approaches adapted to transparent but dispersive media with negligible losses have been also developed based on different techniques such as the slowly varying envelope approximation~\cite{Drumond1990} or the quasi modal expansion method which is valid near resonance for polaritons~\cite{Constantinou1993,Milonni1995,Al1996,Garrison2004}. More recently, losses were included in the theory by adding phenomenologically some optical dissipation channels in the light propagation path~\cite{Jeffers1993,Barnett1998}. Such a method was successfully used for the modeling of Casimir forces in dissipative media~\cite{Genet2003} and surface plasmon polaritons~\cite{Tame2008,Ballester2009,Ballester2010}.\\ \indent Moreover, the most fundamental progress was probably done when Huttner and Barnett,  and others~\cite{Huttner1991,Huttner1992a,Huttner1992b,Huttner1992c,Matloob1995,Matloob1996,Barnett1995} proposed a self-consistent canonical quantization procedure for an homogeneous and causal dielectric medium by coupling photonic degrees of freedom with mechanical oscillator variables acting as thermal baths. The method, based on the pioneer works by Fano and Hopfield \cite{Fano1956,Hopfield1958} (see also Ref.~\cite{Huang1951}), was subsequently extended to several inhomogeneous systems including anisotropic and magnetic properties~\cite{Wubs2001,Suttorp2004a,Suttorp2004b,Suttorp2007,Bhat2006,Judge2013,Philbin2010}. In parallel to these theoretical works  based on the standard canonical quantization method, a different and powerful axis of research appeared after the work by Gruner and Welsch~\cite{Gruner1995,Gruner1996} (see also Ref.\cite{Yeung1996}) based on the quantum Langevin noise approach used in cavity QED (i.e., Quantum ElectroDynamics)~\cite{Cohen}. The method is also known as the Green tensor method~\cite{Gruner1996} since it relies on efficient Green dyadic techniques used nowadays in nano-photonics and plasmonics~\cite{Girard1996,Novotny}.  This `Langevin-noise' approach, which actually extends earlier `semi classical' researches based on the fluctuation-dissipation theorem by Lifshitz and many others in the context of Casimir and optical forces~\cite{Lifshitz1956,Ginzburg,Rytov,Milonnibook,Callen1951,Rosa2010,Agarwal1975,Sipe1984,Carminati1999,Mulet,Rousseau,Henkel}, was successfully applied in the recent years to many issues concerning photonics~\cite{Scheel1998,Dung1998,Dung2000,Scheel2001,Matloob1999,Matloob2004,Fermani2006,Raabe2007,Amooshahi2008,Scheelreview2008} and nano-plasmonics where dissipation can not be neglected~\cite{Dzotjan2010,Cano2011,Hummer2013,Chen2013,Delga2014,Hakami2014,Choquette2012,Grimsmo2013,Rousseaux2016}. In this context the relationship between the Huttner-Barnett approach on the one side and the Langevin noise method on the other side has attracted much attention in the last years, and several works attempted to demonstrate the validity of the Langevin Noise method from a rigorous Hamiltonian perspective which is more in a agreement with the canonical Huttner-Barnett approach~\cite{Wubs2001,Suttorp2004a,Suttorp2004b,Suttorp2007,Bhat2006,Judge2013,Philbin2010,Rosa2010}.\\
\indent The aim of this work is to revisit these derivations of the equivalence between the Langevin  noise and Hamitonian method and to show that some unphysical assumptions actually limit the domain of validity of the previous attempts. More precisely, as we will show in this work, the analysis and derivations always included some hypothesis concerning causality and boundary conditions which actually lead to circularity in the deductions and are not applicable to the most general inhomogeneous systems used in nano-optics. Specifically, these derivations, like the fluctuation-dissipation reasoning in Lifshitz and Rytov works~\cite{Lifshitz1956,Ginzburg,Rytov,Milonnibook}, give too much emphasis on the material origin of quantum fluctuations for explaining macroscopic quantum electrodynamics  in continuous media. However, as it was already pointed out in the 1970's~\cite{Senitzky1973,Ackerhalt1973,Milonni1973,Dalibard1982,Milonnibook}, one must include with an equal footings  both field and matter fluctuations in  a self consistent QED  Hamiltonian in order to preserve rigorously unitarity and causality~\cite{Milonni1982,Milonnibook}. While this doesn't impact too much the homogeneous medium case considered by Huttner and Barnett~\cite{Huttner1992a} it is crucial to analyze further the inhomogeneous medium problem in order to give a rigorous foundation to  the Gruner and Welsch theory~\cite{Gruner1996} based on fluctuating currents. This is the central issue tackled in the present work.\\
\indent The layout of this paper is as follows: In Section II we review the Lagrangian method developed in our previous work~\cite{previous} based on an alternative dual formalism for describing the Huttner-Barnett model. In this section we summarize the essential elements of the general Lagrangian and Hamiltonian model necessary for the present study. In particular we present the fundamental issue about the correct definition of Hamiltonian which will be discussed at length in this article. In Section III we provide a quantitative discussion of the Huttner-Barnett model for an homogeneous dielectric medium. We discuss a modal expansion into plane waves and separate explicitly the electromagnetic field into classical eigenmodes and noise related Langevin's  modes. We show that both contributions are necessary for preserving unitarity and time symmetry. We consider limit cases such as the ideal Hopfield-Fano polaritons \cite{Fano1956,Hopfield1958} without dissipation and the weakly dissipative polariton modes considered  by Milonni and others~\cite{Constantinou1993,Milonni1995,Al1996,Garrison2004}. We discuss the physical interpretation of the Hamiltonian  of the whole system and interpret the various contributions with respect to the  Langevin noise method and to the loss-less  Hopfield-Fano limit.  In section IV we generalize our analysis to the inhomogeneous medium case by using a Green dyadic formalism in both the frequency and time domain.  We demonstrate that in  general it is necessary to keep both pure photonic  and material fluctuations to preserve the unitarity and time symmetry of the quantum evolution. We conclude with a discussion about the physical meaning of the Hamiltonian in presence of inhomogeneities and interpret the various terms associated with  photonic and material modes.        

\section{The Huttner-Barnett model and the dual Lagrangian formalism}
In Ref.~\cite{previous} we developed a new Lagrangian formalism adapted to QED in dielectric media without magnetic properties. Here we will use this model to derive our approach but a standard treatment based on the minimal coupling scheme~\cite{Cohen2} or the Power-Zienau~\cite{Zienau} transformation would lead to similar results.   
We start with the dual Lagrangian density:
\begin{eqnarray}
\mathcal{L}=\frac{\mathbf{B}^2-\mathbf{D}^2}{2}+ \mathbf{F}\cdot\boldsymbol{\nabla}\times\mathbf{P} -\frac{\mathbf{P}^2}{2}+\mathcal{L}_M\label{1}
\end{eqnarray}
where $\mathbf{B}(\mathbf{x},t)$ and $\mathbf{D}(\mathbf{x},t)$ are the magnetic and displacement fields respectively. In this formalism the usual magnetic potential $\textbf{A}$, defined such as $\mathbf{B}=\boldsymbol{\nabla}\times\mathbf{A}$, is replaced by the dual electric potential $\mathbf{F}$ (in the `Coulomb' gauge $\boldsymbol{\nabla}\cdot\mathbf{F}(\mathbf{x},t)=0$) defined by
\begin{eqnarray}
\mathbf{B}(\mathbf{x},t)=\frac{1}{c}\partial_t\mathbf{F}(\mathbf{x},t), &\mathbf{D}(\mathbf{x},t)=\boldsymbol{\nabla}\times\mathbf{F}(\mathbf{x},t).\label{3}
\end{eqnarray}
implying 
\begin{eqnarray}
\boldsymbol{\nabla}\times\mathbf{B}(\mathbf{x},t)=\frac{1}{c}\partial_t\mathbf{D}(\mathbf{x},t), &\boldsymbol{\nabla}\cdot\mathbf{D}(\mathbf{x},t)=0.\label{4}
\end{eqnarray}
The material part $\mathcal{L}_M$ of the Lagrangian density in Eq.~\ref{1} reads
\begin{eqnarray}
\mathcal{L}_M=\int_{0}^{+\infty}d\omega\frac{(\partial_t\mathbf{X}_\omega)^2-\omega^2\mathbf{X}_\omega^2}{2}.\label{2}
\end{eqnarray}
with $\mathbf{X}_\omega(\mathbf{x},t)$ the material oscillator fields describing the Huttner-Barnett bath coupled to the electromagnetic field.  
The coupling depends  on the polarization density which is defined by 
\begin{eqnarray}
\mathbf{P}(\mathbf{x},t)=\int_{0}^{+\infty}d\omega\sqrt{\frac{2\sigma_\omega(\mathbf{x})}{\pi}}\mathbf{X}_\omega(\mathbf{x},t)\label{5}
\end{eqnarray} where the coupling function $\sigma_\omega(\mathbf{x})\geq 0$ defines the conductivity of the medium at the harmonic pulsation $\omega$. 
From Eq.~\ref{1} and Euler-Lagrange equations we deduce the dynamical laws for the electromagnetic field 
 \begin{eqnarray}
\boldsymbol{\nabla}\times\mathbf{E}(\mathbf{x},t)=-\frac{1}{c}\partial_t\mathbf{B}(\mathbf{x},t), &\boldsymbol{\nabla}\cdot\mathbf{B}(\mathbf{x},t)=0\label{6}
\end{eqnarray}
with the electric field $\mathbf{E}(\mathbf{x},t)=\mathbf{D}(\mathbf{x},t)-\mathbf{P}(\mathbf{x},t)$. Similarly for the material oscillators we have:
\begin{eqnarray}
\partial_t^2\mathbf{X}_\omega(\mathbf{x},t)+\omega^2\mathbf{X}_\omega(\mathbf{x},t)=\sqrt{\frac{2\sigma_\omega(\mathbf{x})}{\pi}}\mathbf{E}(\mathbf{x},t),\label{8}
\end{eqnarray}We point out that the Lagrangian density in Eq.~\ref{1} includes a term $-\frac{\mathbf{P}^2}{2}$ which is necessary for the derivation of the dynamical laws for the material fields $\mathbf{X}_\omega$ \cite{previous}.
Furthermore, to complete the QED canonical quantization procedure of the material field  we introduce the lowering $\mathbf{f}_\omega(\mathbf{x},t)$ and rising $\mathbf{f}^\dagger_\omega(\mathbf{x},t)$ operators for the bosonic material field from  the relation  $\mathbf{f}_\omega(\mathbf{x},t)=\frac{i\partial_t\mathbf{X}_\omega(\mathbf{x},t)+\omega\mathbf{X}_\omega(\mathbf{x},t)}{\sqrt{2\hbar\omega}}$. 
As explained in Ref.~\cite{previous} by using the equal time commutation relations between the canonical variables $\mathbf{X}_\omega(\mathbf{x},t)$ and $\partial_t\mathbf{X}_\omega(\mathbf{x},t)$, we deduce the fundamental rules
\begin{eqnarray}
[\mathbf{f}_\omega(\mathbf{x},t),\mathbf{f}^\dagger_{\omega'}(\mathbf{x}',t)]=\delta(\omega-\omega')\delta^3(\mathbf{x}-\mathbf{x'})\textbf{I}. \label{33}
\end{eqnarray} (with $\textbf{I}=\mathbf{\hat{x}}\otimes\mathbf{\hat{x}}+\mathbf{\hat{y}}\otimes\mathbf{\hat{y}}+\mathbf{\hat{z}}\otimes\mathbf{\hat{z}}$ the unit dyad) and $[\mathbf{f}_\omega(\mathbf{x},t),\mathbf{f}_{\omega'}(\mathbf{x}',t)]=[\mathbf{f}^\dagger_\omega(\mathbf{x},t),\mathbf{f}^\dagger_{\omega'}(\mathbf{x}',t)]=0$ allowing a clear interpretation of $\mathbf{f}_\omega(\mathbf{x},t)$ and $\mathbf{f}^\dagger_{\omega}(\mathbf{x},t)$ as lowering and rising operators for the bosonic states associated with the matter oscillators.\\
\indent Moreover,  Eqs.~\ref{5},\ref{8} can be formally integrated  leading to 
\begin{eqnarray}
\mathbf{P}(\mathbf{x},t)=\mathbf{P}^{(0)}(\mathbf{x},t)+\int_{0}^{t-t_0}\chi(\mathbf{x},\tau)d\tau\mathbf{E}(\mathbf{x},t-\tau)\label{18}
\end{eqnarray}
where $t_0$ is an initial time and  where $\mathbf{P}^{(0)}(\mathbf{x},t)$ is a fluctuating dipole density distribution defined by:
\begin{eqnarray}
\mathbf{P}^{(0)}(\mathbf{x},t)=\int_{0}^{+\infty}d\omega\sqrt{\frac{2\sigma_\omega(\mathbf{x})}{\pi}}\mathbf{X}^{(0)}_\omega(\mathbf{x},t)\nonumber\\
=\int_0^{+\infty}d\omega\sqrt{\frac{\hbar\sigma_{\omega}(\mathbf{x})}{\pi\omega}}[\mathbf{f}^{(0)}_{\omega}(\mathbf{x},t)
+\mathbf{f}^{\dagger(0)}_{\omega}(\mathbf{x},t)] \nonumber\\ \label{35}
\end{eqnarray} with $\mathbf{X}^{(0)}_\omega(\mathbf{x},t)=\cos{(\omega (t-t_0))}\mathbf{X}_\omega(\mathbf{x},t_0)+\sin{(\omega (t-t_0))}\partial_t\mathbf{X}_\omega(\mathbf{x},t_0)/\omega$ and where by definition $\mathbf{f}^{(0)}_{\omega}(\mathbf{x},t)=\mathbf{f}_{\omega}(\mathbf{x},t_0)e^{-i\omega(t-t_0)}$.
We therefore have \begin{eqnarray}
\mathbf{D}(\mathbf{x},t)=\mathbf{E}(\mathbf{x},t)+\mathbf{P}(\mathbf{x},t)=\mathbf{P}^{(0)}(\mathbf{x},t)\nonumber\\+\mathbf{E}(\mathbf{x},t)+\int_{0}^{t-t_0}d\tau\chi(\mathbf{x},\tau)\mathbf{E}(\mathbf{x},t-\tau),\label{21}
\end{eqnarray}  which is reminiscent of the general linear response theory used in thermodynamics~\cite{Zwanzig}.  We point out that the term $\int_{0}^{t-t_0}\chi(\mathbf{x},\tau)d\tau\mathbf{E}(\mathbf{x},t-\tau)$ can be seen as an induced dipole density.  However, as we will show in the next section  the electric field itself is decomposed into a purely fluctuating term $\mathbf{E}^{(0)}(\mathbf{x},t)$ and a scattered field $\mathbf{E}^{(s)}(\mathbf{x},t)$ which depends on the density $\mathbf{P}^{(0)}$. Therefore,  the contribution $\int_{0}^{t-t_0}\chi(\mathbf{x},\tau)d\tau\mathbf{E}(\mathbf{x},t-\tau)$
to $\mathbf{P}$ is also decomposed into a pure photon-fluctuation term $\int_{0}^{t-t_0}\chi(\mathbf{x},\tau)d\tau\mathbf{E}^{(0)}(\mathbf{x},t-\tau)$ and an induced term $\int_{0}^{t-t_0}\chi(\mathbf{x},\tau)d\tau\mathbf{E}^{(s)}(\mathbf{x},t-\tau)$ related to material fluctuations $\mathbf{P}^{(0)}$.\\
\indent  Importantly, the linear susceptibility $\chi(\mathbf{x},\tau)$ which is defined by
 \begin{eqnarray}
\chi(\mathbf{x},\tau)=\int_{0}^{+\infty}d\omega\frac{2\sigma_\omega(\mathbf{x})}{\pi}\frac{\sin{\omega\tau}}{\omega}\Theta(\tau)\label{20}
\end{eqnarray} characterizes completely the dispersive and dissipative dielectric medium. We can show that the permittivity $\widetilde{\varepsilon}(\mathbf{x},\omega)=1+\int_0^{+\infty}d\tau\chi(\mathbf{x},\tau)e^{i\omega \tau}$ is an analytical function in the upper part of the complex plane $\omega=\omega'+i\omega''$, i.e., $\omega''>0$, provided $\chi(\mathbf{x},\tau)$ is finite for any time $\tau\geq 0$. From this we deduce the symmetry $\widetilde{\varepsilon}(\mathbf{x},-\omega)^\ast=\widetilde{\varepsilon}(\mathbf{x},\omega^\ast)$ and it is possible to derive the general Kramers-Kr\"{o}nig relations existing between  the real part $\textrm{Re}[\widetilde{\varepsilon}(\mathbf{x},\omega)]\equiv\widetilde{\varepsilon}'(\mathbf{x},\omega)$ and the imaginary part  $\textrm{Imag}[\widetilde{\varepsilon}(\mathbf{x},\omega)]\equiv\widetilde{\varepsilon}''(\mathbf{x},\omega)$ of the dielectric permittivity. Therefore, the Huttner-Barnett model characterized by the conductivity $\sigma_\omega(\mathbf{x})$ is fully causal and can be applied to describe any inhomogeneous dielectric media in the linear regime.\\
\indent The central issue of the present paper concerns the definition of the Hamiltonian  $H(t)$ in the Huttner-Barnett model.  We remind that in Ref.~\cite{previous} we derived the result: 
\begin{eqnarray}
H(t)=\int d^3\mathbf{x}\frac{:\mathbf{B}^2+\mathbf{E}^2:}{2}+H_M\nonumber\\ \label{10}
\end{eqnarray} with $H_M(t)=\int d^3\mathbf{x}\int_{0}^{+\infty}d\omega\frac{:(\partial_t\mathbf{X}_\omega)^2+\omega^2\mathbf{X}_\omega^2:}{2}$ where $:[...]:$ means, as usually, a normally-ordered product for removing the infinite zero-point energy.  Inserting the definition for $\mathbf{f}_\omega(\mathbf{x},t)$ obtained earlier  we get  for the material part
\begin{eqnarray}
H_M(t)=\int d^3\mathbf{x}\int_{0}^{+\infty}d\omega\hbar\omega\mathbf{f}^\dagger_\omega(\mathbf{x},t)\mathbf{f}_\omega(\mathbf{x},t). \label{34a}
\end{eqnarray} which has the standard structure for oscillators (i.e., without the infinite zero-point energy).\\
\indent However, Hutner and Barnett~\cite{Huttner1992a} after diagonalizing their Hamiltonian found that the total evolution is described in the homogeneous medium case by 
  \begin{eqnarray}
H_M^{(0)}(t)=\int d^3\mathbf{x}\int_{0}^{+\infty}d\omega\hbar\omega\mathbf{f}^{\dagger(0)}_\omega(\mathbf{x},t)\mathbf{f}^{(0)}_\omega(\mathbf{x},t) \label{34fluct}
\end{eqnarray}  
While as we will see this is actually a correct description `for all practical purpose' in a homogeneous dissipative medium for large class of physical boundary conditions, this is  in general not acceptable in order to preserve time-symmetry and unitarity in the full Hilbert space for interacting matter and light.  The general method based on Langevin forces and noises avoided quite generally mentioning that difficult point. We emphasize that while the conclusions presented in Refs.~\cite{Yeung1996,Gruner1995,Gruner1996} is accepted by more or less all authors on the subject~\cite{Scheel1998,Dung1998,Dung2000,Scheel2001,Matloob1999,Matloob2004,Fermani2006,Raabe2007,Amooshahi2008,Dzotjan2010,Cano2011,Hummer2013,Chen2013,Delga2014,Hakami2014,Choquette2012,Grimsmo2013,Rousseaux2016} they have been some some few dissident views (see Refs.~\cite{Stefano2000,Stefano2002}) claiming, that in the context of an input-output formalism, the Langevin noise formalism is not complete unless we consider as well fluctuations of the free photon modes (see also the replies with  an opposite perspective in Refs.~\cite{Khanbekyan2003,Khanbekyan2005}). In the present work we will generalize and give a rigrous QED like Hamiltonian foundations to the prescriptions of Refs.~\cite{Stefano2000,Stefano2002} and we will show that it is actually necessary to include a full description of photonic and material quantum excitations in order to preserve unitarity. In order to appreciate this fact further we will first consider the problem associated with quantization of the electromagnetic field.

\section{The quantization of electromagnetic waves in a homogenous dielectric medium} 
\subsection{The general modal expansion}
\indent We first introduce the paradigmatic  homogeneous medium case considered  initially by Huttner and Barnett~\cite{Huttner1992a}, i.e., with $\chi(\mathbf{x},\tau)=\chi(\tau)$.  We start with Faraday's law: $\boldsymbol{\nabla}\times\mathbf{E}(\mathbf{x},t)=-\frac{1}{c}\partial_t\mathbf{B}(\mathbf{x},t)$  rewritten according to Eq.~\ref{21} as:
\begin{eqnarray}
\boldsymbol{\nabla}\times\mathbf{D}(\mathbf{x},t)=-\frac{1}{c}\partial_t\mathbf{B}(\mathbf{x},t)+\boldsymbol{\nabla}\times\mathbf{P}^{(0)}(\mathbf{x},t)\nonumber\\-\frac{1}{c}\int_0^{t-t_0}d\tau\chi(\tau)\partial_{t-\tau}\mathbf{B}(\mathbf{x},t-\tau).\label{90}
\end{eqnarray} 
 Inserting  Eq.~\ref{4} and using the Coulomb (transverse) gauge condition we get:
\begin{eqnarray}
\frac{1}{c^2}\partial_t^2\mathbf{F}(\mathbf{x},t)-\boldsymbol{\nabla}^2\mathbf{F}(\mathbf{x},t)-\boldsymbol{\nabla}\times\mathbf{P}^{(0)}(\mathbf{x},t)\nonumber\\+\frac{1}{c^2}\int_0^{t-t_0}d\tau\chi(\tau)\partial_{t-\tau}^2\mathbf{F}(\mathbf{x},t-\tau)=0.\label{91}
\end{eqnarray}
We use the modal expansion method developed in Ref.~\cite{previous} and write 
\begin{eqnarray}
\mathbf{F}(\mathbf{x},t)=\sum_{\alpha,j} q_{\alpha,j}(t)\boldsymbol{\hat{\epsilon}}_{\alpha,j}\Phi_\alpha(\mathbf{x})\label{40}
\end{eqnarray} with $\alpha$ a generic label for  the wave vector $\mathbf{k}_\alpha$,  $\Phi_\alpha(\mathbf{x})=e^{i\mathbf{k}_\alpha\cdot\mathbf{x}}/\sqrt{V}$ (here we consider as it is usually done  the periodical `Box' Born-von Karman expansion in the rectangular box of volume  $V$), $j=1$ or 2, labels the two transverse polarization states with unit vectors $\boldsymbol{\hat{\epsilon}}_{\alpha,1}=\mathbf{k}_\alpha\times\mathbf{\hat{z}}/|\mathbf{k}_\alpha\times\mathbf{\hat{z}}|$, and $\boldsymbol{\hat{\epsilon}}_{\alpha,2}=\hat{\mathbf{k}}_\alpha\times\boldsymbol{\hat{\epsilon}}_{\alpha,1}$ (conventions and details are given in Appendix A of Ref. \cite{previous}). Inserting Eq.~\ref{40} into Eq.~\ref{91} we obtain the dynamical equation:  
 \begin{eqnarray}
\ddot{q}_{\alpha,j}(t)+\int_0^{t-t_0}d\tau\chi(\tau)\ddot{q}_{\alpha,j}(t-\tau)+\omega_\alpha^2q_{\alpha,j}(t)=S_{\alpha,j}^{(0)}(t)\nonumber\\ \label{92}
\end{eqnarray}
with the time dependent source term
\begin{eqnarray}
S_{\alpha,j}^{(0)}(t)=c^2\int d^3\mathbf{x}\boldsymbol{\nabla}\times\mathbf{P}^{(0)}(\mathbf{x},t)\cdot\boldsymbol{\hat{\epsilon}}_{\alpha,j}\Phi_\alpha^\ast(\mathbf{x}).\label{93}
\end{eqnarray}
To solve this equation we use the Laplace transform of the fields which is defined below.\\
\indent We are interested in the evolution for $t\geq t_0$ of a field $A(t)$ which Fourier transform is not necessary well mathematically defined since the field is not  going to zero fast enough for $t\rightarrow+\infty$ (e.g., a fluctuating current or field). The method followed here is to consider the forward Laplace transform of the different evolution equations (such an approach was also used by Suttorp by mixing both forward and backward Laplace's transforms~\cite{Suttorp2004a}).  To deal with this problem we first change the time  $t$ variable  in $t'=t-t_0$ and define $A'(t')=A(t)$.
We define the (forward) Laplace transform  of $A'(t')$ as:
\begin{eqnarray}
\overline{A'}(p)=\int_{0}^{+\infty}dt'e^{-pt'} A'(t')=\int_{t_0}^{+\infty}dte^{-p(t-t_0)} A(t)\nonumber\\ \label{77}
\end{eqnarray}  with $p=\gamma-i\omega$ ($\omega$ a real number and $\gamma\geq0$). The presence of the term $e^{-\gamma t'}$ ensures the convergence.  We will not here introduce the backward Laplace transform $\int_{-\infty}^{t_0}dt e^{+p(t-t_0)} A(t)=\int_{-t_0}^{+\infty}du e^{-p(u+t_0)} A(-u)$ since the time $t_0$ is arbitrary and can be sent into the remote past if needed. \\
As it is well known, the (forward) Laplace transform is connected to the usual Fourier transform since we have
 \begin{eqnarray}
\overline{A'}(\gamma-i\omega)=2\pi\widetilde{\underline{A}}(\omega)e^{-i\omega t_0}\label{78}
\end{eqnarray}
where $\widetilde{\underline{A}}(\omega)=\int_{-\infty}^{+\infty}\frac{dt}{2\pi}\underline{A}(t)e^{i\omega  t}$ is the Fourier transform of $\underline{A}(t)=A(t)\Theta(t-t_0)e^{-\gamma (t-t_0)}$ with respect to $t$.\\
\indent Now for the specific problem considered here we obtain the separation  $q_{\alpha,j}(t)=q_{\alpha,j}^{(s)}(t)+q_{\alpha,j}^{(0)}(t)$. The $(0)$ contribution corresponds to what classically we call a sum of eigenmodes supported by the medium (i.e. with $\mathbf{P}^{(0)}=0$) while the $(s)$ term is the fluctuating  field generated by the Langevin source $\mathbf{P}^{(0)}(t)$. More explicitly, we have for the source term:
\begin{eqnarray}
q_{\alpha,j}^{(s)}(t)=\int_0^{t-t_0}d\tau H_\alpha(\tau)S_{\alpha,j}^{(0)}(t-\tau)\label{94}
\end{eqnarray} where we use Eq.~\ref{93}.
The propagator function $H_\alpha(\tau)$ is expressed as a Bromwich contour: 
\begin{eqnarray}
H_\alpha(\tau)=\int_{\gamma-i\infty}^{\gamma+i\infty}\frac{i dp}{2\pi}\frac{e^{p\tau}}{\omega_\alpha^2+(1+\bar{\chi}(p))p^2}\nonumber\\
\label{95}
\end{eqnarray} where $\bar{\chi}(p)$ is defined as $\overline{\chi}(\mathbf{x},p)=\int_{0}^{+\infty}d\tau e^{-p\tau} \chi(\mathbf{x},\tau)$. 
We remind that a Bromwich integral by definition will vanish for $\tau<0$  so that on the left side we actually mean  $\theta(\tau)H_\alpha(\tau)$. \\
\indent The Function $H_\alpha(\tau)$ has some remarkable properties which should be emphasized here.  We first introduce the `zeros' of $\omega_\alpha^2-\tilde{\varepsilon}(\omega)\omega^2$, i.e., the set of roots $\Omega_{\alpha,m}^{(\pm)}$ solutions of $\Omega_{\alpha,m}^{(\pm)}\sqrt{\tilde{\varepsilon}(\Omega_{\alpha,m}^{(\pm)})}\pm\omega_\alpha=0$.  From the causal properties of $\tilde{\varepsilon}$ we have $\tilde{\varepsilon}(-\omega^\ast)=\tilde{\varepsilon}(\omega)^\ast$ and therefore we deduce ${\Omega_{\alpha,m}^{(\pm)}}^\ast=-\Omega_{\alpha,m}^{(\mp)}$ implying that  the `+' and `-' roots are not independent. The important fact is that the roots are located in the lower complex plane associated with a negative imaginary part of the frequency~\cite{Huttner1992a} (this is proven in Appendix A). Now, as shown in Appendix B the  integral in Eq.~\ref{95} can be computed  by contour integration in the complex plane after closing the contour with a semi-circle in the lower plane and using the Cauchy residue theorem. We get: \begin{eqnarray}
H_\alpha(\tau)=\sum_{m}\frac{-1}{2i\omega_\alpha}\frac{e^{-i\Omega_{\alpha,m}^{(-)}\tau}}{\frac{\partial(\omega\sqrt{\tilde{\varepsilon}(\omega)})}{\partial \omega}|_{\Omega_{\alpha,m}^{(-)}}} +cc.\label{98}
\end{eqnarray} We also get  $H_\alpha(\tau)=0$ for $\tau\leq 0$ (after integration in the upper plane and considering in  detail the case $\tau=0$). Clearly, the function $H_\alpha(\tau)$ is here expanded into a sum of modes which define the polaritons of the problem (this is shown explicitly in the next subsection).  Here, the normal mode frequency  $\Omega_{\alpha,m}^{(-)}$ are complex numbers ensuring the damped nature of the waves in the future direction. We emphasize that in our knowledge this kind of formula has never been discussed before. The expansion in Eq.~\ref{98} is however rigorous and generalizes the quasimodal approximations used  in the weak dissipation regime and discussed in subsection III(F). \\ 
 \indent Similarly the source-free term $q_{\alpha,j}^{(0)}(t)$ reads: 
\begin{eqnarray}
q_{\alpha,j}^{(0)}(t)=U_\alpha(t-t_0)\dot{q}_{\alpha,j}(t_0)+\dot{U}_\alpha(t-t_0)q_{\alpha,j}(t_0)\nonumber\\ \label{100}
\end{eqnarray} with the new propagator function \begin{eqnarray}
U_\alpha(\tau)=\int_{\gamma-i\infty}^{\gamma+i\infty}\frac{i dp}{2\pi}\frac{(1+\bar{\chi}(p))e^{p\tau}}{\omega_\alpha^2+(1+\bar{\chi}(p))p^2}\nonumber\\
=\sum_{m}\frac{-\tilde{\varepsilon}(\Omega_{\alpha,m}^{(-)})}{2i\omega_\alpha}\frac{e^{-i\Omega_{\alpha,m}^{(-)}\tau}}{\frac{\partial(\omega\sqrt{\tilde{\varepsilon}(\omega)})}{\partial \omega}|_{\Omega_{\alpha,m}^{(-)}}} +cc.\label{101}
\end{eqnarray} Like for $H_\alpha$ we get $U_\alpha(\tau)=0$ for $\tau\leq 0$. Additionally, the boundary condition at $t=t_0$ (i.e., $q_{\alpha,j}^{(0)}(t_0)=q_{\alpha,j}(t_0)$) imposes
$\frac{d}{d\tau}U_\alpha(\tau)|_{\tau=0}=1$ (see Appendix B).\\
\subsection{The classical eigenmodes}
\indent The electromagnetic field can be calculated using the expansion Eq.~\ref{94} or \ref{100}.
First, the mathematical and physical structure of the free-field is seen by using the modal expansion: \begin{eqnarray}
\mathbf{F}^{(0)}(\mathbf{x},t)=\sum_{\alpha,j}[U_\alpha(t-t_0)\dot{q}_{\alpha,j}(t_0)\nonumber\\+\dot{U}_\alpha(t-t_0)q_{\alpha,j}(t_0)]\boldsymbol{\hat{\epsilon}}_{\alpha,j}\Phi_\alpha(\mathbf{x})\nonumber\\
=\sum_{\alpha,j,m}\frac{-\tilde{\varepsilon}(\Omega_{\alpha,m}^{(-)})}{2i\omega_\alpha}\frac{e^{-i\Omega_{\alpha,m}^{(-)}(t-t_0)}}{\frac{\partial(\omega\sqrt{\tilde{\varepsilon}(\omega)})}{\partial \omega}|_{\Omega_{\alpha,m}^{(-)}}}[\dot{q}_{\alpha,j}(t_0)\nonumber\\-i\Omega_{\alpha,m}^{(-)}q_{\alpha,j}(t_0)]\boldsymbol{\hat{\epsilon}}_{\alpha,j}\Phi_\alpha(\mathbf{x}) +hcc.\label{117}
\end{eqnarray} 
where we used the symmetries  of the modal expansion~\cite{footnote}  together with $\Omega_{\alpha,m}^{(-)}=\Omega_{-\alpha,m}^{(-)}$. From this we directly obtain: 
\begin{eqnarray}
\mathbf{D}^{(0)}(\mathbf{x},t)=i\sum_{\alpha,j}\frac{\omega_\alpha}{c}[U_\alpha(t-t_0)\dot{q}_{\alpha,j}(t_0)\nonumber\\+\dot{U}_\alpha(t-t_0)q_{\alpha,j}(t_0)]\boldsymbol{\hat{k}}_{\alpha}\times\boldsymbol{\hat{\epsilon}}_{\alpha,j}\Phi_\alpha(\mathbf{x})\nonumber\\
=\sum_{\alpha,j,m}\frac{-\tilde{\varepsilon}(\Omega_{\alpha,m}^{(-)})}{2c}\frac{e^{-i\Omega_{\alpha,m}^{(-)}(t-t_0)}}{\frac{\partial(\omega\sqrt{\tilde{\varepsilon}(\omega)})}{\partial \omega}|_{\Omega_{\alpha,m}^{(-)}}}[\dot{q}_{\alpha,j}(t_0)\nonumber\\-i\Omega_{\alpha,m}^{(-)}q_{\alpha,j}(t_0)]\boldsymbol{\hat{k}}_{\alpha}\times\boldsymbol{\hat{\epsilon}}_{\alpha,j}\Phi_\alpha(\mathbf{x}) +hcc. \label{118}
\end{eqnarray} 
and similarly for the magnetic field:
\begin{eqnarray}
\mathbf{B}^{(0)}(\mathbf{x},t)=\frac{1}{c}\sum_{\alpha,j}[\dot{U}_\alpha(t-t_0)\dot{q}_{\alpha,j}(t_0)\nonumber\\+\ddot{U}_\alpha(t-t_0)q_{\alpha,j}(t_0)]\boldsymbol{\hat{\epsilon}}_{\alpha,j}\Phi_\alpha(\mathbf{x})\nonumber\\
=\sum_{\alpha,j,m}\frac{\Omega_{\alpha,m}^{(-)}\tilde{\varepsilon}(\Omega_{\alpha,m}^{(-)})}{2c\omega_\alpha}\frac{e^{-i\Omega_{\alpha,m}^{(-)}(t-t_0)}}{\frac{\partial(\omega\sqrt{\tilde{\varepsilon}(\omega)})}{\partial \omega}|_{\Omega_{\alpha,m}^{(-)}}}[\dot{q}_{\alpha,j}(t_0)\nonumber\\-i\Omega_{\alpha,m}^{(-)}q_{\alpha,j}(t_0)]\boldsymbol{\hat{\epsilon}}_{\alpha,j}\Phi_\alpha(\mathbf{x}) +hcc.
\label{119}
\end{eqnarray} 
The (transverse) electric field  associated with this free solutions can also  be obtained from the definition $\mathbf{D}^{(0)}(\mathbf{x},t)=\mathbf{E}^{(0)}(\mathbf{x},t)+\int_{0}^{t-t_0}d\tau\chi(\tau)\mathbf{E}^{(0)}(\mathbf{x},t-\tau)$.  We have thus $\overline{\mathbf{D}'}^{(0)}(\mathbf{x},p)=(1+\bar{\chi}(p))\overline{\mathbf{E}'}^{(0)}(\mathbf{x},p)$ and considering the Laplace transform in Eqs.~\ref{100} and \ref{101} we can express the electric field  as a function of $H_\alpha(\tau)$, i.e., 
 \begin{eqnarray}
\mathbf{E}^{(0)}(\mathbf{x},t)=i\sum_{\alpha,j}\frac{\omega_\alpha}{c}[H_\alpha(t-t_0)\dot{q}_{\alpha,j}(t_0)\nonumber\\+\dot{H}_\alpha(t-t_0)q_{\alpha,j}(t_0)]\boldsymbol{\hat{k}}_{\alpha}\times\boldsymbol{\hat{\epsilon}}_{\alpha,j}\Phi_\alpha(\mathbf{x})\nonumber\\
=\sum_{\alpha,j,m}\frac{-1}{2c}\frac{e^{-i\Omega_{\alpha,m}^{(-)}(t-t_0)}}{\frac{\partial(\omega\sqrt{\tilde{\varepsilon}(\omega)})}{\partial \omega}|_{\Omega_{\alpha,m}^{(-)}}}[\dot{q}_{\alpha,j}(t_0)\nonumber\\-i\Omega_{\alpha,m}^{(-)}q_{\alpha,j}(t_0)]\boldsymbol{\hat{k}}_{\alpha}\times\boldsymbol{\hat{\epsilon}}_{\alpha,j}\Phi_\alpha(\mathbf{x}) +hcc. \label{120}
\end{eqnarray}   From Eqs.~\ref{119} and \ref{120} we see that $\boldsymbol{\nabla}\times\mathbf{E}^{(0)}(\mathbf{x},t)=-\partial_t\mathbf{B}^{(0)}(\mathbf{x},t)/c$ in agreement with Maxwell's equation for a free-field (the other Maxwell's equations are also automatically fulfilled by definition).\\ 
\indent Importantly, in the vacuum case where $\chi(\tau)\rightarrow 0$ we have $\Omega_{\alpha,m}^{(-)}\rightarrow \omega_\alpha$ and we find that the vacuum fields are given by
\begin{eqnarray}
\mathbf{F}^{(v)}(\mathbf{x},t)=\sum_{\alpha,j} ic\sqrt{\frac{\hbar}{2\omega_\alpha}}c_{\alpha,j}^{(v)}(t)\boldsymbol{\hat{\epsilon}}_{\alpha,j}\Phi_\alpha(\mathbf{x})+ hcc.\nonumber\\
\mathbf{D}^{(v)}(\mathbf{x},t)=\sum_{\alpha,j} -\sqrt{\frac{\hbar \omega_\alpha}{2}}c_{\alpha,j}^{(v)}(t)\hat{\mathbf{k}}_\alpha\times\boldsymbol{\hat{\epsilon}}_{\alpha,j}\Phi_\alpha(\mathbf{x})+ hcc.\nonumber\\
\mathbf{B}^{(v)}(\mathbf{x},t)=\sum_{\alpha,j} \sqrt{\frac{\hbar \omega_\alpha}{2}}c_{\alpha,j}^{(v)}(t)\boldsymbol{\hat{\epsilon}}_{\alpha,j}\Phi_\alpha(\mathbf{x})+ hcc..\nonumber\\ \label{57b}
\end{eqnarray}
with $c_{\alpha,j}^{(v)}(t)=c_{\alpha,j}(t_0)e^{-i\omega_\alpha (t-t_0)}$ as expected. In the  general case however causality imposes that the imaginary part of $\Omega_{\alpha,m}^{(-)}$  is negative. Therefore the optical modes labeled by  $\alpha$, $j$  and $m$ are damped in time (the only exception being of course the vacuum case where the only contribution  to the field arises from the source-free term $(0)$  since the $(s)$ terms vanishes together with $\mathbf{P}^{(0)}$).   As a consequence, if $t-t_0\rightarrow+\infty$ the free terms vanish asymptotically. In particular, if $t_0\rightarrow-\infty$ (corresponding to initial conditions fixed in the infinite remote past)  we can omit for all practical purposes the contribution of  $\mathbf{F}^{(0)}$, $\mathbf{D}^{(0)}$, and  $\mathbf{B}^{(0)}$ to the field observed at any finite time $t$ (unless we are in the vacuum) . This is indeed what was implicitly done by Huttner and Barnet~\cite{Huttner1992a} and Gruner and Welsch~\cite{Gruner1996} and that is why for all calculational needs they completely omitted the discussion of the $\mathbf{F}^{(0)}$, $\mathbf{D}^{(0)}$, and  $\mathbf{B}^{(0)}$ fields. However, for preserving the unitarity of the full evolution one must necessarily include  both  $(0)$ and $(s)$ terms. While this problem is apparently only technical we will see in the following its importance for inhomogeneous systems. \\
\subsection{The fluctuating Langevin modes}
\indent The previous discussion concerning the omission of the $(0)$ source free terms is very important since it explains the mechanism at work in the Huttner Barnet model~\cite{Huttner1992a}. To clarify that point further we now express the scattered field  $(s)$ using a Green tensor formalism  (In Appendix C we introduce alternative descriptions based on  vectorial and scalar potentials).  We first observe that from $\mathbf{D}^{(s)}(\mathbf{x},t)=\boldsymbol{\nabla}\times\mathbf{F}^{(s)}(\mathbf{x},t)$  we obtain after integration by parts:
\begin{eqnarray}
\mathbf{D}^{(s)}(\mathbf{x},t)=\int_{\gamma-i\infty}^{\gamma+i\infty}\frac{i dp}{2\pi}\int d^3\mathbf{x'} \mathbf{S}_\chi(\mathbf{x},\mathbf{x'},ip)
\nonumber\\ \cdot \overline{\mathbf{P}'}^{(0)}(\mathbf{x'},p)e^{p(t-t_0)}\label{rehh}
\end{eqnarray} with the dyadic propagator~\cite{Girard1996,Novotny,Gruner1995,Gruner1996}:
\begin{eqnarray}
\mathbf{S}_\chi(\mathbf{x},\mathbf{x'},ip)=\sum_{\alpha,j}\frac{\omega_\alpha^2\Phi_\alpha(\mathbf{x})\Phi_\alpha^\ast(\mathbf{x'})\boldsymbol{\hat{\epsilon}}_{\alpha,j}\otimes\boldsymbol{\hat{\epsilon}}_{\alpha,j}}{\omega_\alpha^2+(1+\bar{\chi}(p))p^2}\nonumber\\ \label{S}
\end{eqnarray}  The meaning of the  tensor $\mathbf{S}_\chi(\mathbf{x},\mathbf{x'},ip)$, which depends on $\chi$, becomes more clear if we introduce  the Green tensor $\mathbf{G}_\chi(\mathbf{x},\mathbf{x'},ip)$ solution of 
\begin{eqnarray}
\boldsymbol{\nabla}\times\boldsymbol{\nabla}\times\mathbf{G}_\chi(\mathbf{x},\mathbf{x'},ip)+\frac{p^2}{c^2}(1+\bar{\chi}(p))\mathbf{G}_\chi(\mathbf{x},\mathbf{x'},ip)\nonumber\\
=\mathbf{I}\delta^{3}(\mathbf{x}-\mathbf{x'})=\boldsymbol{\delta}(\mathbf{x}-\mathbf{x'})\nonumber\\ \label{cool}
\end{eqnarray} We observe that if we separate the tensor into a transverse and longitudinal part we get  $\mathbf{G}_{\chi}(\mathbf{x},\mathbf{x'},ip)=\mathbf{G}_{\chi,\bot}(\mathbf{x},\mathbf{x'},ip)+\mathbf{G}_{\chi,||}(\mathbf{x},\mathbf{x'},ip)$  with for the transverse dyadic green function
\begin{eqnarray}
\mathbf{G}_{\chi,\bot}(\mathbf{x},\mathbf{x'},ip)=\sum_{\alpha,j}\frac{c^2\Phi_\alpha(\mathbf{x})\Phi_\alpha^\ast(\mathbf{x'})\boldsymbol{\hat{\epsilon}}_{\alpha,j}\otimes\boldsymbol{\hat{\epsilon}}_{\alpha,j}}{\omega_\alpha^2+(1+\bar{\chi}(p))p^2}\label{Gper}
\end{eqnarray}
and for the longitudinal part \begin{eqnarray}
\frac{p^2}{c^2}(1+\bar{\chi}(p))\mathbf{G}_{\chi,||}(\mathbf{x},\mathbf{x'},ip)
=\boldsymbol{\delta_{||}}(\mathbf{x}-\mathbf{x'})\label{Gpar}
\end{eqnarray} with the unit longitudinal dyadic distribution $\boldsymbol{\delta_{||}}(\mathbf{x}-\mathbf{x'})=\sum_{\alpha}\hat{\mathbf{k}}_\alpha\otimes\hat{\mathbf{k}}_\alpha\Phi_\alpha^\ast(\mathbf{x'})\Phi_\alpha(\mathbf{x})$. We have also the important relations between the tensor $\mathbf{S}_\chi(\mathbf{x},\mathbf{x'},ip)$ and $\mathbf{G}_\chi(\mathbf{x},\mathbf{x'},ip)$:
\begin{eqnarray}
\mathbf{S}_\chi(\mathbf{x},\mathbf{x'},ip)=-\frac{p^2}{c^2}(1+\bar{\chi}(p))\mathbf{G}_{\chi,\bot}(\mathbf{x},\mathbf{x'},ip)\nonumber\\+\boldsymbol{\delta_{\bot}}(\mathbf{x}-\mathbf{x'})\nonumber\\
=-\frac{p^2}{c^2}(1+\bar{\chi}(p))\mathbf{G}_{\chi}(\mathbf{x},\mathbf{x'},ip)+\boldsymbol{\delta}(\mathbf{x}-\mathbf{x'})\nonumber\\=\boldsymbol{\nabla}\times\boldsymbol{\nabla}\times\mathbf{G}_\chi(\mathbf{x},\mathbf{x'},ip)\nonumber\\ \label{135}
\end{eqnarray}
 Now if we write $\overline{\mathbf{D'}}(\mathbf{x},p)=\overline{\mathbf{E'}}(\mathbf{x},p)+\overline{\mathbf{P'}}(\mathbf{x},p)=(1+\bar{\chi}(p))\overline{\mathbf{E'}}(\mathbf{x},p)+\overline{\mathbf{P'}}^{(0)}(\mathbf{x},p)$ and introduce the relation  $\overline{\mathbf{E'}}(\mathbf{x},p)=\overline{\mathbf{E'}}^{(0)}(\mathbf{x},p)+\overline{\mathbf{E'}}^{(s)}(\mathbf{x},p)$ with $\overline{\mathbf{E'}}^{(0)}(\mathbf{x},p)$ the free mode given by Eq.~\ref{120} and $\overline{\mathbf{E'}}^{(s)}(\mathbf{x},p)$ the source field, induced by $\overline{\mathbf{P}'}^{(0)}(\mathbf{x'},p)$, which is given by 
\begin{eqnarray}
\overline{\mathbf{E}'}^{(s)}(\mathbf{x},p)=
-\int d^3\mathbf{x'} \frac{p^2}{c^2}\mathbf{G}_\chi(\mathbf{x},\mathbf{x'},ip) \cdot
\overline{\mathbf{P}'}^{(0)}(\mathbf{x'},p).\nonumber\\ \label{cool2}
\end{eqnarray}
we get $\overline{\mathbf{D'}}^{(s)}(\mathbf{x},p)=(1+\bar{\chi}(p))\overline{\mathbf{E'}}^{(s)}(\mathbf{x},p)+\overline{\mathbf{P'}}^{(0)}(\mathbf{x},p)$, i.e., Eq.~\ref{rehh} as it could be checked directly after comparing Eq.~ \ref{S} with Eqs.~\ref{Gper} and \ref{Gpar}. Others important relations between the dyadic formalism and scalar Green function are given in Appendix D. \\ 
\indent Importantly, we can write all scattered field as a function of the rising and  lowering operators $\mathbf{f}^{\dagger(0)}_\omega(\mathbf{x},t)$, $\mathbf{f}^{(0)}_\omega(\mathbf{x},t)$. In order to give explicit expressions we use the Fourier transform notations (i.e., with $p=\gamma-i\omega$ with $\gamma\rightarrow 0^+$) and the relation~\cite{previous}
\begin{eqnarray}
\widetilde{\mathbf{P}}^{(0)}(\mathbf{x},\omega)=\int_0^{+\infty}d\omega'\sqrt{\frac{\hbar\sigma_{\omega'}(\mathbf{x})}{\pi\omega'}}[\mathbf{f}^{(0)}_{\omega'}(\mathbf{x},t_0)\nonumber\\ \cdot e^{i\omega't_0}\delta(\omega-\omega')
+\mathbf{f}^{\dagger(0)}_{\omega'}(\mathbf{x},t_0)e^{-i\omega't_0}\delta(\omega+\omega')]\nonumber\\ \label{36}\end{eqnarray}
We obtain:
\begin{eqnarray}
\mathbf{D}^{(s)}(\mathbf{x},t)=\sum_{\alpha,j}\int_{0}^{+\infty}d\omega\frac{\omega_\alpha^2\Phi_\alpha(\mathbf{x})\boldsymbol{\hat{\epsilon}}_{\alpha,j}}{\omega_\alpha^2-\tilde{\varepsilon}(\omega)\omega^2}\sqrt{\frac{\hbar\sigma_\omega}{\pi\omega}}f^{(0)}_{\omega,\alpha,j}(t)\nonumber\\
+hcc. \nonumber\\\label{rehhD}
\end{eqnarray}
\begin{eqnarray}
\mathbf{E}_\bot^{(s)}(\mathbf{x},t)=\sum_{\alpha,j}\int_{0}^{+\infty}d\omega\frac{\omega^2\Phi_\alpha(\mathbf{x})\boldsymbol{\hat{\epsilon}}_{\alpha,j}}{\omega_\alpha^2-\tilde{\varepsilon}(\omega)\omega^2}\sqrt{\frac{\hbar\sigma_\omega}{\pi\omega}}f^{(0)}_{\omega,\alpha,j}(t)\nonumber\\
+hcc. \nonumber\\\label{rehhE}
\end{eqnarray}where $f^{(0)}_{\omega,\alpha,j}(t)$ is a lowering operator associated with the transverse fluctuating field and defined as 
\begin{eqnarray}
f^{(0)}_{\omega,\alpha,j}(t)=\int d^3\mathbf{x'}\Phi_\alpha^\ast(\mathbf{x'})\boldsymbol{\hat{\epsilon}}_{\alpha,j}\cdot\mathbf{f}^{(0)}_\omega(\mathbf{x},t)
\end{eqnarray}
 such that from Eq.~\ref{33} we get the mode commutator 
\begin{eqnarray}[f^{(0)}_{\omega,\alpha,j}(t),f^{\dagger(0)}_{\omega',\beta,k}(t)]=\delta_{\alpha,\beta}\delta_{j,k}\delta(\omega-\omega')\label{commuting}
\end{eqnarray} and the harmonic time evolution $f^{(0)}_{\omega,\alpha,j}(t)=f^{(0)}_{\omega,\alpha,j}(t_0)e^{-i\omega_\alpha (t-t_0)}$ if we impose the initial condition  $f^{(0)}_{\omega,\alpha,j}(t_0)=f_{\omega,\alpha,j}(t_0)=\int d^3\mathbf{x'}\Phi_\alpha^\ast(\mathbf{x'})\boldsymbol{\hat{\epsilon}}_{\alpha,j}\cdot\mathbf{f}_\omega(\mathbf{x},t_0)$.\\
\indent For the longitudinal electric field we deduce  from Eq.~\ref{Gpar} and Eq.~\ref{cool2} $\overline{\mathbf{E'}}_{||}^{(s)}(\mathbf{x},p)=-\frac{\overline{\mathbf{P'}}_{||}^{(0)}(\mathbf{x},p)}{(1+\bar{\chi}(p))}$ (in agreement with the definition $\overline{\mathbf{D'}}_{||}(\mathbf{x},p)=0$) and therefore:
 \begin{eqnarray}
\mathbf{E}_{||}^{(s)}(\mathbf{x},t)=-\sum_{\alpha}\int_{0}^{+\infty}d\omega\frac{\Phi_\alpha(\mathbf{x})\boldsymbol{\hat{k}}_{\alpha}}{\tilde{\varepsilon}(\omega)}\sqrt{\frac{\hbar\sigma_\omega}{\pi\omega}}f^{(0)}_{\omega,\alpha,||}(t)\nonumber\\
+hcc. \nonumber\\\label{rehhEpar}
\end{eqnarray} We have $f^{(0)}_{\omega,\alpha,||}(t)=\int d^3\mathbf{x'}\Phi_\alpha^\ast(\mathbf{x'})\boldsymbol{\hat{k}}_{\alpha}\cdot\mathbf{f}^{(0)}_\omega(\mathbf{x},t)$ and the commutator $[f^{(0)}_{\omega,\alpha,||}(t),f^{\dagger(0)}_{\omega',\beta,||}(t)]=\delta_{\alpha,\beta}\delta_{j,k}\delta(\omega-\omega')$ and the time evolution $f^{(0)}_{\omega,\alpha,||}(t)=f^{(0)}_{\omega,\alpha,||}(t_0)e^{-i\omega_\alpha (t-t_0)}$ with similar initial condition as for the transverse field. Regrouping these definition we have obviously \begin{eqnarray}
\mathbf{f}^{(0)}_\omega(\mathbf{x},t)=\sum_{\alpha}\boldsymbol{\hat{k}}_{\alpha}\Phi_\alpha(\mathbf{x})f^{(0)}_{\omega,\alpha,||}(t)\nonumber\\+\sum_{\alpha,j}\boldsymbol{\hat{\epsilon}}_{\alpha,j}\Phi_\alpha(\mathbf{x})f^{(0)}_{\omega,\alpha,j}(t)\end{eqnarray}
\indent Furthermore, we can easily show  that we have also $\overline{\mathbf{B}'}^{(s)}(\mathbf{x},p)=\int d^3\mathbf{x'} \frac{p}{c}\boldsymbol{\nabla}\times\mathbf{G}_\chi(\mathbf{x},\mathbf{x'},ip)\cdot \overline{\mathbf{P}'}^{(0)}(\mathbf{x'},p)$ leading to  
\begin{eqnarray}
\mathbf{B}^{(s)}(\mathbf{x},t)=\sum_{\alpha,j}\int_{0}^{+\infty}d\omega\frac{\omega c\mathbf{k}_\alpha\times\boldsymbol{\hat{\epsilon}}_{\alpha,j}\Phi_\alpha(\mathbf{x})}{\omega_\alpha^2-\tilde{\varepsilon}(\omega)\omega^2}\nonumber\\ \cdot\sqrt{\frac{\hbar\sigma_\omega}{\pi\omega}}f^{(0)}_{\omega,\alpha,j}(t)
+hcc. \nonumber\\\label{rehhB}
\end{eqnarray} 
The description of the scattered field $(s)$ given here corresponds exactly to what Huttner and Barnett~\cite{Huttner1992a} called the quantized field obtained after generalizing the diagonalization procedure of Fano and Hopfield\cite{Fano1956,Hopfield1958}. Here we justify these modes by using the Laplace transform method and by taking the limit $t_0\rightarrow- \infty$ explicitly. This means that we neglect the contribution of the $(0)$ transverse field which is infinitely damped at time $t$. Importantly $\mathbf{P}^{(0)}(\mathbf{x},t)$ doesn't vanish since the time evolution of $\mathbf{f}^{(0)}_{\omega}(\mathbf{x},t)$ in Eq.~\ref{35} is harmonic. 
\subsection{A discussion on causality and time-symmetry}
\indent It is important to  further comment about causality and on the structure of the total field as a sum of $(0)$ and $(s)$ modes.  The $(0)$ (classical polariton) modes are indeed exponentially damped in the future direction meaning that a privileged temporal direction apparently holds in this model. This would mean that we somehow break the time symmetry of the problem. However, since the evolution equations are fundamentally time symmetric  this should clearly not  be possible. Similarly, propagators such as $\mathbf{G}_\chi(\mathbf{x},\mathbf{x'},ip)$  are also spatially damped at large distance,  see Eq.~\ref{103}, since we have terms like $\sim\frac{e^{i\omega\sqrt{\tilde{\varepsilon}(\omega)}|\mathbf{x}-\mathbf{x'}|/c}}{4\pi |\mathbf{x}-\mathbf{x'}|}$. This also seems to imply a privileged time direction and would lead to a kind of paradox.  However, we should remind that only the sum $(0)$ + $(s)$ has a physical meaning and this sum must preserves time symmetry. Indeed, we remind that time reversal applied to electrodynamics implies that  if  $\mathbf{E}(t)$, $\mathbf{B}(t)$, and $\mathbf{X}_\omega(t)$ is a solution of the coupled set of equations given in section II then the time reversed solutions~\cite{Jackson1999}  $\mathbf{E}_T(t)=\mathbf{E}(-t)$, $\mathbf{B}_T(t)=-\mathbf{B}(-t)$, and $\mathbf{X}_{\omega,T}(t)=\mathbf{X}_{\omega}(-t)$ is also defining  a solution of the same equations (we have also $\mathbf{F}_T(t)=\mathbf{F}(-t)$ and $\mathbf{P}_T(t)=\mathbf{P}(-t)$). Now, considering the dipole density evolution we get from Eq.~\ref{18} after some manipulations:  
 \begin{eqnarray}
\mathbf{P}_T(\mathbf{x},t)=\mathbf{P}_T^{(0)}(\mathbf{x},t)+\int_{t+t_0}^{0}\chi(\mathbf{x},-\tau)d\tau\mathbf{E}_T(\mathbf{x},t-\tau)\nonumber\\ \label{18bis}
\end{eqnarray} where $\mathbf{P}_T^{(0)}$ is defined as $\mathbf{P}^{(0)}$ (see  the definition Eq.~\ref{35}) but with $\mathbf{X}^{(0)}_\omega(\mathbf{x},t)=\cos{(\omega (t-t_0))}\mathbf{X}_\omega(\mathbf{x},t_0)+\frac{\sin{(\omega (t-t_0))}}{\omega}\partial_t\mathbf{X}_\omega(\mathbf{x},t_0)$ replaced by $\mathbf{X}^{(0)}_{\omega,T}(\mathbf{x},t)=\cos{(\omega (t+t_0))}\mathbf{X}_{\omega,T}(\mathbf{x},-t_0)+\frac{\sin{(\omega (t+t_0))}}{\omega}\partial_t\mathbf{X}_{\omega,T}(\mathbf{x},-t_0)$. The presence of the time $-t_0$ everywhere has a clear meaning. Indeed, from special relativity choosing  a time reference frame  with $t'=-t$  (passive  time reversal transformation) implies that the causal evolution of $\mathbf{P}$  defined for $t\geq t_0$ will become an anticausal evolution defined for $t'\leq t'_0=-t_0$. Going back to the active time reversal transformation  Eq.~\ref{18bis} we see that the new  linear susceptibility $\chi_T(\mathbf{x},\tau)=\chi(\mathbf{x},-\tau)$ is given by  
\begin{eqnarray}
\chi(\mathbf{x},-\tau)=-\int_{0}^{+\infty}d\omega\frac{2\sigma_\omega(\mathbf{x})}{\pi}\frac{\sin{\omega\tau}}{\omega}\Theta(-\tau)\label{20bis}
\end{eqnarray} ($\Theta(-\tau)$ is explicitly written in order to emphasize the anticausal structure.
However, since we have $\widetilde{\chi_T}(\omega)=\int_{-\infty}^{+\infty}\frac{dt}{2\pi}e^{i\omega t} \chi_T(t)=\int_{-\infty}^{+\infty}\frac{dt}{2\pi}e^{i\omega t} \chi(-t)$ and $\chi(-t)=\chi^\ast(-t)$ 
we have $\widetilde{\chi_T}(\omega)=\int_{-\infty}^{+\infty}\frac{dt}{2\pi}e^{i\omega t}\chi^\ast(-t)=\int_{-\infty}^{+\infty}\frac{du}{2\pi}e^{-i\omega u}\chi^\ast(u)=\widetilde{\chi}^\ast(\omega)$. This means that the new permittivity (with poles in the upper half complex frequency space) is now associated with growing anticausal modes since $\widetilde{\varepsilon_{T}}''(\omega)<0$. The (active) time reversed evolution is defined for $t<-t_0$ so that we indeed  get modes decaying into the past direction while growing into the future direction. This becomes even more clear for the time reversal evolution of the electromagnetic field  given by the separation in $(0)$ and $(s)$ modes. The time reversal applied on the $(0)$ field  such as $\mathbf{E}_T^{(0)}(t)$ corresponding to classical polaritons  is now  involving  frequency $-\Omega_{\alpha,m}^{(-)}$ instead  of $\Omega_{\alpha,m}^{(-)}$. This leads to reversed temporal evolution such as $e^{i\Omega_{\alpha,m}^{(-)}(t+t_0)}$ and  
$e^{-i\Omega_{\alpha,m}^{(-)\ast}(t+t_0)}$ associated with growing waves in the future direction (since the  new poles are now in the upper half complex frequency space). More generally, we have shown (see Appendix C) that the full evolution of either $(0)$ or $(s)$ fields is completely defined by the knowledge of the Green function and propagators such as \begin{equation}\Delta_\chi(\tau,|\mathbf{x}-\mathbf{x'}|)=c^2\sum_{\alpha}H_\alpha(\tau)\Phi_\alpha^\ast(\mathbf{x'})\Phi_\alpha(\mathbf{x})\end{equation}  (see Eq.~\ref{106}) with the causal function $H_\alpha(\tau)$ given by the Bromwich Fourier integral  $\int_{-\infty}^{+\infty}\frac{d\omega}{2\pi}\frac{e^{-i\omega\tau}}{\omega_\alpha^2-\tilde{\varepsilon}(\omega)\omega^2}$. Now, like for the susceptibility $\chi_T(t)$ time reversal leads to a new propagator    
$\Delta_{\chi,T}(\tau,|\mathbf{x}-\mathbf{x'}|)=\Delta_\chi(-\tau,|\mathbf{x}-\mathbf{x'}|)$ and therefore by a reasoning equivalent to the previous one 
to \begin{equation}\Delta_{\chi,T}(\tau,|\mathbf{x}-\mathbf{x'}|)=c^2\sum_{\alpha}H^\ast_\alpha(-\tau)\Phi_\alpha^\ast(\mathbf{x'})\Phi_\alpha(\mathbf{x})\end{equation} involving the complex conjugate $H^\ast_\alpha(-\tau)$ with a Fourier expansion $\int_{-\infty}^{+\infty}\frac{d\omega}{2\pi}\frac{e^{-i\omega\tau}}{\omega_\alpha^2-\tilde{\varepsilon}^\ast(\omega)\omega^2}$ which again involves the anticausal permittivity $\widetilde{\varepsilon}^\ast(\omega)$. This naturally leads to growing waves in the future direction and to Green function spatially growing as$\sim\frac{e^{i\omega\sqrt{\tilde{\varepsilon}\ast(\omega)}|\mathbf{x}-\mathbf{x'}|/c}}{4\pi |\mathbf{x}-\mathbf{x'}|}$. The full equivalence between the two representation of the total field $(s)+(0)$ is completed if first we observe that the initial  conditions at $t_0$ (i.e., $X_\omega(t_0)$, $\mathbf{E}(t_0)$, etc...) are now replaced by 'final' boundary condition at $t_f=-t_0$ (i.e., $X_{\omega,T}(-t_0)$, $\mathbf{E}_T(-t_0)$, etc...). Second, since
the $-t_0$ value is arbitrary we can send it into the remote future if we want and we will have thus a evolution expressed in term of growing modes for all times $t\leq t_f$.  The value of the field at such a boundary is of course arbitrary, so that if we want the two representation can describe the same field if for the final field at $t_f$ we take the field evolving from $t_0$ using the usual  causal evolution from past to future.  This equivalence is of course reminiscent from  the dual representations obtained using  either retarded or advanced waves~\cite{Feynman,Davies,Zeh}.  Indeed, the total electric field  $\mathbf{E}$ can always be separated into $\mathbf{E}_{in}+\mathbf{E}_{ret}$ or equivalently  $\mathbf{E}_{out}+\mathbf{E}_{adv}$ where $ret$ and  $adv$ label the retarded and advanced source fields respectively and $in$ and $out$ label the homogeneous `free' fields coming from the  past and from the future respectively. This implies that only a specific choice of boundary conditions in the past or future  can lead to a completely causal evolution and therefore  time symmetry in not broken in the evolution equations but only through a specification of the boundary conditions. In other worlds, in agreement with the famous Loschmidt and Poincar\'e objections to Boltzmann, strictly deriving  time irreversibility from an intrinsically time reversible dynamics is obviously impossible without additional postulates. This point  was fully recognized already by Boltzmann long ago and was the basis for his statistical interpretation  of the second law of thermodynamics~\cite{Zeh}.\\    
\subsection{The Fano-Hopfield diagonalization procedure}
In order to further understand the implication of the Huttner-Barnett model~\cite{Huttner1992a} we should discuss how the full Hamiltionan finally reads in this formalism. This is central since the Langevin noise equations developed by Gruner and Welsch~\cite{Gruner1996} use only the $H_M^{(0)}(t)$ Hamitonian. In the Huttner-Barnett model the full dynamics is determined by the complete knowledge of the dipole density $\textbf{P}^{(0)}(t)$ so that the results of Gruner and Welsch~\cite{Gruner1996} should be in principle justifiable. However, if we are not careful, this will ultimately  break unitary and time-symmetry. In order to understand the physical mechanism at work one should first clarify the relation existing between the Hutner-Barnett model, using $\textbf{P}^{(0)}(t)$ as a fundamental field, and the historical Hopfield-Fano approach~\cite{Fano1956,Hopfield1958} for defining discrete polariton modes as normal coordinate solutions of the full Hamiltonian.\\    
\indent We remind that the historical method for dealing with polariton is indeed based on the pioneer work by Fano and Hopfield~\cite{Fano1956,Hopfield1958} for diagonalizing  the full Hamiltonian. The procedure is actually reminiscent of the classical problem of finding normal coordinates and normal (real valued) eigen frequencies  associated with free vibrations of a system of linearly coupled harmonic oscillators. The classical diagonalization method~\cite{Goldstein} relies on the resolution of secular equations already studied by Laplace.  Actually, the first `semiclassical' treatment  made by Born and Huang~\cite{Huang1951} is mathematically correct and equivalent to the one made later by  Fano and Hopfield even though the relation between those formalisms is somehow hidden behind the mathematical symbols.
 The Full strategy for finding normal coordinates and frequencies becomes clear if we use Fourier transforms of the various fields for the problem under study. Indeed, a Fourier expansion of a field component $A(t)=\int_{-\infty}^{+\infty} d\Omega \widetilde{A}(\Omega)e^{-i\Omega} $ will  obviously define the needed harmonic expansion. In the problems considered by Born, Huang, Fano, and Hopfield,  the Fourier spectrum $\widetilde{A}(\Omega)$ is a sum of Dirac distributions $\delta(\Omega\mp\Omega_n)$ peaked on the real valued eigenvibrations $\Omega_n$. This specific situation needs a complete discussion since the Hopfield-Fano model~\cite{Fano1956,Hopfield1958} leads to an exact diagonalization of the full Hamiltonian $H(t)$. This will in turn makes clear some fundamental relations with the Laplace Transform formalism used in the previous subsections.\\
\indent We start with the Fourier transformed equations 
\begin{eqnarray}
\boldsymbol{\nabla}\times\boldsymbol{\nabla}\times\widetilde{\textbf{E}}(\Omega)-\frac{\Omega^2}{c^2}\widetilde{\textbf{E}}(\Omega)=\frac{\Omega^2}{c^2}\widetilde{\textbf{P}}(\Omega)\label{A}
\end{eqnarray} 
\begin{eqnarray}
(\omega^2-\Omega^2)\widetilde{\textbf{X}_\omega}(\Omega)=\sqrt{\frac{2\sigma_\omega}{\pi}}\widetilde{\textbf{E}}(\Omega)\label{B}
\end{eqnarray}  and 
  \begin{eqnarray}
\int_0^{+\infty}d\omega\sqrt{\frac{2\sigma_\omega}{\pi}}\widetilde{\textbf{X}_\omega}(\Omega)=\widetilde{\textbf{P}}(\Omega)\label{C}
\end{eqnarray}  With the constraint $\widetilde{\textbf{E}}(\Omega)=\widetilde{\textbf{E}}(-\Omega)^\ast$, $\widetilde{\textbf{X}_\omega}(\Omega)=\widetilde{\textbf{X}_\omega}(-\Omega)^\ast$ coming from the real valued nature of the fields.     From Eq.~\ref{B} we get using the properties of distributions:
 \begin{eqnarray}
\widetilde{\textbf{X}_\omega}(\Omega)=P[\frac{1}{\omega^2-\Omega^2}]\sqrt{\frac{2\sigma_\omega}{\pi}}\widetilde{\textbf{E}}(\Omega)+\widetilde{\textbf{X}_\omega}^{(sym)}(\Omega)\label{D}
\end{eqnarray}    with
\begin{eqnarray}
\widetilde{\textbf{X}_\omega}^{(sym)}(\Omega)=\sqrt{\frac{\hbar}{2\omega}}[f_\omega^{(sym)}\delta(\omega-\Omega)+f_\omega^{(sym)\ast}\delta(\omega+\Omega)]\nonumber\\ \label{E}
\end{eqnarray}
and where we used the reality constraint and introduced  constants of motions $f_\omega^{(sym)}$, $f_\omega^{(sym)\ast}$ which will becomes annihilation and creation operators in the second quantized formalism.  The principal value can be conveniently written $P[\frac{1}{\omega^2-\Omega^2}]=\frac{1}{\omega^2-(\Omega+i0^+)^2}-\frac{i\pi}{2\omega}[\delta(\omega-\Omega)-\delta(\omega+\Omega)]$ or equivalently $P[\frac{1}{\omega^2-\Omega^2}]=\frac{1}{\omega^2-(\Omega-i0^+)^2}+\frac{i\pi}{2\omega}[\delta(\omega-\Omega)-\delta(\omega+\Omega)]$. This leads to two different representations of Eq.~\ref{D}:  
\begin{eqnarray}
\widetilde{\textbf{X}_\omega}(\Omega)= \frac{1}{\omega^2-(\Omega+i0^+)^2}\sqrt{\frac{2\sigma_\omega}{\pi}}\widetilde{\textbf{E}}(\Omega)+\widetilde{\textbf{X}_\omega}^{(in)}(\Omega)\nonumber\\
= \frac{1}{\omega^2-(\Omega-i0^+)^2}\sqrt{\frac{2\sigma_\omega}{\pi}}\widetilde{\textbf{E}}(\Omega)+\widetilde{\textbf{X}_\omega}^{(out)}(\Omega) \nonumber\\  \label{F}
\end{eqnarray} where $\widetilde{\textbf{X}_\omega}^{(in)}(\Omega)$ and $\widetilde{\textbf{X}_\omega}^{(out)}(\Omega)$ can also be written like  Eq.~\ref{E}, i.e., respectively as $\sqrt{\frac{\hbar}{2\omega}}[f_\omega^{(in)}\delta(\omega-\Omega)+f_\omega^{(in)\ast}\delta(\omega+\Omega)]$ or $\sqrt{\frac{\hbar}{2\omega}}[f_\omega^{(out)}\delta(\omega-\Omega)+f_\omega^{(out)\ast}\delta(\omega+\Omega)]$. This discussion is reminiscent of the different representation given in Section III D involving retarded, advanced or time symmetric modes.  Of course the usual causal representation is  $(in)$ which is obtained from the definition of  $f_\omega^{(0)}(t)$ given in section III at the limit $t_0\rightarrow-\infty$. However, all the descriptions are rigorously  equivalent. It is also easy to show that we have $f_\omega^{(sym)}=\frac{f_\omega^{(in)}+f_\omega^{(out)}}{2}$ which justifies why we called this field symmetrical. It corresponds to a representation of the problem mixing boundary conditions in the future and the past in a symmetrical way like it was used for instance by Feynman and Wheeler in their description of electrodynamics~\cite{Feynman} as discussed in Section III D.\\
\indent Now after inserting the causal representation of Eq.~\ref{F} in Eq.~\ref{C} and then into Eq.~\ref{A} we get
 \begin{eqnarray}
\boldsymbol{\nabla}\times\boldsymbol{\nabla}\times\widetilde{\textbf{E}}(\Omega)-\frac{\Omega^2}{c^2}\widetilde{\varepsilon}(\Omega)\widetilde{\textbf{E}}(\Omega)=\frac{\Omega^2}{c^2}\widetilde{\textbf{P}}^{(in)}(\Omega)\label{G}
\end{eqnarray}  with $\widetilde{\textbf{P}}^{(in)}(\Omega)=\int_0^{+\infty}d\omega\sqrt{\frac{2\sigma_\omega}{\pi}}\widetilde{\textbf{X}_\omega}^{(in)}(\Omega)$ and 
\begin{eqnarray}
\widetilde{\varepsilon}(\omega)=1+\int_0^{+\infty}d\tau\chi(\tau)e^{i\omega \tau} \label{perm}
\end{eqnarray}
This causal permittivity $\widetilde{\varepsilon}(\Omega)$ being given by the Huttner-Barnett model \cite{Huttner1992a,previous}  the secular equations $\omega_\alpha^2=\Omega^2\widetilde{\varepsilon}(\Omega)$ for transverse modes have no root in the upper complex frequency half-plane and in particular on the real frequency axis (the longitudinal term is discussed below). This means that, unlike Eq.~\ref{D}, Eq.~\ref{G} has in general no Dirac term corresponding to  independent eigenmodes.  The electric field is thus  represented by a source term
\begin{eqnarray}
\widetilde{\textbf{E}}(\mathbf{x},\Omega)=\frac{\Omega^2}{c^2}\int d^3\mathbf{x'} \mathbf{G}_\chi(\mathbf{x},\mathbf{x'},\Omega)\cdot
\widetilde{\mathbf{P}}^{(in)}(\mathbf{x'},\Omega)\label{glopiglopa}
\end{eqnarray}  obtained like in the previous subsection using a causal Green function. The absence of free normal modes for the electric field is of course reminiscent from the rapid decay of the free modes $(0)$ when $t_0\rightarrow-\infty$ as discussed before. The representation given here doesn't distinguish between transverse and longitudinal fields but this should naturally occur since we have the constraint $\boldsymbol{\nabla}\cdot\widetilde{\textbf{E}}(\Omega)=-\boldsymbol{\nabla}\cdot\widetilde{\textbf{P}}(\Omega)$ which implies $\widetilde{\textbf{E}}_{||}(\Omega)=-\widetilde{\textbf{P}}_{||}(\Omega)$. Together with Eq.~\ref{A} we thus get 
\begin{eqnarray}\widetilde{\textbf{E}}_{||}(\Omega)=-\widetilde{\textbf{P}}_{||}(\Omega)=-\frac{\widetilde{\textbf{P}}_{||}^{(in)}(\Omega)}{\widetilde{\varepsilon}(\Omega)}.\label{long}\end{eqnarray}
Eq.~\ref{long} is actually reminiscent of the charge screening by $\widetilde{\varepsilon}(\Omega)$. Here we used the fact that $\widetilde{\varepsilon}(\Omega)$ has no root on the real axis otherwise the imaginary part of  the permittivity should vanish and the the medium would be lossless at the frequency $\Omega$ a fact which is prohibited by physical consideration about irreversibility~\cite{Lifshitzbook}. This reasoning  is rigorously not valid  at $\Omega=0$  since the imaginary part of the permittivity  is a odd function on the real axis. But then in general to have a root at $\Omega=0$ it would require that the real part  of the permittivity vanishes as well and this is not allowed from usual permittivity model  (see Eq.~\ref{B4}) which makes therefore this possibility very improbable.\\ 
\indent The previous reasoning is clearly formally equivalent to the ones obtained in the previous subsection and in both case the field $\widetilde{\mathbf{P}}^{(in)}(\mathbf{x'},\Omega)$ or $\textbf{P}^{(0)}(t)$ completely determine the electromagnetic evolution. Still, there are exceptions for instance in the Drude Lorentz model with $\widetilde{\varepsilon}(\Omega)=1+\frac{\omega_p^2}{\omega_0^2-(\Omega+i0^+)^2}$ which forms the basis for the Hopfield~\cite{Hopfield1958} polariton model. This model is rigorously not completely lossless since we have  $\widetilde{\varepsilon}(\Omega)=1+P[\frac{\omega_p^2}{\omega_0^2-\Omega^2}]+\frac{i\pi\omega_p^2}{2\omega_0}(\delta(\Omega-\omega_0)-\delta(\Omega+\omega_0))$ corresponding to a singular absorption peak.\\
\indent Moreover, in this Hopfield model~\cite{Hopfield1958}, which is a limit case of the Huttner Barnett model~\cite{model,Huttner1992a}, we get the exact evolution equation \begin{equation}\partial_t^2\textbf{P}+\omega_0^2\textbf{P}=\omega_p^2\textbf{E}\end{equation} which in the case of the longitudinal  modes leads to solving the secular equation $(\omega_0^2-\Omega^2)\widetilde{\textbf{P}}_{||}(\Omega)=-\omega_p^2\widetilde{\textbf{P}}_{||}(\Omega)$. It has the solution   
\begin{equation}\widetilde{\textbf{P}}_{||}(\Omega)=\boldsymbol{\beta}\delta(\Omega-\omega_L)+\boldsymbol{\beta}^\ast\delta(\Omega+\omega_L)\end{equation} where $\omega_L=\sqrt{\omega_0^2+\omega_p^2}$ isthe longitudinal plasmon frequency. However relating this result to Eq.~\ref{long} requires careful calculations since  $\widetilde{\varepsilon}(\Omega)$ is here a highly singular distribution. Indeed, applying Eq.~\ref{long} will lead to find solutions of 
$\widetilde{\varepsilon}(\Omega)\widetilde{\textbf{P}}_{||}(\Omega)=\widetilde{\textbf{P}}_{||}^{(in)}(\Omega)$  with 
$\widetilde{\textbf{P}}_{||}^{(in)}(\Omega)=\boldsymbol{\alpha}\delta(\Omega-\omega_0)+\boldsymbol{\alpha}^\ast\delta(\Omega+\omega_0)$ with $\boldsymbol{\alpha}(\mathbf{x})$ a longitudinal vector field. This singular $(in)$ field at $\Omega=\pm\omega_0$ seems to imply that polaritons are resonant at such frequency in contradiction with the result leading to $\Omega=\pm\omega_L$ for such polariton modes. Now, due to the presence of absorption peaks at $\Omega=\pm\omega_0$ we have near this singular points  $\boldsymbol{\alpha}=\frac{i\pi\omega_p^2}{2\omega_0}\widetilde{\textbf{P}}_{||}(\omega_0)$ where $\widetilde{\textbf{P}}_{||}(\omega_0)$ is supposed to be regular~\cite{footnotebis}. Moreover,  outside this narrow absorption band the medium is effectively lossless  and  instead of Eq.~\ref{long} we have  $(1+\frac{\omega_p^2}{\omega_0^2-\Omega^2})\widetilde{\textbf{P}}_{||}(\Omega)=0$ which has the singular solution $\widetilde{\textbf{P}}_{||}(\Omega)=\boldsymbol{\beta}\delta(\Omega-\omega_L)+\boldsymbol{\beta}^\ast\delta(\Omega+\omega_L)$ with $\boldsymbol{\beta}(\mathbf{x})$ a longitudinal vector field . Importantly, from the hypothesis of regularity  at $\pm\omega_0$ we have  $\widetilde{\textbf{P}}_{||}(\omega_0)=0$ and therefore $\boldsymbol{\alpha}=0$ which means that $\widetilde{\textbf{P}}_{||}^{(in)}(\Omega)=0$ everywhere. The Lorentz-Drude model leads therefore to genuine longitudinal polaritons eigenfrequencies $\pm\omega_L$ solutions of  $\widetilde{\varepsilon}(\omega_L)=0$. We emphasize that the same result could be obtained using the Laplace transform method. We have indeed 
$\overline{\mathbf{E'}}_{||}^{(s)}(\mathbf{x},p)=-\frac{\overline{\mathbf{P'}}_{||}^{(0)}(\mathbf{x},p)}{1+\omega_p^2/(p^2+\omega_0^2)}$ with  $\overline{\mathbf{P'}}_{||}^{(0)}(\mathbf{x},p)=
\frac{p\mathbf{P}_{||}(\mathbf{x},t_0)+\partial_{t_0}\mathbf{P}_{||}(\mathbf{x},t_0)}{\omega_0^2+p^2}$.  Therefore we can rewrite the  longitudinal scattered field as $\overline{\mathbf{E'}}_{||}^{(s)}(\mathbf{x},p)=-\frac{p\mathbf{P}_{||}(\mathbf{x},t_0)+\partial_{t_0}\mathbf{P}_{||}(\mathbf{x},t_0)}{\omega_L^2+p^2}$.  Moreover since there is no longitudinal $(0)$ electric field and since $\textbf{P}_{||}=-\textbf{E}_{||}$ we have $\textbf{P}_{||}(t)=\cos{(\omega_L(t-t_0))}\mathbf{P}_{||}(t_0)+\sin{(\omega_L(t-t_0))}\partial_{t_0}\mathbf{P}_{||}(t_0)$ which shows that the genuine longitudinal polariton oscillates at the frequency $\omega_L$ as expected.\\
\indent The transverse polariton modes of the Hopfield model are obtained in a similar way from Eq.~\ref{G} with  $\widetilde{\textbf{P}}_{\bot}^{(in)}(\Omega)=\boldsymbol{\gamma}\delta(\Omega-\omega_0)+\boldsymbol{\gamma}^\ast\delta(\Omega+\omega_0)$ with $\boldsymbol{\gamma}(\mathbf{x})$ a transverse vector field. 
 As explained in the appendix G solving the problem with a plane wave expansion labeled by $\alpha$ and $j$ leads again to a secular equation $\omega_\alpha^2-\Omega_{\alpha,\pm}2\widetilde{\varepsilon}(\Omega_{\alpha,\pm})=0$ for the two transverse modes ($\pm$) giving a quartic dispersion relation
\begin{eqnarray}
\omega_\alpha^2\omega_0^2-\Omega_{\alpha,\pm}^2(\omega_\alpha^2+\omega_L^2)+\Omega_{\alpha,\pm}^4=0
\end{eqnarray}   with two usual Hopfield solutions \begin{eqnarray}\Omega_\pm=\frac{\sqrt{[\omega_\alpha^2+\omega_L^2\pm\sqrt{((\omega_\alpha^2+\omega_L^2)^2-4\omega_\alpha^2\omega_0^2)}]}}{\sqrt{2}}.\nonumber\\ \end{eqnarray} The two-modes field have now the structure  
\begin{eqnarray}
\widetilde{\textbf{E}}(\mathbf{x},\Omega)=\sum_{\alpha,j,\pm}\widetilde{E}_{\alpha,j,\pm}(\Omega)\boldsymbol{\hat{\epsilon}}_{\alpha,j}\Phi_\alpha(\mathbf{x}) 
\end{eqnarray} with $\widetilde{E}_{\alpha,j,\pm}(\Omega)=\phi_{\alpha,j,\pm}\delta(\Omega-\Omega_{\alpha,\pm})+\eta_j\phi_{-\alpha,j,\pm}^\ast\delta(\Omega+\Omega_{\alpha,\pm})$ with $\phi_{\alpha,j,\pm}$ an amplitude coefficient for the mode (see Appendix E and \cite{footnote} for a derivation).\\
\indent One of the most important issue in the context of the Hopfield model concerns the Hamiltonian. Indeed, in this model the full Hamiltonian Eq.~\ref{10}  $H(t)=\int d^3\mathbf{x}\frac{\mathbf{B}^2+\mathbf{E}^2}{2}+H_M$ reads 
\begin{eqnarray}
H(t)=\int d^3\mathbf{x}[\frac{\mathbf{B}^2+\mathbf{E}^2}{2}+\frac{(\partial_t\mathbf{P})^2+\omega_0^2\mathbf{P}^2}{2\omega_p^2}].\label{polaritonhopfield}
\end{eqnarray}
If we isolate first the longitudinal term we get 
 \begin{eqnarray}
H_{||}(t)=\int d^3\mathbf{x}[\frac{\mathbf{P}_{||}^2}{2}+\frac{(\partial_t\mathbf{P}_{||})^2+\omega_0^2\mathbf{P}_{||}^2}{2\omega_p^2}]\nonumber\\=\int d^3\mathbf{x}[\frac{(\partial_t\mathbf{P}_{||})^2+\omega_L^2\mathbf{P}_{||}^2}{2\omega_p^2}]=2\frac{\omega_L^2}{\omega_p^2}\int d^3\mathbf{x}\boldsymbol{\beta}^\ast\boldsymbol{\beta}
\end{eqnarray}
We can of course introduce a Fourier transform of the dipole field as $\mathbf{P}_{||}=\sum_{\alpha}P_{\alpha}(\Omega)\boldsymbol{\hat{k}}_{\alpha}\Phi_\alpha(\mathbf{x})$  and new polariton fields amplitudes $\beta_\alpha=\int d^3\mathbf{x}\boldsymbol{\beta}(\mathbf{x})\cdot\boldsymbol{\hat{k}}_{\alpha}\Phi_\alpha(\mathbf{x})$. We thus have $H_{||}(t)=2\frac{\omega_L^2}{\omega_p^2}\sum_\alpha\beta_\alpha^\ast\beta_\alpha$. This expression of the Hamiltonian is standard for normal coordinates expansion in linearly coupled harmonic oscillators.\\
\indent Furthermore, using commutators like Eq.~\ref{33} one deduce $[\mathbf{P}(\mathbf{x},t),\partial_t \mathbf{P}(\mathbf{x}',t)]=i\hbar\delta^3(\mathbf{x}'-\mathbf{x})\textbf{I}$ and other similar ones.   In the Fourier space we thus obtain 
$[P_\alpha(t),\dot{P}_\beta(t)]=i\hbar\delta_{\alpha,\beta}$ which lead after straightforward transformation to the commutators $[f_{\alpha,||}(t),f_{\beta,||}^\dagger(t)]=\delta_{\alpha,\beta}$, $[f_{\alpha,||}(t),f_{\beta,||}(t)]=[f_{\alpha,||}^\dagger(t),f_{\beta,||}^\dagger(t)]=0.$ with $\beta_\alpha=\omega_p\sqrt{\frac{\hbar}{2\omega_L}}f_{\alpha,||}$ (the time dependence in the Heisenberg picture means $f_{||,\beta}(t)=f_{||,\beta}e^{-i\omega_L t}$).  This naturally leads to the Hopfield-Fano Hamiltonian expansion for longitudinal polaritons: 
\begin{eqnarray}
H_{||}(t)=\hbar\omega_L\sum_\alpha f_{\alpha,||}^\dagger(t) f_{\alpha,||}(t).
\end{eqnarray}
 \indent A similar analysis can be handled for the transverse polaritons modes but the calculation is a bit longer (see Appendix E). To summarize this calculation in few words: using a Fourier expansion of the different  primary transverse field operators  in Eq.~\ref{polaritonhopfield} we get after some manipulations the Hopfield-Fano expansion~\cite{Fano1956,Hopfield1958}  
\begin{eqnarray}
H_{\bot}(t)=2\sum_{\alpha,j,\pm}(1+\frac{\omega_0^2}{\omega_p^2}(\frac{\omega_\alpha^2}{\Omega_{\alpha,\pm}^2}-1)^2)\phi_{\alpha,j,\pm}^\dagger\phi_{\alpha,j,\pm}\nonumber\\
=\sum_{\alpha,j,\pm}\hbar\Omega_{\alpha,\pm}\alpha f_{\alpha,j,\pm}^\dagger(t) f_{\alpha,j\pm}(t)\end{eqnarray} with the operator
$f_{\alpha,j,\pm}(t)=f_{\alpha,j,\pm}e^{-i\Omega_{\alpha,\pm}t}$ obeying the usual commutation rules for rising and lowering operators. Again this result is expected in a modal expansion using normal coordinates and again the same result could be alternatively obtained using the Laplace transform method. 
\indent To summarize, the approach developed previously using the Laplace transform formalism agrees with the normal coordinates methods based on the Fourier expansion in the frequency domain.    Both approaches lead to the conclusion that for an homogeneous medium the  various electromagnetic and material fields are completely determined by the knowledge of the matter oscillating dipole density  $\textbf{P}^{(in)}(t)$ (Fourier's method) or $\textbf{P}^{(0)}(t)$ (Laplace's method).  In the limit  $t_0\rightarrow-\infty$  both approach are equivalent   and there is no contribution of the free field in a homogenous dissipative medium (the residual $\textbf{E}^{(0)}$,$\textbf{B}^{(0)}$ is exponentially damped in the regime $t_0\rightarrow-\infty$).  We also showed that if losses in the Huttner Barnett model are sharply confined in the frequency domain we can find exact polaritonic modes which agree with the historical method developed by Fano and Hopfield~\cite{Fano1956,Hopfield1958}. These modes fully diagonalize  the Hamitonian $H(t)$. While we focused our study on the particular Drude Lorentz model the result is actually generalizable~\cite{Kumar1993} to homogenous media with conductivity $\sigma(\Omega)=\sum_n\frac{\pi\omega_{p,n}^2}{2}(\delta(\Omega-\omega_{0,n})+\delta(\Omega+\omega_{0,n}))$ which lead to a permittivity  
\begin{eqnarray}
\widetilde{\varepsilon}(\Omega)=1+\sum_n P[\frac{\omega_{p,n}^2}{\omega_{0,n}^2-\Omega^2}]+i\frac{\sigma(\Omega)}{\Omega}.\label{dangerous}
\end{eqnarray}  In particular the Hamiltonian can in these special cases be written as a sum of harmonic oscillator terms corresponding to the different longitudinal and transversal polariton modes. We thus write $H_{||}=\sum_{m,\alpha}\hbar\Omega_{\alpha,m} f_{\alpha,m,||}^\dagger f_{\alpha,m,||}$ and $H_\bot=\sum_{m',\alpha,j}\hbar\Omega'_{\alpha,m'} f_{\alpha,j,m'}^\dagger f_{\alpha,j,m'}$ where $m$ and $m'$ label the discrete longitudinal and transverse polaritons modes. However, for a more general  Huttner-Barnett model where the permittivity is only constrained  by Kramers-Kronig relations such a simple interpretation is not possible and the Hamiltonian is not fully diagonalized. The additional physical requirement~\cite{Lifshitzbook} imposing that the imaginary part of the permittivity should be rigorously positive valued, i.e., $\widetilde{\varepsilon}''(\Omega)>0$, also prohibits these exceptional cases  which should therefore only appear as ideal limits with loss confined in infinitely narrow absorption bands. However as we will show in the next subsection the lossless idealization represents a good approximation for a quite general class of medium with weak dissipation. The previous results of Hopfield and Fano have still a physical meaning and are for example used with success for the description of planar cavity polaritons~\cite{Tassone1990,Gerace2007,Juzeliunas1996,Todorov2014,Todorov2015,Tokman2015,Bhat2006}. \\
\subsection{The approximately transparent medium case: Milonni's approach}
\indent It is particularly relevant to consider what happens in the Huttner Barnett approach if we relax a bit the demanding  constraints of the original Hopfield-Fano model based on Eq.~\ref{dangerous}. For this we consider a medium with low loss such as the medium can be considered with a good approximation as transparent in a given spectral band where the field is supposed to be limited. This approach was introduced by Milonni~\cite{Milonni1995} and is based on the Hamiltonian obtained long ago by Brillouin~\cite{Brillouin} and later by Landau and Lifschitz~\cite{Lifshitzbook,Ginzburg,Rosa2010} for dispersive  but slowly absorbing media.  The main idea is to replace the electromagnetic  energy density $(\textbf{E}^2+\textbf{B}^2)/2$ in the full Hamiltonian by a term like    
$(\frac{d\omega_c\tilde{\varepsilon}(\omega_c)}{d\omega_c}\textbf{E}^2+\textbf{B}^2)/2$ where $\tilde{\varepsilon}(\omega_c)$ is the approximately real valued permittivity of the field at the central pulsation $\omega_c$ with which the wave-packet propagates. Since this approach has been successfully applied to quantize polaritons~\cite{Constantinou1993,Al1996,Garrison2004,Rosa2010} or surface plasmons~\cite{Archambault2010,Yang2015} it is particularly interesting to justify it in the context of the more rigorous Huttner-Barnett approach developed here. In the mean time this will justify the use of Hopfield-Fano approach as an effective method applicable for the low loss regime which is a good assumption in most dielectric (excluding metals supporting lossy plasmon modes).\\
\indent From Poynting's theorem, it is usual in macroscopic electromagnetism to isolate the work density $W_e=\textbf{E}\cdot\partial_t\textbf{D}$ such as the  energy conservation  reads 
\begin{eqnarray}
\partial_t u =W_e+\partial_t(\frac{\mathbf{B}^2}{2})=-\boldsymbol{\nabla}\cdot(c\mathbf{E}\times\mathbf{B})\label{12new}.
\end{eqnarray}  
By direct integration we thus get the usual formula  for the time derivative of the total energy $H(t)=\int d^3\mathbf{x}u(\mathbf{x},t)$ such as:
\begin{eqnarray}
\frac{d}{dt} H(t)=\frac{d}{dt}(\int d^3\mathbf{x}\frac{\mathbf{B}^2}{2})+\int d^3\mathbf{x}\textbf{E}\cdot\partial_t\textbf{D}\nonumber\\
=\frac{d}{dt}(\int d^3\mathbf{x}\frac{\mathbf{B}^2}{2})+\int d^3\mathbf{x}\textbf{E}_\bot\cdot\partial_t\textbf{D}
\end{eqnarray}  
  which cancels if the fields decay sufficiently at spatial infinity (assumption which will be done in the following) as it can be proven after using the Poynting vector divergence and Stokes theorem. We now consider a temporal integration window $\delta t$ to compute the average derivative
	\begin{eqnarray}
\frac{\int_{\delta t} dt'\int d^3\mathbf{x}\textbf{E}_\bot(t')\cdot\partial_t\textbf{D}(t')}{\delta t}+\frac{\delta}{\delta t}(\int d^3\mathbf{x}\frac{\mathbf{B}^2}{2})\simeq 0, \nonumber\\ \label{energybri}
\end{eqnarray} where $\delta\int d^3\mathbf{x}\frac{\mathbf{B}^2}{2}$ means the  magnetic energy variation during the time $\delta t$ and $\int_{\delta t}d t'[...]=\int_{t}^{t+\delta t}dt'[...]$  is an integration domain from an initial time $t$ to a final time $t+\delta t$. The next step is to Fourier expand the field in the frequency domain  and we write $\textbf{E}_\bot=\textbf{E}_\bot^{(+)}+\textbf{E}_\bot^{(-)}$  where the positive frequency part of the field is defined as $\textbf{E}_\bot^{(+)}(t)=\int_0^{+\infty}d\omega\widetilde{\textbf{E}_\bot}(\omega)e^{-i\omega t}$ (the negative frequency part is then $\textbf{E}_\bot^{(-)}(t)=[\textbf{E}_\bot^{(+)}(t)]^\dagger$). We use similar notation for the displacement field and we introduce  the Fourier field $\widetilde{\textbf{D}}(\omega)$.   In order to achieve the integration Eq.~\ref{energybri} the temporal window will be supposed sufficiently large compared to the typical period $2\pi/\omega_c$ of the light pulse. This allows us to simplify the calculation and most contributions cancel out during the integration~\cite{Brillouin,Lifshitzbook,Rosa2010}.  Additionally to perform the calculation we assume that we have $\widetilde{\textbf{D}}(\omega)=\widetilde{\varepsilon}(\omega)\widetilde{\textbf{E}_\bot}(\omega)$. This is a usual formula in classical physics where the term $\textbf{P}^{(0)}$ is supposed equal to zero, but here we  are dealing with a quantized theory and we can not omit this term. Furthermore, we showed that in the Huttner-Barnett model when $t_0\rightarrow-\infty$ only the $(s)$ contributions discussed in section III C remain~\cite{Huttner1992a}. Assuming therefore this regime the transverse field $\textbf{D}^{(s)}$, and $\textbf{E}^{(s)}$ are fully expressed as a function of operators $f^{(0)}_{\omega,\alpha,j}(t)$.  Eqs.~\ref{rehhD} and \ref{rehhE} allow us to define the Fourier fields:       
	\begin{eqnarray}
\widetilde{\mathbf{D}}^{(s)}(\mathbf{x},\omega)=\sum_{\alpha,j}\frac{\omega_\alpha^2\Phi_\alpha(\mathbf{x})\boldsymbol{\hat{\epsilon}}_{\alpha,j}}{\omega_\alpha^2-\tilde{\varepsilon}(\omega)\omega^2}\sqrt{\frac{\hbar\sigma_\omega}{\pi\omega}}f^{(0)}_{\omega,\alpha,j}(t_0)e^{i\omega_\alpha t_0} \nonumber\\ \label{glopglop}
\end{eqnarray}
\begin{eqnarray}
\widetilde{\mathbf{E}}_\bot^{(s)}(\mathbf{x},\omega)=\sum_{\alpha,j}\frac{\omega^2\Phi_\alpha(\mathbf{x})\boldsymbol{\hat{\epsilon}}_{\alpha,j}}{\omega_\alpha^2-\tilde{\varepsilon}(\omega)\omega^2}\sqrt{\frac{\hbar\sigma_\omega}{\pi\omega}}f^{(0)}_{\omega,\alpha,j}(t_0)e^{i\omega_\alpha t_0},\nonumber\\ \label{reglop} 
\end{eqnarray} for $\omega>0$ (for $\omega<0$ we have $\widetilde{\mathbf{D}}^{(s)}(\mathbf{x},\omega)=\widetilde{\mathbf{D}}^{(s)\dagger}(\mathbf{x},-\omega)$ where  $\widetilde{\mathbf{D}}^{(s)}(\mathbf{x},-\omega)$ is given by Eq.~\ref{glopglop} at the positive frequency $-\omega$. Similar symmetries and properties hold for the electric and magnetic fields). We thus see that the relation 	$\widetilde{\textbf{D}}(\omega)\simeq\widetilde{\varepsilon}(\omega)\widetilde{\textbf{E}_\bot}(\omega)$ is approximately fulfilled if we consider only frequencies near a resonance at $\omega_\alpha^2=\tilde{\varepsilon}(\omega)\omega^2$ (the spectral distribution in Eq.~\ref{glopglop} is thus extremely peaked since losses are weak). This  dispersion relation occurs for transverse polaritons modes $\omega\simeq \Omega_{\alpha,m}$ (neglecting the imaginary part) and if the wave packet of spectral extension $\delta \omega\sim 1/\delta t$ is centered on such a wavelength we can replace with a good approximation  $\omega_\alpha^2$  by $\tilde{\varepsilon}(\omega)\omega^2$ in the numerator of 	Eq.~\ref{glopglop} leading thus to $\widetilde{\textbf{D}}(\omega)\simeq\widetilde{\varepsilon}(\omega)\widetilde{\textbf{E}_\bot}(\omega)$. After this assumption the calculation can be done like in classical textbooks~\cite{Brillouin,Lifshitzbook} and Eq.~\ref{energybri} becomes (this usual calculation will not be repeated here):
	\begin{eqnarray}
\frac{\delta}{\delta t} H(t)=\frac{\delta}{\delta t}(\int d^3\mathbf{x}[\frac{d\omega_c\tilde{\varepsilon}(\omega_c)}{d\omega_c}\textbf{E}_\bot^{(-)}\textbf{E}_\bot^{(+)}+\frac{\mathbf{B}^2}{2}])=0, \nonumber\\ \label{energytri}
\end{eqnarray}   where $\omega_c$ denotes now the transverse polariton frequency $\omega_c\simeq\textrm{Re}[\Omega_{\alpha,m}]$ for the homogeneous medium considered here.	In this formula the imaginary part of $\tilde{\varepsilon}(\omega_c)$ is systematically neglected in agreement with the reasoning discussed  for instance in Ref.~\cite{Lifshitzbook}. Alternatively Eq.~\ref{energytri} could be rewritten using the time average $\langle\mathbf{B}^2\rangle\simeq 2\textbf{E}_\bot^{(-)}\textbf{E}_\bot^{(+)}$ in order to get the classical Brillouin formula for the electric energy density in the medium but this will not be useful here. Moreover, we introduce the mode operators:
 \begin{eqnarray}
E_{\alpha,j,m}^{(+)}(t)=\int_{\delta \omega_{\alpha,m}} d\omega\frac{\omega^2}{\omega_\alpha^2-\tilde{\varepsilon}(\omega)\omega^2}\sqrt{\frac{\hbar\sigma_\omega}{\pi\omega}}f^{(0)}_{\omega,\alpha,j}(t).\nonumber\\ 
\label{modepolar}\end{eqnarray} where $\delta \omega_{\alpha,m}$ is a frequency window centered on the polariton pulsation $\omega_c\simeq\textrm{Re}[\Omega_{\alpha,m}]:=\Omega'_{\alpha,m}$.
 Now, if we suppose that the electromagnetic field is given by a sum of such transverse modes (without overlap of the frequency domains $\delta \omega_{\alpha,m}$) then Eq.~\ref{energytri} reads:
\begin{eqnarray}
\frac{\delta}{\delta t} H(t)=\frac{\delta}{\delta t}(\sum_{\alpha,j,m}[\frac{d(\Omega'_{\alpha,m}\tilde{\varepsilon}(\Omega'_{\alpha,m}))}{d\Omega'_{\alpha,m}}\nonumber\\ +\frac{\omega_\alpha^2}{{\Omega'_{\alpha,m}}^2}]E_{\alpha,j,m}^{(-)}E_{\alpha,j,m}^{(+)})=0, \label{energyquadri}
\end{eqnarray} 
 where the contribution  $\frac{\omega_\alpha^2}{{\Omega'_{\alpha,m}}^2}$ arises from a modal expansion of the magnetic field and from using the resonance condition in the numerator of Eq.~\ref{rehhB} (which involves $\omega\omega_\alpha\simeq\omega^2\omega_\alpha/\Omega_{\alpha,m}$).\\ 
\indent  What is also fundamental here is that we have the commutators (the derivation in the complex plane in given in Appendix F):
 \begin{eqnarray}
[E_{\alpha,j,m}^{(+)}(t),E_{\beta,l,n}^{(-)}(t)]=\delta_{\alpha,\beta}\delta_{j,l}\delta_{m,n}\frac{\hbar\Omega_{\alpha,m}}{2}\frac{d\Omega_{\alpha,m}^2}{d\omega_\alpha^2}.\nonumber\\ \label{window}
\end{eqnarray} and $[E_{\alpha,j,m}^{(+)}(t),E_{\beta,l,n}^{(+)}(t)]=[E_{\alpha,j,m}^{(-)}(t),E_{\beta,l,n}^{(-)}(t)]=0$. These relation imply the existence of effective rising and lowering operators $f_{\alpha,j,m}^{(+)}(t)$ for polaritons defined by $E_{\alpha,j,m}^{(+)}(t)=\sqrt{\left(\frac{\hbar\Omega_{\alpha,m}}{2}\frac{d\Omega_{\alpha,m}^2}{d\omega_\alpha^2}\right)}f_{\alpha,j,m}(t)$.\\ 
\indent  These relations were phenomenologically obtained by Milonni~\cite{Milonni1995,Garrison2004} after quantizing Brillouin's energy formula. Here we justify this result from the ground using the Huttner-Barnett formalism.  Importantly, after defining the optical index of the polariton mode $n_{\alpha,m}\simeq\sqrt{\tilde{\varepsilon}(\Omega'_{\alpha,m})}$ we can rewrite  $\frac{d(\Omega'_{\alpha,m}\tilde{\varepsilon}(\Omega'_{\alpha,m}))}{d\Omega'_{\alpha,m}}+\frac{\omega_\alpha^2}{{\Omega'_{\alpha,m}}^2}$  as $2n_{\alpha,m}c/v_{g}(\Omega'_{\alpha,m})$ where $v_{g}(\Omega'_{\alpha,m})$ is the group velocity of the mode defined by $d\Omega'_{\alpha,m}/d k_\alpha$. This allows us to rewrite the operators as   $E_{\alpha,j,m}^{(+)}(t)=\sqrt{\left(\frac{\hbar\Omega_{\alpha,m}}{2}\frac{v_{g}(\Omega'_{\alpha,m})}{n_{\alpha,m}c}\right)}f_{\alpha,j,m}(t)$ (since $\frac{v_{g}(\Omega'_{\alpha,m})}{n_{\alpha,m}c}=\frac{d\Omega_{\alpha,m}^2}{d\omega_\alpha^2}$) and finally to have :
\begin{eqnarray}
\frac{\delta}{\delta t} H(t)=\frac{\delta}{\delta t}(\sum_{\alpha,j,m}\hbar\Omega_{\alpha,m}f_{\alpha,j,m}^\dagger f_{\alpha,j,m})=0. \label{energy5}
\end{eqnarray} The total energy is thus defined as $ H(t)=\sum_{\alpha,j,m}\hbar\Omega_{\alpha,m}f_{\alpha,j,m}^\dagger f_{\alpha,j,m}$  which is a constant of motion defined up to an arbitrary additive constant. This formula involves only the transverse modes so that actually it gives the energy $H_\bot(t)$ associated with the transverse polariton modes in weekly dissipative medium and represents a generalization of Hopfield-Fano results as an effective but approximative model. \\
\indent Few remarks are here necessary. First,  the model proposed here relies on the assumption that the fields is a sum of wave packets spectrally non overlapping. This hypothesis which was also made by Garrison and Chio~\cite{Garrison2004}  was then called  the `quasi multimonochromatic' approximation. This assumption is certainly not necessary since Milonni's model includes as a limit the  rigorous Hopfield- Fano model~\cite{Fano1956,Hopfield1958} which doesn't rely on such an assumption.  In order to justify further Milonni's approach~\cite{Milonni1995} and relax the hypothesis made it is enough to observe first, that in Eq.~\ref{glopglop} the approximation  $\widetilde{\textbf{D}}(\omega)\simeq\widetilde{\varepsilon}(\omega)\widetilde{\textbf{E}_\bot}(\omega)$ is quite robust even if the fields is spectrally very broad. Indeed, since losses are here supposed to be very weak the resonance will practically cancel out  if $\omega$ differs significantly of a values where the condition $\omega_\alpha^2=\tilde{\varepsilon}(\omega)\omega^2$ occurs.  Second, if we insert formally Eqs.~\ref{glopglop} and Eq.~\ref{reglop} with the previous assumption into Eq.~\ref{energytri} then instead of the term $\frac{\delta}{\delta t}(\int d^3\mathbf{x}\frac{d\omega_c\tilde{\varepsilon}(\omega_c)}{d\omega_c}\textbf{E}_\bot^{(-)}\textbf{E}_\bot^{(+)})$ in Eq.~\ref{energytri} we get a term  $\frac{\delta}{\delta t}(\int_0^{+\infty}d\omega\int_0^{+\infty}d\omega'\int d^3\mathbf{x}\frac{d\omega\tilde{\varepsilon}^\ast(\omega)}{d\omega}\widetilde{\textbf{E}_\bot}(\omega)\widetilde{\textbf{E}_\bot}^\ast(\omega')e^{i(\omega'-\omega)t})$. This contribution is in general more complicated because $\frac{d\omega\tilde{\varepsilon}^\ast(\omega)}{d\omega}$ depends on $\omega$. However, using explicitly Eqs.~\ref{glopglop} and Eq.~\ref{reglop} and specially the Fourier expansion in plane waves we see that for the specific fields considered here Eq.~\ref{energyquadri} still holds.  This means that we can again introduce polariton operators  $E_{\beta,l,n}^{(+)}(t)$, and $E_{\beta,l,n}^{(-)}(t)$ defined by  Eq.~\ref{modepolar}.  As previously these operators depend on a frequency window $\delta \omega_{\alpha,m}$ and here these are introduced quite formally for taking into account the fact that the resonance $\frac{1}{\omega_\alpha^2-\tilde{\varepsilon}(\omega)\omega^2}$ is extremely peaked near the different polariton frequencies $\Omega_{\alpha,m}$. A product like $\frac{1}{\omega_\alpha^2-\tilde{\varepsilon}(\omega)\omega^2}\frac{1}{\omega_\alpha^2-\tilde{\varepsilon}^\ast(\omega')\omega'^2}$ occurring in the integration will thus not contribute unless the frequency $\omega'$ and $\omega$ are in a given window $\delta \omega_{\alpha,m}$. Eq.~\ref{energyquadri} thus results as a very good practical approximation.\\ 
\indent Remarkably, as observed in Eq.~\ref{window}  (and explained in Appendix E) the commutator does not depend explicitly on the size of these windows (which are only supposed to be small compared to the separation between the different mode frequencies and large enough to include the resonance peaks as explained in Appendix E). Therefore, this allows us to renormalize these operators as before by introducing the same rising and lowering polariton operators $f_{\alpha,j,m}$ such as  Eq.~\ref{energy5} and  $ H_\bot(t)=\sum_{\alpha,j,m}\hbar\Omega_{\alpha,m}f_{\alpha,j,m}^\dagger f_{\alpha,j,m}$ hold identically. We thus have completed the justification of the Milonni's approach for dielectric medium with weak absorption~\cite{Milonni1995}.\\ 
\indent An other remark concerns the longitudinal electric field which was omitted here since it does not play an active role in pulse propagation through the medium. We have indeed $\int d^3\mathbf{x}\textbf{E}\cdot\partial_t\textbf{D}=\int d^3\mathbf{x}\textbf{E}_\bot\cdot\partial_t\textbf{D}$ so that the reasoning was only done on the transverse modes. However, this was not necessary and one could have kept the longitudinal electric field all along the reasoning. Since the transverse part is a constant as we showed before, this should be the case for the longitudinal part as well since $H(t)-H_\bot(t)$ is also an integral of motion. Now, a reasoning similar to the previous one for transverse waves will lead to the Brillouin formula for the longitudinal electric energy:
\begin{eqnarray}
\frac{\delta}{\delta t} H_{||}(t)=\frac{\delta}{\delta t}(\int d^3\mathbf{x}\int d^3\mathbf{x}\textbf{E}_{||}\cdot\partial_t\textbf{D}\nonumber\\
\simeq \frac{\delta}{\delta t}(\sum_{\alpha,m}\frac{d(\Omega'_{\alpha,m}\tilde{\varepsilon}(\Omega'_{\alpha,m}))}{d\Omega'_{\alpha,m}}E_{\alpha,m,||}^{(-)}E_{\alpha,m,||}^{(+)})=0
\end{eqnarray}   where the  longitudinal polariton modes are defined by: 
  \begin{eqnarray}
E_{\alpha,m}^{(+)}(t)=\int_{\delta \omega_{\alpha,m}} d\omega\frac{-1}{\tilde{\varepsilon}(\omega)}\sqrt{\frac{\hbar\sigma_\omega}{\pi\omega}}f^{(0)}_{\omega,\alpha,||}(t). 
\label{modepolarlong}\end{eqnarray}  In this formalism the longitudinal polariton frequencies $\Omega'_{\alpha,m}$ are the solutions of $\tilde{\varepsilon}'(\Omega'_{\alpha,m})\simeq0$ (where losses are again supposed to be weak). The commutator can be defined using a method equivalent to Eq.~\ref{window} and we get: 
     \begin{eqnarray}
[E_{\alpha,m,||}^{(+)}(t),E_{\beta,n,||}^{(-)}(t)]=\delta_{\alpha,\beta}\delta_{m,n}\frac{\hbar}{|M_{\alpha,m}|}. \label{windowlong}
\end{eqnarray}  with $M_{\alpha,m}=\frac{d\tilde{\varepsilon}'(\omega)}{d\omega}|_{\Omega'_{\alpha,m}}$.
After defining the lowering polariton operator as $E_{\alpha,m,||}^{(+)}(t)=f_{\alpha,m,||}(t)\sqrt{\left(\frac{\hbar}{|M_{\alpha,m}|}\right)}$ we thus obtain:   
    \begin{eqnarray}
\frac{\delta}{\delta t} H_{||}(t)\simeq \frac{\delta}{\delta t}(\sum_{\alpha,m}\hbar\Omega'_{\alpha,m}f_{\alpha,m,||}^{\dagger}f_{\alpha,m,||})=0
\end{eqnarray}  as it should be.  Milonni's approach~\cite{Milonni1995} leads therefore to an effective justification of longitudinal polaritons as well and this includes the Hopfield-Fano~\cite{Hopfield1958} model as a limiting case when losses are vanishing outside infinitely narrow absorption  bands.\\
\indent A final important remark should be done since it concerns the general significance of the scattered field (s) in the lossless limit. Indeed, we see from Eq.~\ref{modepolar} that the transverse mode operators $E_{\alpha,j,m}^{(+)}(t)$ and  $E_{\alpha,j,m}^{(+)}(t)$  rigorously vanish in the limit $\sigma_\omega\rightarrow 0$ (this is not true for longitudinal operators Eq.~\ref{modepolarlong} which are physically linked to bound  and Coulombian fields). In agreement with the sub-section III C we thus conclude that in the vacuum limit one should consider the $(0)$ fields as the only surviving contribution. However,  we also see that for all practical needs  if the losses are weak  but not equal to zero then  by imposing  $t_0\rightarrow -\infty$ the $(0)$ terms should cancel and only will survive a scattered term which will formally looks as a free photon in a bulk medium with optical index $n_\omega\simeq\sqrt{\varepsilon_\omega}$. Therefore, we justify the formal canonical quantization procedure used  by Milonni and others~\cite{Milonni1995,Constantinou1993,Al1996,Garrison2004,Rosa2010,Archambault2010,Yang2015} which reduces to the historical quantization methods in the (quasi-) non dispersive limit~\cite{Jauch1948,Glauber1991}. However, this can only be considered as an approximation and therefore the original claim presented in Ref.~\cite{Gruner1995} that the scattered field $(s)$ is sufficient for justifying the exact limit  $\sigma_\omega\rightarrow 0$ without the $(0)$ term was actually unfounded. As we will see this will become specially relevant when we will generalize the Langevin noise approach to an inhomogeneous medium.  
              				
\subsection{The energy conservation puzzle and the interpretation of the Hamiltonian for an homogeneous dielectric medium}    
\indent The central issue in this work is to interpret  the physical meaning of quantized polariton modes in the general Huttner Barnett framework of Section II and this will go far beyond the limiting Hopfield-Fano~\cite{Fano1956,Hopfield1958} or Milonni's approaches~\cite{Milonni1995} which are valid in restricted  conditions when losses and/or dispersion are weak enough. For the present purpose we will focus on the Homogeneous medium case (the most general inhomogeneous medium case is analyzed in the next section). It is fundamental to compare the mode structure of Eqs.~\ref{rehhD}, \ref{rehhE}, \ref{rehhEpar}, and \ref{rehhB}  on the one side and the mode structure of Eqs.~\ref{118},  \ref{119} and \ref{120} on the other side which are  associated respectively  with the free modes `(0)' and the scattered modes `(s)'. The `(0)' modes  are the eigenstates of the classical propagation problem when we can cancel the fluctuating term $\mathbf{P}^{(0)}$. This is however not allowed in QED since we are now considering operators in the Hilbert space and one cannot omit these terms without breaking unitarity. Inversely the scattered modes are the modes which were considered by  Huttner and Barnett~\cite{Huttner1992a}.  Moreover only these `(s)' modes survive here  if the initial time $t_0$ is sent into the remote past, i.e., if  $t_0\rightarrow -\infty$. For all operational needs it is therefore justified to omit altogether the `(0)' mode contribution in the homogeneous medium case. Clearly, this was the choice made by Gruner and Welsch~\cite{Gruner1996} and later by more or less all authors working on the subject (see however Refs.~\cite{Stefano2000,Stefano2002}) which accepted this rule even for non-homogeneous media.  If we accept this axiom then the Hamiltonian of the problem seems to reduce to $H_M^{(0)}$, i.e. in the homogeneous medium case, to  
\begin{eqnarray}
H_M^{(0)}(t)=\int d^3\mathbf{x}\int_{0}^{+\infty}d\omega\hbar\omega\mathbf{f}^{\dagger(0)}_\omega(\mathbf{x},t)\mathbf{f}^{(0)}_\omega(\mathbf{x},t)\nonumber\\
 = \sum_{\alpha,j}\int_0^{+\infty}d\omega\hbar\omega f^{\dagger(0)}_{\omega,\alpha,j}(t),f^{(0)}_{\omega,\alpha,j}(t) \nonumber\\
+\sum_{\alpha}\int_0^{+\infty}d\omega\hbar\omega f^{\dagger(0)}_{\omega,\alpha,||}(t),f^{(0)}_{\omega,\alpha,||}(t) \label{30c}
\end{eqnarray} which depends only on the fluctuating operators $f^{(0)}_{\omega,\alpha,||}(t)$, $f^{\dagger(0)}_{\omega,\alpha,||}(t)$ in agreement with the Langevin force/noise approach of Gruner and Welsch~\cite{Gruner1996}.\\
\indent Now, there is apparently a paradox: the complete Hamiltonian of the system is in agreement with Eq.~\ref{10} given by 
\begin{eqnarray}
H(t)=\int d^3\mathbf{x}:\frac{\mathbf{B}(\mathbf{x},t)^2+\mathbf{E}(\mathbf{x},t)^2}{2}:\nonumber\\ +\int d^3\mathbf{x}\int_{0}^{+\infty}d\omega\hbar\omega\mathbf{f}^\dagger_\omega(\mathbf{x},t)\mathbf{f}_\omega(\mathbf{x},t) \label{158n}
\end{eqnarray} Can we show that $H(t)$ is actually equivalent to $H_M^{(0)}(t)$?  This is indeed the case  for all practical purposes at least for the Homogeneous medium case treated by Huttner and Barnett\cite{Huttner1992a} that we analyzed in details before. To see that remember that $H(t)$ is actually a constant of motion therefore we should have $H(t)=H(t_0)$.  This equality reads also  $H(t)=\int d^3\mathbf{x}:\frac{\mathbf{B}(\mathbf{x},t_0)^2+\mathbf{E}(\mathbf{x},t_0)^2}{2}:+H_M(t_0)$.  Now, the central point here is to use the boundary condition at time $t_0$ which imply  $\mathbf{f}_\omega(\mathbf{x},t_0)=\mathbf{f}^{(0)}_\omega(\mathbf{x},t_0)$. Furthermore, since the time evolution of $\mathbf{f}^{(0)}_\omega$ is harmonic we have $\mathbf{f}^{\dagger(0)}_\omega(\mathbf{x},t)\mathbf{f}^{(0)}_\omega(\mathbf{x},t)=\mathbf{f}^{\dagger(0)}_\omega(\mathbf{x},t_0)\mathbf{f}^{(0)}_\omega(\mathbf{x},t_0)$. Altogether these relations imply 
\begin{eqnarray} H_M(t_0)=H_M^{(0)}(t_0)=H_M^{(0)}(t)\end{eqnarray} so that we have: 
\begin{eqnarray}
H(t)=\int d^3\mathbf{x}:\frac{\mathbf{B}(\mathbf{x},t_0)^2+\mathbf{E}(\mathbf{x},t_0)^2}{2}: +H_M^{(0)}(t).\nonumber\\ \label{158o}
\end{eqnarray} 
In other words we get a description  in which the Fock number states associated with the fluctuating operators $\mathbf{f}^{(0)}_{\omega}(\mathbf{x},t)$ or equivalently $f^{(0)}_{\omega,\alpha,j}(t)$, $f^{(0)}_{\omega,\alpha,||}(t)$  diagonalize  not the full Hamiltonian but only a part that we noted $H_M^{(0)}(t)$. However, this is not problematic since the remaining term
\begin{eqnarray}
H_{\textrm{rem.}}(t)=\int d^3\mathbf{x}:\frac{\mathbf{B}(\mathbf{x},t_0)^2+\mathbf{E}(\mathbf{x},t_0)^2}{2}:\label{rema}
\end{eqnarray} 
is also clearly by definition a constant of motion since it only depends on fields at time $t_0$. How can we interpret this constant of motion?  We can clearly rewrite it as
$H_{\textrm{rem.}}(t)=\int d^3\mathbf{x}:\frac{\mathbf{B}(\mathbf{x},t_0)^2+(\mathbf{D}(\mathbf{x},t_0)-\mathbf{P}(\mathbf{x},t_0))^2}{2}:$. Therefore, in the $\textbf{F}$ potential vector formalism defined in Section II, this constant depends on both $c_{\alpha,j}(t_0)$ and $c_{\alpha,j}^\dagger(t_0)$, i.~e., lowering and rising operators associated with the transverse $\mathbf{D}$  and $\mathbf{B}$ fields, and it also depends on the operators $\mathbf{f}^\dagger_\omega(\mathbf{x},t_0)$, $\mathbf{f}_\omega(\mathbf{x},t_0)$ which are associated with the fluctuating dipole distribution at the initial time $t_0$. Furthermore since we have also $\textbf{D}^{(s)}(t_0)=0$, $\textbf{B}^{(s)}(t_0)=0$, $\mathbf{E}^{(s)}(t_0)=-\mathbf{P}^{(0)}(t_0)$ and since $\textbf{D}(t_0)=\textbf{D}^{(0)}(t_0)=\textbf{E}^{(0)}(t_0)$, $\textbf{B}(t_0)=\textbf{B}^{(0)}(t_0)$ we can alternatively write: \begin{eqnarray}H_{\textrm{rem.}}(t)=\int d^3\mathbf{x}:\frac{\mathbf{B}^{(0)}(\mathbf{x},t_0)^2}{2}\nonumber\\+\frac{(\mathbf{D}^{(0)}(\mathbf{x},t_0)-\mathbf{P}^{(0)}(\mathbf{x},t_0))^2}{2}:.\end{eqnarray} 
This means that the remaining term $H_{\textrm{rem.}}(t)$ depends on the knowledge of $\textbf{D}^{(0)}(t_0)$ and $\textbf{B}^{(0)}(t_0)$ electromagnetic free field. Since in the limit $t_0\rightarrow-\infty$  these $(0)$ electromagnetic terms vanish at any finite time $t$ this would justify to consider $H_{\textrm{rem.}}(t)$ as an inoperative constant. However, we have also the contribution of $\mathbf{P}^{(0)}(t_0)$  which play a fundamental role in the determination of $\textbf{D}^{(s)}(t)$, $\textbf{B}^{(s)}(t)$ and $\textbf{E}^{(s)}(t)$ at the finite time $t$. Therefore, while $H_{\textrm{rem.}}(t)$ is a constant of motion it nevertheless contains quantities which will affect the evolution of the surviving $(s)$ fields at time $t$. It is for this reason that we can say that for `all practical purposes only' $H_{\textrm{rem.}}(t)$ is unnecessary and that $H_M^{(0)}(t)$ is sufficient for describing the energy problem.\\
\indent  There are however many remarks to be done here concerning this analysis.   First,  while the Hamiltonian  $H_M^{(0)}(t)$ gives a good view of the energy up to an additive inoperative constant it is the full Hamiltonian which is necessary for deriving the equations of motions from Hamilton's equations or equivalently  from Heisenberg's evolution like $i\hbar\frac{d}{dt}A(t)=[A(t),H(t)]$.   It is also only with $H(t)$ that time symmetry is fully preserved.  In particular do not forget that in deriving the Fano-Hopfield~\cite{Fano1956,Hopfield1958} formalism  we introduced (see Eq.~\ref{G}) the evolution  
$\boldsymbol{\nabla}\times\boldsymbol{\nabla}\times\widetilde{\textbf{E}}(\Omega)-\frac{\Omega^2}{c^2}\widetilde{\varepsilon}(\Omega)\widetilde{\textbf{E}}(\Omega)=\frac{\Omega^2}{c^2}\widetilde{\textbf{P}}^{(in)}(\Omega)$ which depends on the causal  `in' field $\widetilde{\textbf{P}}^{(in)}(\Omega)$ and on the causal permittivity $\widetilde{\varepsilon}(\Omega)$.  Since the knowledge of $\mathbf{P}^{(in)}(t)$ is equivalent to $\mathbf{P}^{(0)}(t)$ in the limit $t_0 \rightarrow-\infty$ the (forward) Laplace transform method is leading to the same result that the usual Fano method and this fits as well with the Gruner and Welsch Langevin's equations~\cite{Gruner1996}. However, instead of Eq.~\ref{G} we could equivalently use the anticausal equation
\begin{eqnarray}
\boldsymbol{\nabla}\times\boldsymbol{\nabla}\times\widetilde{\textbf{E}}(\Omega)-\frac{\Omega^2}{c^2}\widetilde{\varepsilon}^\ast(\Omega)\widetilde{\textbf{E}}(\Omega)=\frac{\Omega^2}{c^2}\widetilde{\textbf{P}}^{(out)}(\Omega)\label{Gbis}
\end{eqnarray}  where $\widetilde{\textbf{P}}^{(out)}(\Omega)$ replaces $\widetilde{\textbf{P}}^{(out)}(\Omega)$ and where the causal permittivity $\widetilde{\varepsilon}(\Omega)$ becomes now $\widetilde{\varepsilon}^\ast(\Omega)$ which is associated with amplification instead of the usual dissipation. The Green integral equation now becomes 
\begin{eqnarray}
\widetilde{\textbf{E}}(\mathbf{x},\Omega)=\frac{\Omega^2}{c^2}\int d^3\mathbf{x'} \mathbf{G}_{\chi_{T}}(\mathbf{x},\mathbf{x'},\Omega)\cdot
\widetilde{\mathbf{P}}^{(out)}(\mathbf{x'},\Omega)\nonumber\\  \label{outoftime}
\end{eqnarray} where $\chi_T(t)$ replaces $\chi(t)$ (see Eq.~\ref{20bis}) and is associated with the anticausal dynamics which is connected to the time-reversed evolution. Both formalism developed with either `in' or `out' fields are completely equivalent  but this shows that we have the freedom to express the scattered field  in term of $\mathbf{f}^{(in)}_{\omega}(\mathbf{x},t)$, and $\mathbf{f}^{(in)\dagger}_{\omega}(\mathbf{x},t)$ or in term of $\mathbf{f}^{(out)}_{\omega}(\mathbf{x},t)$, and $\mathbf{f}^{(out)\dagger}_{\omega}(\mathbf{x},t)$. Since there is in general no fully propagative `free'  electromagnetic modes (if we exclude the lossless medium limit considered by Hopfield and Fano~\cite{Fano1956,Hopfield1958}) then the surviving fields at finite time $t$  obtained either with $t_0\rightarrow-\infty$ or $t_f\rightarrow+\infty$ correspond to decaying or growing waves  in agreement with the results discussed for the Laplace transform methods. The full Hamiltonian $H(t)$ is thus expressed equivalently either as 
 \begin{eqnarray}
H(t)=\int d^3\mathbf{x}:\frac{\mathbf{B}(\mathbf{x},t_0)^2+\mathbf{E}(\mathbf{x},t_0)^2}{2}:\nonumber\\ +\int d^3\mathbf{x}\int_{0}^{+\infty}d\omega\hbar\omega\mathbf{f}^{(in)\dagger}_\omega(\mathbf{x},t)\mathbf{f}^{(in)}_\omega(\mathbf{x},t) 
\end{eqnarray}  (with $\mathbf{f}^{(in)}_\omega(\mathbf{x},t):=\mathbf{f}^{(0)}_\omega(\mathbf{x},t)=\mathbf{f}_\omega(\mathbf{x},t_0)e^{-i\omega(t-t_0)}$ the fluctuating field defined in section II) 
or as 
\begin{eqnarray}
H(t)=\int d^3\mathbf{x}:\frac{\mathbf{B}(\mathbf{x},t_f)^2+\mathbf{E}(\mathbf{x},t_f)^2}{2}:\nonumber\\ +\int d^3\mathbf{x}\int_{0}^{+\infty}d\omega\hbar\omega\mathbf{f}^{(out)\dagger}_\omega(\mathbf{x},t)\mathbf{f}^{(out)}_\omega(\mathbf{x},t) 
\end{eqnarray}  with $\mathbf{f}^{(out)}_\omega(\mathbf{x},t)=\mathbf{f}_\omega(\mathbf{x},t_f)e^{-i\omega(t-t_f)} $ the fluctuating field using using a final boundary condition at time $t_f$. In the limits $t_0\rightarrow-\infty$, $t_f\rightarrow+\infty$  this leads to  \begin{eqnarray}H(t)=H_{in,\textrm{rem.}}(t)\nonumber\\ +\int d^3\mathbf{x}\int_{0}^{+\infty}d\omega\hbar\omega\mathbf{f}^{(in)\dagger}_\omega(\mathbf{x},t)\mathbf{f}^{(in)}_\omega(\mathbf{x},t) \end{eqnarray} or \begin{eqnarray}H(t)=H_{out,\textrm{rem.}}(t)\nonumber\\ +\int d^3\mathbf{x}\int_{0}^{+\infty}d\omega\hbar\omega\mathbf{f}^{(out)\dagger}_\omega(\mathbf{x},t)\mathbf{f}^{(out)}_\omega(\mathbf{x},t)\end{eqnarray} where the remaining terms (see Eq.~\ref{rema}) 
$H_{in,\textrm{rem.}}(t)$ and $H_{out,\textrm{rem.}}(t)$ are two different integrals of motion. Therefore, it shows that the representation chosen by Gruner and Welsch in Ref.~\cite{Gruner1996} is not univocal and that one could reformulate all the theory in  terms of $\widetilde{\textbf{P}}^{(out)}(\Omega)$ instead of $\widetilde{\textbf{P}}^{(in)}(\Omega)$ to respect time symmetry. Furthermore, we emphasize  that while the 
two  Hamiltonians $\int d^3\mathbf{x}\int_{0}^{+\infty}d\omega\hbar\omega\mathbf{f}^{(0)\dagger}_\omega(\mathbf{x},t)\mathbf{f}^{(0)}_\omega(\mathbf{x},t) $  and $\int d^3\mathbf{x}\int_{0}^{+\infty}d\omega\hbar\omega\mathbf{f}^{(f)\dagger}_\omega(\mathbf{x},t)\mathbf{f}^{(f)}_\omega(\mathbf{x},t) $ have formally  the same mathematical structure they are not associated with the same physical electromagnetic fields since Eq.~\ref{glopiglopa} and  Eq.~\ref{outoftime} corresponds respectively to decaying and growing radiated fields. Causality therefore requires to make a choice between two different representations. It is only the choice on a boundary condition in the remote past or future together with thermodynamical considerations which allow us to favor the decaying regime given by Eq.~\ref{glopiglopa}. \\
\indent  A different but related point to be discussed here concerns the Hopfield-Fano limit~\cite{Fano1956,Hopfield1958} for which the general conductivity like  $\sigma(\Omega)=\sum_n\frac{\pi\omega_{p,n}^2}{2}(\delta(\Omega-\omega_{0,n})+\delta(\Omega+\omega_0))$ leads to quasi-lossless permittivity (see Eq.~\ref{dangerous}). In this regime we found  that an exact diagonalization procedure can be handled out leading to genuine transverse and longitudinal polaritons.  These exceptions also fit with the time-symmetry considerations discussed previously since the quasi-absence of absorption makes the problem much more time symmetrical that   in the cases where   the only viable representations involve either  $\widetilde{\textbf{P}}^{(out)}(\Omega)$ or $\widetilde{\textbf{P}}^{(in)}(\Omega)$. Of course the Hopfield-Fano model is an idealization which, in the  context of the Huttner-Barnett framework, gets a clear physical interpretation only as an approximation for low-loss media, as explained in Section III G.  \\
\indent This leads us to a new problem, which is certainly the most important in the present work:  Considering the vacuum limit $\chi(t)\rightarrow 0$ discussed after Eq.~\ref{120} we saw that only the $(0)$ electromagnetic modes survive in this regime and that the $(s)$ modes are killed together with $\textbf{P}^{(0)}(t)\rightarrow 0$. These vacuum modes are completely decoupled from the undamped mechanical oscillators motion $\textbf{X}_\omega(t)=\textbf{X}^{(0)}_\omega(t)$. In this limit time symmetry is of course respected and we see that the full set of eigenmodes diagonalizing the Hamiltonian corresponds to the uncoupled free photons and
free mechanical oscillator motions.  An other way to see that is to use again the Fourier formalism instead of the Laplace transform method. From Eq.~\ref{G} we see that in the vacuum limit it is not the scattered field defined by Eq.~\ref{glopiglopa}  which survives (since $\textbf{P}^{(0)}(t)\rightarrow 0$) but an additional  term corresponding to free space photon modes. While this should be clear after our discussion this point has tremendous consequences if we want to generalize  properly the Huttner-Barnett approach to an inhomogeneous medium.  In such a medium the permittivity $\widetilde{\varepsilon}(\textbf{x},\Omega)$ is position dependent. Now, generally speaking in nanophotonics we consider problems where a dissipative object like a metal particle is confined in a finite region of space surrounded by vacuum.  The susceptibility 
$\chi(\mathbf{x},\tau)$ therefore vanishes outside the object and we expect electromagnetic vacuum modes associated  with free space photon to play a important role in the final analysis.  This should contrast with the Huttner-Barnett case for homogenous medium which supposes an unphysical  infinite dissipative medium supporting bulk polaritons or plasmons. The inhomogeneous polariton case will be treated in the next section.
\section{Quantizing polaritons in an inhomogeneous medium}  
\subsection{The Green dyadic problem in the time domain } 
\indent In order to deal with the most general situation of polaritons in inhomogeneous media we need first to consider the formal separation between source fields and free-space photon mode. The separation used here is clearly different from the one developed in the previous section since we now consider on the one side as source term the total polarization $\mathbf{P}(\mathbf{x},t)$ (see Eq.~\ref{18}) which includes both the fluctuating term $\mathbf{P}^{(0)}(\mathbf{x},t)$ but also the induced polarization $\int_{0}^{t-t_0}\chi(\mathbf{x},\tau)d\tau\mathbf{E}(\mathbf{x},t-\tau)$ and on the other side as source-free terms some general photon modes solutions of Maxwell equations in vacuum. In order not to get confused with the previous notations we now label $(v)$ the vacuum modes and $(d)$ the modes induced by the total dipolar distribution $\mathbf{P}(\mathbf{x},t)$.\\ 
\indent We start with the second order dynamical equation
\begin{eqnarray}
\frac{1}{c^2}\partial_t^2\mathbf{F}(\mathbf{x},t)-\boldsymbol{\nabla}^2\mathbf{F}(\mathbf{x},t)-\boldsymbol{\nabla}\times\mathbf{P}(\mathbf{x},t)=0\label{39}
\end{eqnarray} 
 which can be solved using the method developed for Eq.~\ref{90}. Indeed, by imposing $\chi=0$ in Eq.~\ref{90} and by replacing $\mathbf{P}^{(0)}$  by $\mathbf{P}$,  $\mathbf{F}^{(0)}$ by $\mathbf{F}^{(v)}$, $\mathbf{F}^{(s)}$ by $\mathbf{F}^{(d)}$ and so on in the calculations of Section III we can easily obtain the formalism needed.  Consider first the vacuum fields
$\mathbf{F}^{(v)}$, $\mathbf{D}^{(v)}$  and   $\mathbf{B}^{(v)}$. From Eqs.~\ref{118}, \ref{119} in the limit $\chi(\tau)\rightarrow 0$ we get a plane-wave modal expansion for the free photon field which we write in analogy with Eq.~\ref{57b} as  
\begin{eqnarray}
\mathbf{F}^{(v)}(\mathbf{x},t)=\sum_{\alpha,j} ic\sqrt{\frac{\hbar}{2\omega_\alpha}}c_{\alpha,j}^{(v)}(t)\boldsymbol{\hat{\epsilon}}_{\alpha,j}\Phi_\alpha(\mathbf{x})+ cc.\nonumber\\
\mathbf{D}^{(v)}(\mathbf{x},t)=\sum_{\alpha,j} -\sqrt{\frac{\hbar \omega_\alpha}{2}}c_{\alpha,j}^{(v)}(t)\hat{\mathbf{k}}_\alpha\times\boldsymbol{\hat{\epsilon}}_{\alpha,j}\Phi_\alpha(\mathbf{x})+ cc.\nonumber\\
\mathbf{B}^{(v)}(\mathbf{x},t)=\sum_{\alpha,j} \sqrt{\frac{\hbar \omega_\alpha}{2}}c_{\alpha,j}^{(v)}(t)\boldsymbol{\hat{\epsilon}}_{\alpha,j}\Phi_\alpha(\mathbf{x})+ cc..\nonumber\\ \label{121}
\end{eqnarray} with  the modal expansion coefficients $c_{\alpha,j}^{(v)}(t)=c_{\alpha,j}(t_0)e^{-i\omega_\alpha(t-t_0)}$ showing the harmonic structure of the fields.  Of course this transverse vacuum field satisfies Maxwell's equations without source terms and in particular  $\boldsymbol{\nabla}\times\mathbf{E}^{(v)}(\mathbf{x},t)=-\partial_t\mathbf{B}^{(v)}(\mathbf{x},t)/c$  with   $\mathbf{E}^{(v)}(\mathbf{x},t)=\mathbf{D}^{(v)}(\mathbf{x},t)$.\\
\indent We now study the scattered-fields $\mathbf{E}^{(d)}$, $\mathbf{D}^{(d)}$  and   $\mathbf{B}^{(d)}$. Like for the calculations presented in section III for the homogeneous medium case it appears convenient to use the Green dyadic formalism which is well adapted for nanophotonics studies in particular for numerical computation of fields in complex  dielectric environment where no obvious spatial symmetry are visible. Starting with this formalism we get  for the $(d)$ electric field:
 \begin{eqnarray}
\mathbf{D}^{(d)}(\mathbf{x},t)=\int_{\gamma-i\infty}^{\gamma+i\infty}\frac{i dp}{2\pi} e^{p(t-t_0)}\int d^3\mathbf{x'}\nonumber\\ \mathbf{S}_v(\mathbf{x},\mathbf{x'},ip)
\cdot \overline{\mathbf{P}'}(\mathbf{x'},p). \label{124}
\end{eqnarray} Equivalently, by using the inverse Laplace transform (see Appendix D for details) $
\mathbf{Q}_v(\tau,\mathbf{x},\mathbf{x'})=\int_{\gamma-i\infty}^{\gamma+i\infty}\frac{i dp}{2\pi}e^{p\tau}\mathbf{S}_{v}(\mathbf{x},\mathbf{x'},ip)$, we obtain in the time domain
\begin{eqnarray}
\mathbf{D}^{(d)}(\mathbf{x},t)=\int_0^{t-t_0}d\tau\int d^3\mathbf{x'}\mathbf{Q}_v(\tau,\mathbf{x},\mathbf{x'})\cdot\mathbf{P}(\mathbf{x'},t-\tau).\nonumber\\ \label{125}
\end{eqnarray}
This analysis implies $\mathbf{D}^{(d)}(\mathbf{x},t_0)=0$ and therefore  we have $\mathbf{D}(\mathbf{x},t_0)=\mathbf{D}^{(v)}(\mathbf{x},t_0)$.\\
 \indent Moreover, most studies consider instead of the propagator $\mathbf{S}_v(\mathbf{x},\mathbf{x'},ip)$ the dyadic 
Green function $\mathbf{G}_v(\mathbf{x},\mathbf{x'},ip)$ which is a solution of \begin{eqnarray}
\boldsymbol{\nabla}\times\boldsymbol{\nabla}\times\mathbf{G}_v(\mathbf{x},\mathbf{x'},ip)+\frac{p^2}{c^2}\mathbf{G}_v(\mathbf{x},\mathbf{x'},ip)\nonumber\\
=\boldsymbol{\delta}(\mathbf{x}-\mathbf{x'})=\mathbf{I}\delta^{3}(\mathbf{x}-\mathbf{x'})\nonumber\\ \label{129}
\end{eqnarray} and which is actually connected to $\mathbf{S}_v(\mathbf{x},\mathbf{x'},ip)$ by $\mathbf{S}_v(\mathbf{x},\mathbf{x'},ip)=\mathbf{I}\delta^{3}(\mathbf{x}-\mathbf{x'})-\frac{p^2}{c^2}\mathbf{G}_v(\mathbf{x},\mathbf{x'},ip)$. This dyadic Green function is very convenient  since practical calculations very often involve not the displacement field $\mathbf{D}$ but the electric field $\mathbf{E}=\mathbf{D}-\mathbf{P}$. We obtain:  
\begin{eqnarray}
\mathbf{E}(\mathbf{x},t)=\mathbf{E}^{(v)}(\mathbf{x},t)-\int_{\gamma-i\infty}^{\gamma+i\infty}\frac{i dp}{2\pi} e^{p(t-t_0)}\nonumber\\ \cdot \int d^3\mathbf{x'} \frac{p^2}{c^2}\mathbf{G}_v(\mathbf{x},\mathbf{x'},ip)
\cdot \overline{\mathbf{P}'}(\mathbf{x'},p).  \label{130}
\end{eqnarray}
Alternatively in the time domain we have  for the scattered field $\mathbf{E}^{(d)}(\mathbf{x},t)=\mathbf{E}(\mathbf{x},t)-\mathbf{E}^{(v)}(\mathbf{x},t)$:
\begin{eqnarray}
\mathbf{E}^{(d)}(\mathbf{x},t)=\int_0^{t-t_0}d\tau\int d^3\mathbf{x'}\mathbf{Q}_v(\tau,\mathbf{x},\mathbf{x'})\cdot\mathbf{P}(\mathbf{x'},t-\tau)\nonumber\\-\mathbf{P}(\mathbf{x},t). \nonumber\\
 \label{131}
\end{eqnarray} This can rewritten by using the propagator
$\mathbf{U}_v(\tau,\mathbf{x},\mathbf{x'})=\int_{\gamma-i\infty}^{\gamma+i\infty}\frac{i dp}{2\pi}e^{p\tau}\mathbf{G}_{v}(\mathbf{x},\mathbf{x'},ip)$  together with Eq.~\ref{definiti} as:
  \begin{eqnarray}
\mathbf{E}^{(d)}(\mathbf{x},t) = -\int_0^{t-t_0}d\tau\int d^3\mathbf{x'}\frac{\partial_\tau^2\mathbf{U}_v(\tau,\mathbf{x},\mathbf{x'}) }{c^2}\nonumber\\ \cdot\mathbf{P}(\mathbf{x'},t-\tau)-\mathbf{P}(\mathbf{x},t).  \label{131b}
\end{eqnarray}
\indent Finally, we can also use the dyadic formalism to represent the magnetic field $\mathbf{B}^{(d)}(\mathbf{x},t)$. Starting from  Eq.~\ref{6} which yields $\boldsymbol{\nabla}\times\overline{\mathbf{E}'}(\mathbf{x},p)=-\frac{p}{c}\overline{\mathbf{B}'}(\mathbf{x},p)
+\frac{\mathbf{B}(\mathbf{x},t_0)}{c}$ and therefore 
\begin{eqnarray}
\overline{\mathbf{B}'}(\mathbf{x},p)=\overline{\mathbf{B}'}^{(v)}(\mathbf{x},p)\nonumber\\ +\int d^3\mathbf{x'} \frac{p}{c}\boldsymbol{\nabla}\times\mathbf{G}_v(\mathbf{x},\mathbf{x'},ip)
 \cdot \overline{\mathbf{P}'}(\mathbf{x'},p).\label{134}\end{eqnarray}   where we introduced the definition $\overline{\mathbf{B}'}^{(v)}(\mathbf{x},p)=\frac{\boldsymbol{\nabla}\times\overline{\mathbf{D}'}^{(v)}(\mathbf{x},p)}{-p/c}+\frac{\mathbf{B}(\mathbf{x},t_0)}{p}$ (here we used the condition $\mathbf{B}(\mathbf{x},t_0)=\mathbf{B}^{(v)}(\mathbf{x},t_0)$). In the time domain we thus directly obtain 
\begin{eqnarray}
\mathbf{B}^{(d)}(\mathbf{x},t)=\int_0^{t-t_0}d\tau\int d^3\mathbf{x'}\frac{1}{c}\boldsymbol{\nabla}\times\partial_\tau\mathbf{U}_v(\tau,\mathbf{x},\mathbf{x'})\nonumber\\ \cdot\mathbf{P}(\mathbf{x'},t-\tau)\nonumber\\
\label{135}
\end{eqnarray}
which yields $\mathbf{B}^{(d)}(\mathbf{x},t_0)=0$ as expected.\\  
 \indent We emphasize here once again the fundamental role played by the boundary conditions at $t_0$.  What we showed is that at the initial time $t_0$ we have  $\mathbf{D}^{(d)}(\mathbf{x},t_0)=0$ and thus $\mathbf{D}^{(v)}(\mathbf{x},t_0)=\mathbf{D}(\mathbf{x},t_0)$ which means, by definition of our vacuum modes, $\mathbf{E}^{(v)}(\mathbf{x},t_0)=\mathbf{D}(\mathbf{x},t_0)$. In other words, the electric field associated with vacuum modes equals the total displacement fields at the initial time. This is interesting since it also implies $\mathbf{E}(\mathbf{x},t_0)=\mathbf{E}^{(v)}(\mathbf{x},t_0)-\mathbf{P}(\mathbf{x},t_0)$ (which means $\mathbf{E}^{(d)}(\mathbf{x},t_0)=-\mathbf{P}(\mathbf{x},t_0)$). This can be written after separation into transverse and longitudinal parts  as
 \begin{eqnarray}
\mathbf{E}_\bot(\mathbf{x},t_0)=\mathbf{E}^{(v)}(\mathbf{x},t_0)-\mathbf{P}_\bot(\mathbf{x},t_0)\nonumber\\=\mathbf{D}^{(v)}(\mathbf{x},t_0)-\mathbf{P}_\bot(\mathbf{x},t_0)\nonumber\\ 
\textrm{i.e., }\mathbf{E}_\bot^{(d)}(\mathbf{x},t_0)=-\mathbf{P}_\bot(\mathbf{x},t_0)\label{136}
\end{eqnarray} for the transverse (solenoidal or divergence- free) components
and \begin{eqnarray}
\mathbf{E}_{||}(\mathbf{x},t_0)=-\mathbf{P}_{||}(\mathbf{x},t_0)\label{gourbi}
\end{eqnarray} for the longitudinal (irrotational or curl-free) components. Eq.~\ref{gourbi} is well known in QED since it rigorously agrees with the definition of the longitudinal field obtained in usual Coulomb gauge using the $\textbf{A}$ potential  instead of $\textbf{F}$.
\subsection{Separation between fluctuating and induced current: macroscopic versus microscopic description}
Untill now we didn't specify the form of the dipole density $\mathbf{P}(\mathbf{x},t)$. The separation between source and free terms for the field was therefore analyzed from a microscopic perspective where the diffracted fields $\mathbf{E}^{(d)}(\mathbf{x},t)$ and $\mathbf{B}^{(d)}(\mathbf{x},t)$ was generated by the full microscopic current. In order to generalize the description given in section III for the homogeneous medium we will now use the separation Eq.~\ref{18} of $\mathbf{P}(\mathbf{x},t)$ into a fluctuating term  $\mathbf{P}^{(0)}(\mathbf{x},t)$ and an induced  contribution $\int_{0}^{t-t_0}\chi(\mathbf{x},\tau)d\tau\mathbf{E}(\mathbf{x},t-\tau)$ of essentially classical origin.
Using the Laplace transform we get $\overline{\mathbf{P}'}(\mathbf{x},p)=\overline{\mathbf{P}'}^{(0)}(\mathbf{x},p)+\overline{\chi}(\mathbf{x},p)\overline{\mathbf{E}'}(\mathbf{x},p)$. Now from the previous section we have therefore for the Laplace transform of the electric field  the following Lippman-Schwinger integral equation  
\begin{eqnarray}
\overline{\mathbf{E}'}(\mathbf{x},p)=\overline{\mathbf{D}'}^{(v)}(\mathbf{x},p)
-\int d^3\mathbf{x'} \frac{p^2}{c^2}\mathbf{G}_v(\mathbf{x},\mathbf{x'},ip)
\nonumber\\ \cdot [\bar{\chi}(\mathbf{x}',p)\overline{\mathbf{E}'}(\mathbf{x'},p) +\overline{\mathbf{P}'}^{(0)}(\mathbf{x'},p)].\nonumber\\ \label{149}
\end{eqnarray}
In order to get a meaningful separation of the total field we here define 
\begin{eqnarray}
\overline{\mathbf{E}'}^{(0)}(\mathbf{x},p)=\overline{\mathbf{D}'}^{(v)}(\mathbf{x},p)
-\int d^3\mathbf{x'} \frac{p^2}{c^2}\mathbf{G}_v(\mathbf{x},\mathbf{x'},ip)
\nonumber\\ \cdot \bar{\chi}(\mathbf{x}',p)\overline{\mathbf{E}'}^{(0)}(\mathbf{x'},p) \nonumber\\ \label{150}
\end{eqnarray} and 
\begin{eqnarray}
\mathbf{G}_\chi(\mathbf{x},\mathbf{x''},ip)=\mathbf{G}_v(\mathbf{x},\mathbf{x''},ip)
-\int d^3\mathbf{x'} \frac{p^2}{c^2}\mathbf{G}_v(\mathbf{x},\mathbf{x'},ip)
\nonumber\\ \cdot \bar{\chi}(\mathbf{x}',p)\mathbf{G}_\chi(\mathbf{x'},\mathbf{x''},ip)\nonumber\\ \label{151}
\end{eqnarray}
We have clearly 
\begin{eqnarray}
\boldsymbol{\nabla}\times\boldsymbol{\nabla}\times\mathbf{G}_\chi(\mathbf{x},\mathbf{x'},ip)+\frac{p^2}{c^2}(1+\bar{\chi}(\mathbf{x}',p))\mathbf{G}_\chi(\mathbf{x},\mathbf{x'},ip)\nonumber\\
=\boldsymbol{\delta}(\mathbf{x}-\mathbf{x'})\nonumber\\ \label{152}
\end{eqnarray}
and 
\begin{eqnarray}
\boldsymbol{\nabla}\times\boldsymbol{\nabla}\times\overline{\mathbf{E}'}^{(0)}(\mathbf{x},p)+\frac{p^2}{c^2}(1+\bar{\chi}(\mathbf{x}',p))\overline{\mathbf{E}'}^{(0)}(\mathbf{x},p)=0\nonumber\\ \label{153}
\end{eqnarray}
This allows us to interpret $\mathbf{G}_\chi(\mathbf{x},\mathbf{x'},ip)$ as the green function of the inhomogeneous dielectric medium  while $\overline{\mathbf{E}'}^{(0)}(\mathbf{x},p)$ is a free solution of Maxwell's equation in the dielectric medium in absence of the fluctuating source $\overline{\mathbf{P}'}^{(0)}(\mathbf{x'},p)$. 
By direct replacement of Eqs.~\ref{150}  and \ref{151} into  \ref{149} one get 
\begin{eqnarray}
\overline{\mathbf{E}'}(\mathbf{x},p)=\overline{\mathbf{E}'}^{(0)}(\mathbf{x},p)
-\int d^3\mathbf{x'} \frac{p^2}{c^2}\mathbf{G}_\chi(\mathbf{x},\mathbf{x'},ip)\nonumber\\ \cdot
\overline{\mathbf{P}'}^{(0)}(\mathbf{x'},p).\nonumber\\ \label{154}
\end{eqnarray} meaning that the total field can be seen as the sum of the free solution $\overline{\mathbf{E}'}^{(0)}(\mathbf{x},p)$ and of scattering contribution $\overline{\mathbf{E}'}^{(s)}(\mathbf{x},p)$ induced by the fluctuating source $\overline{\mathbf{P}'}^{(0)}(\mathbf{x'},p)$.\\
\indent In the time domain we get for the electric field
\begin{eqnarray}
\mathbf{E}^{(0)}(\mathbf{x},t)=\mathbf{D}^{(v)}(\mathbf{x},t)-\int_0^{t-t_0}d\tau\int d^3\mathbf{x'}\frac{\partial_\tau^2\mathbf{U}_v(\tau,\mathbf{x},\mathbf{x'}) }{c^2}\nonumber\\\cdot\int_{0}^{t-\tau-t_0} d\tau'\chi(\mathbf{x}',\tau')\mathbf{E}^{(0)}(\mathbf{x}',t-\tau-\tau')\nonumber\\-\int_{0}^{t-t_0}\chi(\mathbf{x},\tau')d\tau'\mathbf{E}^{(0)}(\mathbf{x},t-\tau')\nonumber\\
\label{155}
\end{eqnarray}
and for the magnetic field
\begin{eqnarray}
\mathbf{B}^{(0)}(\mathbf{x},t)=\mathbf{B}^{(v)}(\mathbf{x},t)\nonumber\\+\int_0^{t-t_0}d\tau\int d^3\mathbf{x'}\frac{1}{c}\boldsymbol{\nabla}\times\partial_\tau\mathbf{U}_v(\tau,\mathbf{x},\mathbf{x'})\nonumber\\ \cdot\int_{0}^{t-\tau-t_0} d\tau'\chi(\mathbf{x}',\tau')\mathbf{E}^{(0)}(\mathbf{x}',t-\tau-\tau')\nonumber\\
\label{135bb}
\end{eqnarray} Which are completed by the constitutive relation: $\mathbf{D}^{(0)}(\mathbf{x},t)=\int_{0}^{t-t_0}d\tau\chi(\mathbf{x}',\tau)\mathbf{E}^{(0)}(\mathbf{x},t-\tau)+\mathbf{E}^{(0)}(\mathbf{x},t)$.
Similarly we obtain for the new propagator:
\begin{eqnarray}
\mathbf{U}_\chi(t,\mathbf{x},\mathbf{x''})=\mathbf{U}_v(t,\mathbf{x},\mathbf{x''})-\int_0^{t}d\tau\int d^3\mathbf{x'}\frac{\partial_\tau^2\mathbf{U}_v(\tau,\mathbf{x},\mathbf{x'}) }{c^2}\nonumber\\ \cdot
\int_{0}^{t-\tau}\chi(\mathbf{x}',\tau')d\tau'\mathbf{U}_\chi(t-\tau-\tau',\mathbf{x'},\mathbf{x''})]\nonumber\\
-\int_{0}^{t}d\tau'\chi(\mathbf{x},\tau')\mathbf{U}_\chi(t-\tau',\mathbf{x},\mathbf{x''})\nonumber\\ \label{156}
\end{eqnarray} We have the important properties $\mathbf{U}_\chi(t=0,\mathbf{x},\mathbf{x''})=0$, $\partial_t\mathbf{U}_\chi(t,\mathbf{x},\mathbf{x'})|_{t=0}=c^{2}\delta^{3}(\mathbf{x}-\mathbf{x}')\mathbf{I}$. The total electromagnetic field in the time domain is thus expressed as:
\begin{eqnarray}
\mathbf{E}(\mathbf{x},t)=\mathbf{E}^{(0)}(\mathbf{x},t)-\int_0^{t-t_0}d\tau\int d^3\mathbf{x'}\frac{\partial_\tau^2\mathbf{U}_\chi(\tau,\mathbf{x},\mathbf{x'}) }{c^2}\nonumber\\ \cdot\mathbf{P}^{(0)}(\mathbf{x'},t-\tau)-\mathbf{P}^{(0)}(\mathbf{x},t)\nonumber\\ \label{157}
\end{eqnarray}
\begin{eqnarray}
\mathbf{B}(\mathbf{x},t)=\mathbf{B}^{(0)}(\mathbf{x},t)\nonumber\\ +\int_0^{t-t_0}d\tau\int d^3\mathbf{x'}\frac{1}{c}\boldsymbol{\nabla}\times\partial_\tau\mathbf{U}_\chi(\tau,\mathbf{x},\mathbf{x'})\cdot\mathbf{P}^{(0)}(\mathbf{x}',t-\tau)\nonumber\\
\label{135bbb}
\end{eqnarray} with $\mathbf{E}=\mathbf{E}^{(s)}+\mathbf{E}^{(0)}$, $\mathbf{B}=\mathbf{B}^{(s)}+\mathbf{B}^{(0)}$  Finally, the constitutive relation for the scattered displacement field  $\mathbf{D}^{(s)}=\mathbf{D}-\mathbf{D}^{(0)}$ reads: $\mathbf{D}^{(s)}(\mathbf{x},t)=\int_{0}^{t-t_0}d\tau\chi(\mathbf{x}',\tau)\mathbf{E}^{(s)}(\mathbf{x},t-\tau)+\mathbf{E}^{(s)}(\mathbf{x},t)+ \mathbf{P}^{(0)}(\mathbf{x},t)$. The boundary conditions at $t_0$ together with the field equations determine the full evolution and we have:   
\begin{eqnarray}
\mathbf{E}^{(0)}(\mathbf{x},t_0)=\mathbf{D}^{(0)}(\mathbf{x},t_0)=\mathbf{D}^{(v)}(\mathbf{x},t_0)\nonumber\\
\mathbf{D}^{(s)}(\mathbf{x},t_0)=\mathbf{E}^{(s)}(\mathbf{x},t_0)+\mathbf{P}^{(0)}(\mathbf{x},t_0)=0\nonumber\\
\mathbf{B}^{(s)}(\mathbf{x},t_0)=0\nonumber\\
\mathbf{B}^{(0)}(\mathbf{x},t_0)=\mathbf{B}^{(v)}(\mathbf{x},t_0)=\mathbf{B}(\mathbf{x},t_0).
\end{eqnarray}
\indent We point out that the description of the longitudinal field should be treated independently in this formalism since at any time $t$ we have the constraint $\mathbf{E}_{||}(\mathbf{x},t)=-\mathbf{P}_{||}(\mathbf{x},t)$ which shows that fluctuating current and field are not independent. More precisely, if we insert the constraint $\mathbf{E}_{||}(\mathbf{x},t)=-\mathbf{P}_{||}(\mathbf{x},t)$ into the Lagrangian formalism developed in section II we get a new effective Lagrangian density for the longitudinal field which reads: 
\begin{eqnarray}
\mathcal{L}_{||}(\mathbf{x},t)=\int_{0}^{+\infty}d\omega\frac{(\partial_t\mathbf{X}_{\omega,||}(\mathbf{x},t))^2-\omega^2\mathbf{X}_{\omega,\||}^2(\mathbf{x},t)}{2}\nonumber\\-\frac{\mathbf{P}_{||}^2(\mathbf{x},t)}{2}.\nonumber\\\label{2bb}
\end{eqnarray} From Eq.~\ref{2bb} we get the Euler-Lagrange equation:
\begin{eqnarray}
\partial_t^2\mathbf{X}_{\omega,||}(\mathbf{x},t)+\omega^2\mathbf{X}_{\omega,||}(\mathbf{x},t)=-\sqrt{\frac{2\sigma_\omega(\mathbf{x})}{\pi}}\mathbf{P}_{||}(\mathbf{x},t),\nonumber\\ \label{8b}
\end{eqnarray}
with $\mathbf{P}_{||}(\mathbf{x},t)=\int_{0}^{+\infty}d\omega\sqrt{\frac{2\sigma_\omega(\mathbf{x})}{\pi}}\mathbf{X}_{\omega,||}(\mathbf{x},t)$ and which agrees with Eq.~\ref{8} if the constraint $\mathbf{E}_{||}(\mathbf{x},t)=-\mathbf{P}_{||}(\mathbf{x},t)$ is used. Now, the formal solution of Eq.~\ref{8b} is obtained from Eq.~\ref{17bb}
\begin{eqnarray}
\mathbf{X}_{\omega,||}(\mathbf{x},t)
=\mathbf{X}^{(0)}_{\omega,||}(\mathbf{x},t)\nonumber\\-\sqrt{\frac{2\sigma_\omega(\mathbf{x})}{\pi}}\int_0^{t-t_0}d\tau \frac{\sin{\omega\tau}}{\omega}\mathbf{P}_{||}(\mathbf{x},t-\tau)\label{17bb}
\end{eqnarray} and allows for a separation between a fluctuating term $\mathbf{X}^{(0)}_{\omega,||}(\mathbf{x},t)$ and a source term $\mathbf{X}^{(s)}_{\omega,||}(\mathbf{x},t)$. From this we naturally deduce
\begin{eqnarray}
\mathbf{P}_{||}(\mathbf{x},t)=\mathbf{P}_{||}^{(0)}(\mathbf{x},t)-\int_{0}^{t-t_0}d\tau\chi(\mathbf{x},\tau)\mathbf{P}_{||}(\mathbf{x},t-\tau).\nonumber\\ \label{dde}
\end{eqnarray} Integral Eqs.~\ref{17bb} or \ref{dde} could in principle be solved iteratively in order to find expressions $\mathbf{P}_{||}$ and $\mathbf{X}_{\omega,||}$ which are linear functionals of 
$\mathbf{P}_{||}^{(0)}$ and $\mathbf{X}_{\omega,||}^{(0)}$. Alternatively, this can be done self-consistently using the Laplace transform of Eq.~\ref{dde} which reads   
$\overline{\mathbf{P'}}_{||}(\mathbf{x},p)=\overline{\mathbf{P'}}_{||}^{(0)}(\mathbf{x},p)-\bar{\chi}(\mathbf{x},p)\overline{\mathbf{P'}}_{||}(\mathbf{x},p)$ and leads to: 
\begin{eqnarray}
\overline{\mathbf{P'}}_{||}(\mathbf{x},p)=-\frac{\overline{\mathbf{P'}}_{||}^{(0)}(\mathbf{x},p)}{1+\bar{\chi}(\mathbf{x},p)}.
\end{eqnarray}
In the time domain we have thus 
\begin{eqnarray}
\mathbf{P}_{||}(\mathbf{x},t)=\mathbf{P}_{||}^{(0)}(\mathbf{x},t)-\int_0^{t-t_0}\chi_{eff}(\mathbf{x},\tau)\mathbf{P}_{||}^{(0)}(\mathbf{x},t-\tau)\nonumber\\
\end{eqnarray}
with the effective susceptibility defined as\begin{eqnarray}
\chi_{eff}(\mathbf{x},\tau)=\int_{\gamma-i\infty}^{\gamma+i\infty}\frac{i dp}{2\pi}\frac{e^{p\tau}\bar{\chi}(\mathbf{x},p)}{1+\bar{\chi}(\mathbf{x},p)}
\end{eqnarray} We have  equivalently for the effective susceptibility $\chi_{eff}(\mathbf{x},\tau)=\int_{-\infty}^{+\infty}\frac{d\omega}{2\pi}e^{-i\omega\tau}\frac{\tilde{\varepsilon}(\mathbf{x},\omega)-1}{\tilde{\varepsilon}(\mathbf{x},\omega)}$. After closing the contour in the complex plane we get $\chi_{eff}(\mathbf{x},\tau)=i\sum_{m}\frac{1}{\frac{\partial\tilde{\varepsilon}(\mathbf{x},\omega)}{\partial\omega}|_{\Omega_m}}e^{-i\Omega_m\tau}+cc.$ where the sum is taken over the longitudinal modes solutions of $\tilde{\varepsilon}(\mathbf{x},\Omega_m)=0$ (with $\Omega''_m<0$ and $\Omega'_m>0$ by definition). The frequencies considered here are in general spatially dependent  since the medium is inhomogeneous and are therefore very often difficult to find. In the limit of the homogeneous lossless medium we obtain the Hopfield-Fano~\cite{Fano1956,Hopfield1958} model.\\
\indent We emphasize  that the present description of the polariton field contrast with the integral solution of Eq.~\ref{G}  $\boldsymbol{\nabla}\times\boldsymbol{\nabla}\times\widetilde{\textbf{E}}(\Omega)-\frac{\Omega^2}{c^2}\widetilde{\varepsilon}(\Omega)\widetilde{\textbf{E}}(\Omega)=\frac{\Omega^2}{c^2}\widetilde{\textbf{P}}^{(in)}(\Omega)$ which was obtained for the Homogeneous medium:    
\begin{eqnarray}
\widetilde{\textbf{E}}(\mathbf{x},\Omega)=\frac{\Omega^2}{c^2}\int d^3\mathbf{x'} \mathbf{G}_{\chi}(\mathbf{x},\mathbf{x'},\Omega)\cdot
\widetilde{\mathbf{P}}^{(in)}(\mathbf{x'},\Omega)\nonumber\\
\end{eqnarray} and which included only a scattering contribution $(s)$ due to the cancellation of the $(0)$ term for $t_0\rightarrow-\infty$. Here, we can not neglect or cancel the $(0)$ mode solutions since in general the medium is not necessary  lossy at  spatial infinity. This will be in particular the case for all scattering problems involving a localized system such as a metal or dielectric antenna supporting plasmon-polariton localized modes. However, mostly all studies, inspirited by the success of the Huttner-Barnett model~\cite{Huttner1992a} for the homogeneous lossy medium, and following the Langevin-Noise method proposed Gruner and Welsch~\cite{Gruner1996,Suttorp2004a,Suttorp2004b,Suttorp2007,Philbin2010}, neglected or often even completely omitted  the contribution of the $(0)$ modes. Still, these $(0)$ modes are crucial for preserving the unitarity of the  full matter field dynamics and can not be rigorously omitted. Only in those case where  absorption is present at infinity  we can omit the $(0)$ modes.\\
\indent To clarify this point further consider  a medium made of a spatially homogeneous background susceptibility $\bar{\chi_1}(p)$ and of a localized susceptibility $\bar{\chi_2}(\mathbf{x},p)$
such as $\bar{\chi}(\mathbf{x},p)\rightarrow 0$ at spatial infinity.  The electromagnetic field propagating into the medium with total permittivity   $\bar{\chi}(\mathbf{x},p)=\bar{\chi_1}(p)+\bar{\chi_2}(\mathbf{x},p)$ can be thus formally developed using the Lippeman-Schwinger equation as:
\begin{eqnarray}
\overline{\mathbf{E}'}(\mathbf{x},p)=\overline{\mathbf{E}'}^{(v)}(\mathbf{x},p)
-\int d^3\mathbf{x'} \frac{p^2}{c^2}\mathbf{G}_v(\mathbf{x},\mathbf{x'},ip)\nonumber\\ \cdot
[\bar{\chi}(\mathbf{x}',p)\overline{\mathbf{E}'}(\mathbf{x},p)+\overline{\mathbf{P}'}^{(0)}(\mathbf{x'},p)].\nonumber\\
= \overline{\mathbf{E}'}^{(1)}(\mathbf{x},p)
-\int d^3\mathbf{x'} \frac{p^2}{c^2}\mathbf{G}_{\chi_1}(\mathbf{x},\mathbf{x'},ip)\nonumber\\ \cdot
[\bar{\chi_2}(\mathbf{x}',p)\overline{\mathbf{E}'}(\mathbf{x},p)+\overline{\mathbf{P}'}^{(0)}(\mathbf{x'},p)].\nonumber\\\label{154bb}
\end{eqnarray}  
 where we have defined a new background free field 
\begin{eqnarray}
\overline{\mathbf{E}'}^{(1)}(\mathbf{x},p)=\overline{\mathbf{D}'}^{(v)}(\mathbf{x},p)
-\int d^3\mathbf{x'} \frac{p^2}{c^2}\mathbf{G}_v(\mathbf{x},\mathbf{x'},ip)
\nonumber\\ \cdot \bar{\chi_1}(\mathbf{x}',p)\overline{\mathbf{E}'}^{(1)}(\mathbf{x'},p) \nonumber\\ \label{150b}
\end{eqnarray} and a new Green dyadic tensor for the  background medium
\begin{eqnarray}
\mathbf{G}_{\chi_1}(\mathbf{x},\mathbf{x''},ip)=\mathbf{G}_v(\mathbf{x},\mathbf{x''},ip)
\nonumber\\-\int d^3\mathbf{x'} \frac{p^2}{c^2}\mathbf{G}_v(\mathbf{x},\mathbf{x'},ip)
 \cdot \bar{\chi_1}(\mathbf{x}',p)\mathbf{G}_{\chi_1}(\mathbf{x'},\mathbf{x''},ip).\nonumber\\ \label{151b}
\end{eqnarray} We have naturally $\boldsymbol{\nabla}\times\boldsymbol{\nabla}\times\mathbf{G}_{\chi_1}(\mathbf{x},\mathbf{x'},ip)+\frac{p^2}{c^2}(1+\bar{\chi_1}(\mathbf{x}',p))\mathbf{G}_{\chi_1}(\mathbf{x},\mathbf{x'},ip)=\boldsymbol{\delta}(\mathbf{x}-\mathbf{x'})$ and similarly $\boldsymbol{\nabla}\times\boldsymbol{\nabla}\times\overline{\mathbf{E}'}^{(1)}(\mathbf{x},p)+\frac{p^2}{c^2}(1+\bar{\chi_1}(\mathbf{x}',p))\overline{\mathbf{E}'}^{(1)}(\mathbf{x},p)=0$.\\
\indent Importantly, in  the time domain we can write:  
\begin{eqnarray}
\mathbf{E}(\mathbf{x},t)=\mathbf{E}^{(1)}(\mathbf{x},t)-\int_0^{t-t_0}d\tau\int d^3\mathbf{x'}\frac{\partial_\tau^2\mathbf{U}_{\chi_1}(\tau,\mathbf{x},\mathbf{x'}) }{c^2}\nonumber\\ \cdot[\mathbf{P}^{(0)}(\mathbf{x'},t-\tau) +\int_{0}^{t-\tau}d\tau'\chi_2(\mathbf{x}',\tau')\textbf{E}(\mathbf{x}',t-\tau-\tau')]\nonumber\\ -\mathbf{P}^{(0)}(\mathbf{x},t)+\int_{0}^{t}d\tau\chi_2(\mathbf{x},\tau)\textbf{E}(\mathbf{x},t-\tau).\nonumber\\ \label{15bbb}
\end{eqnarray} We can check that we have $\mathbf{E}(\mathbf{x},t_0)=\mathbf{E}^{(1)}(\mathbf{x},t_0)-\mathbf{P}^{(0)}(\mathbf{x},t_0)$. Moreover, since from Eq.~\ref{150b} (written in the time domain) we have $\mathbf{E}^{(1)}(\mathbf{x},t_0)=\mathbf{D}^{(v)}(\mathbf{x},t_0)$ and since $\mathbf{P}^{(0)}(\mathbf{x},t_0)=\mathbf{P}(\mathbf{x},t_0)$, we deduce that at the initial time $t_0$ $\mathbf{D}(\mathbf{x},t_0)=\mathbf{E}(\mathbf{x},t_0)+\mathbf{P}(\mathbf{x},t_0)=\mathbf{E}^{(1)}(\mathbf{x},t_0)=\mathbf{D}^{(v)}(\mathbf{x},t_0)$ as it should be to agree with the general formalism presented in Section IV A.\\
\indent Now, since the background dissipative medium is not spatially bound the $\mathbf{E}^{(1)}(\mathbf{x},t)$ field  associated with damped modes will vanish if $t_0\rightarrow-\infty$ as explained before. We could therefore be tempted~\cite{Gruner1996,Suttorp2004a,Suttorp2004b,Suttorp2007,Philbin2010} to eliminate from the start $\mathbf{E}^{(1)}(\mathbf{x},t)$ and thus get in the  Laplace transform language the effective formula: 
\begin{eqnarray}
\overline{\mathbf{E}'}(\mathbf{x},p)=
-\int d^3\mathbf{x'} \frac{p^2}{c^2}\mathbf{G}_{\chi_1}(\mathbf{x},\mathbf{x'},ip)\nonumber\\ \cdot
[\bar{\chi_2}(\mathbf{x}',p)\overline{\mathbf{E}'}(\mathbf{x},p)+\overline{\mathbf{P}'}^{(0)}(\mathbf{x'},p)]\nonumber\\ 
=-\int d^3\mathbf{x'} \frac{p^2}{c^2}\mathbf{G}_{\chi}(\mathbf{x},\mathbf{x'},ip)\cdot\overline{\mathbf{P}'}^{(0)}(\mathbf{x'},p)].
\label{154bbb}
\end{eqnarray} with the total Green dyadic function \begin{eqnarray}
\mathbf{G}_{\chi}(\mathbf{x},\mathbf{x''},ip)=\mathbf{G}_{\chi_1}(\mathbf{x},\mathbf{x''},ip)
\nonumber\\-\int d^3\mathbf{x'} \frac{p^2}{c^2}\mathbf{G}_{\chi_1}(\mathbf{x},\mathbf{x'},ip)
 \cdot \bar{\chi_2}(\mathbf{x}',p)\mathbf{G}_{\chi}(\mathbf{x'},\mathbf{x''},ip).\nonumber\\ \label{micro}
\end{eqnarray} obeying to  Eq.~\ref{152} with the total permittivity  $\bar{\chi}(\mathbf{x},p)=\bar{\chi_1}(p)+\bar{\chi_2}(\mathbf{x},p)$. It is straightforward to check that $\mathbf{G}_{\chi}(\mathbf{x},\mathbf{x''},ip)$ satisfies also Eq.~\ref{151} so that it is the same Green function.\\
\indent However, removing $\mathbf{E}^{(1)}(\mathbf{x},t)$ from  the start in
Eq.~\ref{15bbb} would mean  that the boundary conditions at the initial time $t_0$ have been obliviated since we should now necessarily have $\mathbf{D}(\mathbf{x},t_0)=\mathbf{E}(\mathbf{x},t_0)+\mathbf{P}(\mathbf{x},t_0)=\mathbf{E}^{(1)}(\mathbf{x},t_0)=0$. This corresponds to a very specific boundary condition which is certainly allowed in classical physics (where we can put $c_{\alpha,j}(t_0)=0$) but which in the quantum formalism means that  we break the unitarity of the evolution. To say it differently, it means that in the Langevin noise formalism~\cite{Gruner1996,Suttorp2004a,Suttorp2004b,Suttorp2007,Philbin2010} the photon field $\textbf{F}$ is not anymore an independent canonical contribution to the evolution since all electromagnetic fields are induced by the material part. The  Green formalism presented by Gruner and Welsch~\cite{Gruner1996}, and abundantly used since~\cite{Scheel1998,Dung1998,Dung2000,Scheel2001,Matloob1999,Matloob2004,Fermani2006,Raabe2007,Amooshahi2008,Scheelreview2008,Dzotjan2010,Cano2011,Hummer2013,Chen2013,Delga2014,Hakami2014,Choquette2012,Grimsmo2013,Rousseaux2016}, represents therefore an alternative theory which rigorously speaking is not equivalent, contrary to the claim in Refs.~\cite{Suttorp2004a,Suttorp2004b,Suttorp2007,Philbin2010}, to the Lagrangian formalism discussed in section II for the general Huttner-Barnett model~\cite{Huttner1992a}. Our analysis, as already explained in the introduction, agrees with the general studies made in the 1970's and 1980's in QED~\cite{Senitzky1973,Ackerhalt1973,Milonni1973,Dalibard1982,Milonnibook,Milonni1982} since one must include with an equal footings  both the field and matter fluctuations in  a self consistent QED in order to preserve rigorously unitarity and causality.\\
\subsection{Discussions concerning unitarity and Hamiltonians}
\indent Two issues are important to emphasize here. 
First, observe that in the limit where the background susceptibility $\chi_1(\mathbf{x},\tau)$ vanishes then the term  $\mathbf{E}^{(1)}(\mathbf{x},t)=\mathbf{D}^{(v)}(\mathbf{x},t)$ in general does not cancel at any time, and therefore the coupling to photonic modes can not be omitted even in practice from the evolution at finite time $t$. This is is particularly important in nanophotonics where an incident exiting photon field interact with a localized nano-antenna. It is therefore crucial to analyze further the impact of our findings on the quantum dynamics of polaritons in presence of sources such as quantum fluorescent emitters. This will the subject of a subsequent article.\\
\indent The second issue concerns the Hamiltonian definition in the new formalism.  Indeed, the definition of the full system Hamiltonian $H(t)$ was previously given for the homogeneous medium case in section III G. We showed (see Eq.~\ref{158o}) that $H(t)$ is given by 
\begin{eqnarray}
H(t)=\int d^3\mathbf{x}:\frac{\mathbf{B}(\mathbf{x},t_0)^2+\mathbf{E}(\mathbf{x},t_0)^2}{2}: +H_M^{(0)}(t).\nonumber\\ \label{158ob}
\end{eqnarray} 
 where $H_M^{(0)}(t)$ is the material Hamiltonian defined in Eq.~\ref{30c} and which depends only on the free mode operators $\mathbf{f}^{\dagger(0)}_\omega(\mathbf{x},t)$, $\mathbf{f}^{(0)}_\omega(\mathbf{x},t)$. This $H_M^{(0)}(t)=\int d^3\mathbf{x}\int_{0}^{+\infty}d\omega\hbar\omega\mathbf{f}^{\dagger(0)}_\omega(\mathbf{x},t)\mathbf{f}^{(0)}_\omega(\mathbf{x},t)$ is the Hamiltonian considered by the Noise Langevin approach and the remaining term (see Eq.~\ref{rema}) $H_{\textrm{rem.}}(t)=\int d^3\mathbf{x}:\frac{\mathbf{B}(\mathbf{x},t_0)^2+\mathbf{E}(\mathbf{x},t_0)^2}{2}:$ is an additional constant of motion. This constant proved to be irrelevant for all practical purposes since the only surviving electromagnetic fields  (i.e. if $t_0\rightarrow-\infty$) are the induced $(s)$ modes which are generated by the fluctuating dipole density $\mathbf{P}^{(0)}(\mathbf{x},t)$ (see however the different remarks concerning time symmetry at the end of section III G). Now, for the inhomogeneous problem the complete reasoning leading to Eq.~\ref{158ob} is still rigorously valid. The main difference being that in general  the constant of motion  $H_{\textrm{rem.}}(t)=\int d^3\mathbf{x}:\frac{\mathbf{B}(\mathbf{x},t_0)^2+\mathbf{E}(\mathbf{x},t_0)^2}{2}:$ is not irrelevant at all since the $(0)$ electromagnetic modes are not in general vanishing even if $t_0\rightarrow-\infty$.\\
\indent In order to clarify this point we should now physically interpret the term $H_{\textrm{rem.}}(t)$. We first start with the less relevant term in optics: the longitudinal polariton.  Indeed, the longitudinal field is here decoupled from the rest and evolves independently using the Lagrangian density $\mathcal{L}_{||}(\mathbf{x},t)$ defined in Eq.~\ref{2bb} (this should not be necessarily true if the polaritons are coupled to external sources such as fluorescent emitters). We thus get the following Hamiltonian
  \begin{eqnarray}
H_{||}(t)=:\int d^3\mathbf{x}[\int_{0}^{+\infty}d\omega\frac{(\partial_t\mathbf{X}_{\omega,||}(\mathbf{x},t))^2+\omega^2\mathbf{X}_{\omega,\||}^2(\mathbf{x},t)}{2}\nonumber\\+\frac{\mathbf{P}_{||}^2(\mathbf{x},t)}{2}]:.\nonumber\\ \label{2c}
\end{eqnarray}  $H_{||}(t)$  is a constant of motion and can used (with the Hamiltonian formalism) to deduce the evolution equation (see Eq.~\ref{8b}) and the solution Eq.~\ref{17bb}. Since $H_{||}(t)$  is a constant of motion we have $H_{||}(t)=H_{||}(t_0)$ and from the form of the solution we obtain the equivalent formula   
\begin{eqnarray}
H_{||}(t)=\int d^3\mathbf{x}[\int_{0}^{+\infty}d\omega\hbar\omega\mathbf{f}^{\dagger(0)}_{\omega,||}(\mathbf{x},t)\mathbf{f}^{(0)}_{\omega,||}(\mathbf{x},t)\nonumber\\+\frac{:\mathbf{P}_{||}^{(0)}(\mathbf{x},t_0)^2:}{2}]. \label{2d}
\end{eqnarray} where we have clearly by definition 	
		\begin{eqnarray}
H_{rem,||}(t)=\int d^3\mathbf{x}\frac{:\mathbf{P}_{||}^{(0)}(\mathbf{x},t_0)^2:}{2} 
\end{eqnarray}and thus $H_{||}(t)=H_{M,||}^{(0)}(t)+H_{rem,||}(t)$ (with $H_{M,||}^{(0)}(t)=\int d^3\mathbf{x}[\int_{0}^{+\infty}d\omega\hbar\omega\mathbf{f}^{\dagger(0)}_{\omega,||}(\mathbf{x},t)\mathbf{f}^{(0)}_{\omega,||}(\mathbf{x},t)$).
While $H_{rem,||}(t)$ is a constant of motion it is not irrelevant here since the equivalence of Eq.~\ref{2c} and Eq.~\ref{2d} leads to the complete solution Eq.~\ref{17bb} for $\mathbf{X}_{\omega,||}(\mathbf{x},t)$. Oppositely,  taking $H_{M,||}^{(0)}(t)$ and omitting $H_{rem,||}(t)$ would lead to the free solution $\mathbf{X}_{\omega,||}^{(0)}(\mathbf{x},t)$ in contradiction with the dynamical law. This again stresses the importance of keeping all contributions in the evolution and Hamiltonian. \\ 
\indent In order to analyze the transverse field Hamiltonian we should comment further on the difference between the description using the $\textbf{F}$ potential used in this work and the most traditional treatment using  the \textbf{A} potential (in the Coulomb Gauge). Indeed, by analogy with the separation between $(d)$ and $(v)$ modes discussed in section VI A we can using the \textbf{A} potential vector representation get a separation between free-space modes $\mathbf{A}^{(v')}(\mathbf{x},t)$ and source field $\mathbf{A}^{(d')}(\mathbf{x},t)$. Here, we label these modes by an additional prime for reasons which will become clear below. First, the source field contribution $(d')$ is given by 
\begin{eqnarray}
\mathbf{A}^{(d')}(\mathbf{x},t)=\frac{1}{c}\int_0^{t-t_0}d\tau\int d^3\mathbf{x'}\Delta_v(\tau,|\mathbf{x}-\mathbf{x'}|)\nonumber\\ \cdot\mathbf{J}_\bot(\mathbf{x'},t-\tau)\nonumber\\
 \label{138}
\end{eqnarray} and with by definition $\mathbf{J}_\bot(\mathbf{x},t)=\partial_t\mathbf{P}_\bot(\mathbf{x},t)$. Importantly we have $\mathbf{A}^{(d')}(\mathbf{x},t_0)=0$ meaning also $\mathbf{A}^{(v')}(\mathbf{x},t_0)=\mathbf{A}(\mathbf{x},t_0)$.  The free-space modes $(v')$ are easily obtained using a plane wave expansion as 
\begin{eqnarray}
\mathbf{A}^{(v')}(\mathbf{x},t)=\sum_{\alpha,j} -ic\sqrt{\frac{\hbar}{2\omega_\alpha}}a^{(v)}_{\alpha,j}(t)\boldsymbol{\hat{\epsilon}}_{\alpha,j}\Phi_\alpha(\mathbf{x})+ cc.\nonumber\\ \label{139}
\end{eqnarray} where $a^{(v)}_{\alpha,j}(t)=a_{\alpha,j}(t_0)e^{-i\omega_\alpha(t-t_0)}$. Using the $\textbf{A}$ potential we therefore get for the free electromagnetic fields: 
\begin{eqnarray}
\mathbf{B}^{(v')}(\mathbf{x},t)=\sum_{\alpha,j} \sqrt{\frac{\hbar \omega_\alpha}{2}}a^{(v)}_{\alpha,j}(t)\hat{\mathbf{k}}_\alpha\times\boldsymbol{\hat{\epsilon}}_{\alpha,j}\Phi_\alpha(\mathbf{x})+ cc.\nonumber\\
\mathbf{E}^{(v')}_\bot(\mathbf{x},t)=\sum_{\alpha,j} \sqrt{\frac{\hbar \omega_\alpha}{2}}a^{(v)}_{\alpha,j}(t)\boldsymbol{\hat{\epsilon}}_{\alpha,j}\Phi_\alpha(\mathbf{x})+ cc.\nonumber\\ \label{140nnn}
\end{eqnarray}
These transverse fields satisfy Maxwell's equation in vacuum like the free-fields given by Eq.~\ref{121} do as well.  However, it should now be clear that this two sets of free-fields given either by Eq.~\ref{140nnn} or  Eq.~\ref{121} are not equivalent. To see that we must express the source field $\mathbf{E}^{(d')}(\mathbf{x},t)$. The longitudinal contribution $\mathbf{E}_{||}^{(d')}(\mathbf{x},t)$ is given by the  usual instantaneous  Coulomb field \begin{eqnarray}
\mathbf{E}_{||}^{(d')}(\mathbf{x},t)=-\boldsymbol{\nabla}(V(\mathbf{x},t))\nonumber\\
=\boldsymbol{\nabla}\left(\int d^3\mathbf{x'}\frac{\boldsymbol{\nabla'}\cdot\mathbf{P}(\mathbf{x}',t)}{4\pi|\mathbf{x}-\mathbf{x}'|}\right)\nonumber\\=-\mathbf{P}_{||}(\mathbf{x},t)
\label{141}
\end{eqnarray} Since there is no other longitudinal contribution this leads to $\mathbf{E}_{||}(\mathbf{x},t_0)=-\mathbf{P}_{||}(\mathbf{x},t_0)
$ which is Eq.~\ref{gourbi}. As mentioned already this is the usual result.  However the most important term here is the transverse source field: $\mathbf{E}_\bot^{(d')}(\mathbf{x},t)=-\frac{1}{c}\partial\mathbf{A}^{(d')}(\mathbf{x},t)$. We get for this term
   \begin{eqnarray}
\mathbf{E}_\bot^{(d')}(\mathbf{x},t)=-\frac{1}{c^2}\int_0^{t-t_0}d\tau\int d^3\mathbf{x'}\Delta_v(\tau,|\mathbf{x}-\mathbf{x'}|)\nonumber\\ \cdot\partial_{t-\tau}^2\mathbf{P}_\bot(\mathbf{x'},t-\tau)\nonumber\\
 -\frac{1}{c^2}\int d^3\mathbf{x'}\Delta_v(t-t_0,|\mathbf{x}-\mathbf{x'}|)\cdot\partial_{t_0}\mathbf{P}_\bot(\mathbf{x'},t_0)\nonumber\\ \label{142nnn}
\end{eqnarray} which is also written as:  
\begin{eqnarray}
\mathbf{E}_\bot^{(d')}(\mathbf{x},t)=-\frac{1}{c^2}\int_0^{t-t_0}d\tau\int d^3\mathbf{x'}\mathbf{U}_{v,\bot}(\tau,\mathbf{x},\mathbf{x'})\nonumber\\ \cdot\partial_{t-\tau}^2\mathbf{P}(\mathbf{x'},t-\tau)\nonumber\\
 -\frac{1}{c^2}\int d^3\mathbf{x'}\mathbf{U}_{v,\bot}(t-t_0,\mathbf{x},\mathbf{x'}|)\cdot\partial_{t_0}\mathbf{P}(\mathbf{x'},t_0)\nonumber\\ \label{142}
\end{eqnarray}
  Equivalently we have in the frequency domain:
\begin{eqnarray}
\mathbf{E}_\bot^{(d')}(\mathbf{x},t)=-\int_{\gamma-i\infty}^{\gamma+i\infty}\frac{i dp}{2\pi} e^{p(t-t_0)}\int d^3\mathbf{x'}\mathbf{G}_v(\mathbf{x},\mathbf{x'},ip)\nonumber\\
\cdot [\frac{p^2}{c^2}\overline{\mathbf{P}'}(\mathbf{x'},p)-p\mathbf{P}_\bot(\mathbf{x'},t_0)] \nonumber\\ \label{143}
\end{eqnarray}
This  transverse scattered field vanishes at $t_0$ : 
\begin{eqnarray}
\mathbf{E}_\bot^{(d')}(\mathbf{x},t_0)=0 \label{144}
\end{eqnarray}
and therefore at this initial time it differs  by an amount $-\mathbf{P}_\bot(\mathbf{x},t_0)$ from the scattered field  given by Eq.~\ref{136}. We thus get $\mathbf{E}^(\mathbf{x},t_0)=\mathbf{E}_\bot^{(v')}(\mathbf{x},t_0)-\mathbf{P}_{||}\mathbf{x},t_0)$. Importantly, by comparing Eq.~\ref{136} and Eq.~\ref{144} we obtain a relation  for the free-space modes in the two representations using   either the $\textbf{F}$ or $\textbf{A}$ potential vectors:
\begin{eqnarray}
\mathbf{E}_\bot^{(v')}(\mathbf{x},t_0)=\mathbf{E}_\bot^{(v)}(\mathbf{x},t_0)-\mathbf{P}_\bot(\mathbf{x},t_0) \nonumber\\ \label{145}
\end{eqnarray}  This is reminiscent from the relation $\mathbf{E}_\bot(\mathbf{x},t_0)=\mathbf{D}(\mathbf{x},t_0)-\mathbf{P}_\bot(\mathbf{x},t_0)$.  It shows that while the two fields $\mathbf{E}_\bot^{(v')}$ and $\mathbf{E}_\bot^{(v)}$ are solutions of the same Maxwell's equations in vacuum  they are not defined by the same initial conditions.   We must therefore be extremely careful when   dealing with the modes in order not to get confused with the solutions chosen. We also mention that the scattered magnetic field $\mathbf{B}^{(d')}$ is given by
 \begin{eqnarray}
\mathbf{B}^{(d')}(\mathbf{x},t)=\boldsymbol{\nabla}\times\int_0^{t-t_0}d\tau\int d^3\mathbf{x'}\frac{1}{c}\Delta_v(\tau,|\mathbf{x}-\mathbf{x'}|)\nonumber\\ \cdot\partial_{t-\tau}\mathbf{P}_\bot(\mathbf{x'},t-\tau)\nonumber\\
=\int_0^{t-t_0}d\tau\int d^3\mathbf{x'}\frac{1}{c}\boldsymbol{\nabla}\times\mathbf{U}_v(\tau,\mathbf{x},\mathbf{x'})\nonumber\\ \cdot\partial_{t-\tau}\mathbf{P}(\mathbf{x'},t-\tau)\nonumber\\
 \label{146}
\end{eqnarray} or equivalently by  
\begin{eqnarray}
\mathbf{B}^{(d')}(\mathbf{x},t)=\int_0^{t-t_0}d\tau\int d^3\mathbf{x'}\frac{1}{c}\boldsymbol{\nabla}\times\partial_\tau\mathbf{U}_v(\tau,\mathbf{x},\mathbf{x'})\nonumber\\ \cdot\mathbf{P}(\mathbf{x'},t-\tau)-\int d^3\mathbf{x'}\frac{1}{c}\boldsymbol{\nabla}\times\mathbf{U}_v(t-t_0,\mathbf{x},\mathbf{x'})\cdot\mathbf{P}(\mathbf{x'},t_0)\nonumber\\
\label{147}
\end{eqnarray}
In the frequency  domain this gives:   
 \begin{eqnarray}
\mathbf{B}^{(d')}(\mathbf{x},t)=\int_{\gamma-i\infty}^{\gamma+i\infty}\frac{i dp}{2\pi} e^{p(t-t_0)}\int d^3\mathbf{x'} \boldsymbol{\nabla}\times\mathbf{G}_v(\mathbf{x},\mathbf{x'},ip)
\nonumber\\  \cdot (\frac{p}{c}\overline{\mathbf{P}'}(\mathbf{x'},p)-\frac{\mathbf{P}(\mathbf{x'},t_0)}{c})\nonumber\\
\label{148}
\end{eqnarray} A comparison with the formulas obtained in the  subsection VI A shows that $\mathbf{B}^{(d')}(\mathbf{x},t)$ differs from $\mathbf{B}^{(d)}(\mathbf{x},t)$ but that at time $t_0$ both vanish so that $\mathbf{B}^{(d')}(\mathbf{x},t_0)=\mathbf{B}^{(d)}(\mathbf{x},t_0)=0$. It implies that $\mathbf{B}^{(v')}(\mathbf{x},t_0)=\mathbf{B}(\mathbf{x},t_0)=\mathbf{B}^{(v)}(\mathbf{x},t_0)$ so that while $\mathbf{B}^{(v')}(\mathbf{x},t)$ differs from $\mathbf{B}^{(v)}(\mathbf{x},t)$ for $t>t_0$ they become equal at the initial time $t_0$. Again, this stresses the difference between the representations based either on $\textbf{F}$ or $\textbf{A}$.\\
\indent Now, this description using  $\textbf{A}$ leads to a clear interpretation of $H_{rem,\bot}(t)=\int d^3\mathbf{x}:\frac{\mathbf{B}(\mathbf{x},t_0)^2+\mathbf{E}_\bot(\mathbf{x},t_0)^2}{2}:$. Indeed, at time $t_0$ only the $(v')$ solution survives for the transverse part of the field. Importantly, the set of free-space solutions $(v')$ actually depends on lowering and rising operators $a_{\alpha,j}^{(v')}(t)$, 
$a_{\alpha,j}^{\dagger(v')}(t)$ defined  such that $[a_{\alpha,j}^{(v')}(t),a_{\beta,k}^{\dagger(v')}(t)]=\delta_{\alpha,\beta}\delta_{j,k}$ and   $a_{\alpha,j}^{(v')}(t)=a_{\alpha,j}^{(v')}(t_0)e^{-i\omega_\alpha(t-t_0)}$  with $a_{\alpha,j}^{(v')}(t_0)=a_{\alpha,j}(t_0)$.  Therefore, by a reasoning  similar to the one leading to Eq.~\ref{158o} we deduce 
\begin{eqnarray}
H_\bot(t)=\int d^3\mathbf{x}\int_{0}^{+\infty}d\omega\hbar\omega\mathbf{f}^{\dagger(0)}_{\omega,\bot}(\mathbf{x},t)\mathbf{f}^{(0)}_{\omega,\bot}(\mathbf{x},t)\nonumber\\
+\sum_{\alpha,j}\hbar\omega_\alpha a_{\alpha,j}^{\dagger(v')}(t)a_{\alpha,j}^{(v')}(t)\label{energytransverse}
\end{eqnarray}
We clearly here get a physical interpretation of the remaining term $H_{rem,\bot}(t)$ as a energy sum over the transverse photon modes propagating in free space. These free photons are calculated using the $\textbf{A}$ potential vector.  From Eq.~\ref{145} we know that these modes differ in general  from those in the $\textbf{F}$ potential vector since $\mathbf{E}_\bot^{(v')}(\mathbf{x},t_0)$ is not identical to $\mathbf{E}_\bot^{(v)}(\mathbf{x},t_0)$   unless the polarization  density $\mathbf{P}_\bot(\mathbf{x},t_0)$ cancels (which is the case in vacuum).\\
\indent Now, in classical physics the meaning of expansion Eq.~\ref{energytransverse} is clear: it corresponds to a diagonalization of the Hamiltonian in term of normal coordinates, i.e., like for classical mechanics \cite{Goldstein}, and similarly to the Huang, Fano, Hopfield procedure for polaritons~\cite{Huang1951,Fano1956,Hopfield1958}. In QED the problem is different since, as explained in details in Ref.~\cite{previous} fields like $a_{\alpha,j}(t)$ and $\mathbf{f}_{\omega,\bot}(\mathbf{x},t)$ (and their Hermitian conjugate variables)  do not commute, unlike it is for $c_{\alpha,j}(t)$ and $\mathbf{f}_{\omega,\bot}(\mathbf{x},t)$. It is thus not possible to find common eigenstates of the $A^{(v')}$ operators for photons (in the usual representation) and for  $X^{(0)}_\omega$  associated with the material fluctuations. This is not true for the representation using $F^{(v)}$ and $X^{(0)}_\omega$ operators but now the full Hamiltonian is not fully diagonalized as seen from Eq.~\ref{158ob}. Only if one neglect the remaining term $H_{\textrm{rem.}}(t)$, like it was done in Refs.~\cite{Gruner1996,Suttorp2004a,Suttorp2004b,Suttorp2007,Philbin2010}, can we diagonalize the Hamiltonian. However, then we get the troubles concerning unitarity, causality and time symmetry discussed along this manuscript.  
\section{General conclusion and perspectives}  
The general formalism discussed in this article using the $\textbf{F}$ potential provide a natural way for dealing with QED in dispersive and dissipative media. It is based on a canonical quantization procedure  generalizing the early work of Huttner and Barnett~\cite{Huttner1991,Huttner1992a,Huttner1992b,Huttner1992c,Matloob1995,Matloob1996,Barnett1995} for polaritons in Homogeneous media. The method is unambiguous as far as we conserve all terms associated  with free photons $\textbf{F}^{(0)}$ and material fluctuations $\textbf{P}^{(0)}$ for describing the quantum evolution. In particular, in order to preserve the full unitarity and the time symmetry of the coupled system of equations we have to include in the evolution terms associated with fluctuating electromagnetic modes $\textbf{E}^{(0)}$, $\textbf{B}^{(0)}$ which have a classical interpretation as polariton eigenmodes and can not in general be omitted if the medium is spatially localized in vacuum. We also discussed alternative representation based on  the potential $\textbf{A}$ instead of $\textbf{F}$. At the end both representations are clearly equivalent and could be used for generalizing the present theory to other linear media including  tensorial anisotropy, magnetic properties, and constitutive equations coupling $\textbf{E}$ and $\textbf{B}$ (magneto/electric media). Moreover, the most important finding of the present article concerns the comparisons  between the generalized Huttner-Barnett approach, advocated here, which involves both photonic and material independent degrees of freedom, and  the Langevin-noise method proposed initially by Gruner and Welsch~\cite{Gruner1996} which involves only the material degrees of freedom associated with fluctuating currents.  We showed that rigorously speaking  the Langevin-noise method is not equivalent to the full  Hamiltonian  QED evolution coupling photonic and material fields. Only in the regime where the dissipation of the bulk surrounding medium is non vanishing at spatial infinity could we, i.e., for all practical purposes, identify the two theories. However, even with such assumptions the Langevin noise model is breaking time-symmetry since it considers only decaying modes while the full Hamiltonian theory used in our work accepts also growing waves associated with anti-thermodynamic processes. We claim that this is crucial in nanophotonics/plasmonics where quantum emitters, spatially localized, are coupled to photonic and material modes available in the complex environment, e.g., near nano-antennas in vacuum (i.e. in a spatial domain where losses are vanishing at infinity). Since most studies consider the interaction between molecules or quantum-dots and plasmon/polaritons using the Langevin noise approach we think that it is urgent to clarify and clean up the problem by analyzing the coupling regime using the full Hamiltonian evolution advocated in the present work. Finally, we suggest that this work could impact the interpretation and discussion of pure QED effects such as the Casimir force or the Lamb shift which are strongly impacted by polariton and plasmon modes.  All this will be the subject for future works and therefore the present detailed analysis is expected to play an important role in nanophotonics and plasmonics for both the classical and quantum regimes.     

\section{Acknowledgments}
This work was supported by Agence Nationale de la Recherche (ANR), France,
through the SINPHONIE (ANR-12-NANO-0019) and PLACORE (ANR-13-BS10-0007) grants. The author gratefully acknowledges several discussions with G. Bachelier.  
\appendix
\section{Absence of Zeros in the upper half plane}
The relation
\begin{eqnarray}
\omega_\alpha^2-\tilde{\varepsilon}(\omega)\omega^2=0 \label{B1}
\end{eqnarray}
admits zeros $\Omega_{\alpha,m}^{(\pm)}$ as postulated in the text.  Writing  $\omega=\omega'+i\omega''$ one of such zero  and $\tilde{\varepsilon}(\omega)=\varepsilon'+\varepsilon''$ the condition Eq.~\ref{B1} means:
\begin{eqnarray}
(\omega'^2-\omega''^2)\varepsilon'-2\omega'\omega''\varepsilon''=\omega_\alpha^2\nonumber\\
(\omega'^2-\omega''^2)\varepsilon''+2\omega'\omega''\varepsilon'=0\label{B2}
\end{eqnarray} from which we deduce after eliminating $(\omega'^2-\omega''^2)$
\begin{eqnarray}
2\omega'\omega''\frac{\varepsilon'^2+\varepsilon''^2}{\varepsilon''}=0\label{B3}
\end{eqnarray} Therefore a necessary but not sufficient condition for having zeros is that if $\omega'\omega''>0$ for such a zero then $\varepsilon''<0$ while if $\omega'\omega''<0$ then $\varepsilon''<0$.  Actually we also see from  Eq.~\ref{B2} that the
zero are allowed to be located along the real or imaginary axis  of the complex $\omega$ plane if $\varepsilon''=0$ along these axes. This is in general not possible for a large class of permittivity function. Consider for example the quite general causal permittivity, i.e., satisfying the Kramers-Kr\"onig relation, defined by: \begin{eqnarray}
\tilde{\varepsilon}(\omega)=1+\sum_n\frac{f_n}{\omega_n^2-(\omega+i\gamma_n)^2}\label{B3}
\end{eqnarray} with $f_n,\gamma_n>0$.
Then we have also
 \begin{eqnarray}
\tilde{\varepsilon}(\omega'+i\omega'')=1\nonumber\\+\sum_n f_n\frac{\omega_n^2-\omega'^2+(\omega''+\gamma_n)^2+2i\omega'(\gamma_n+\omega'')}{(\omega_n^2-\omega'^2+(\omega''+\gamma_n)^2)^2+4\omega'^2(\gamma_n+\omega'')^2}\nonumber\\ \label{B4}
\end{eqnarray} Clearly, $\varepsilon''>0$ if $\omega'>0$ and $\omega''\geq0$ in contradiction with the necessary condition for zeros existence mentioned before.  This reasoning is valid in one quarter plane but now, if $\omega'+i\omega''$ is a zero $-\omega'+i\omega''$ is also a zero.   Therefore, the absence of a zero in  the quarter plane $\omega'>0$, $\omega''\geq0$ implies the absence of zero in the second quarter plane $\omega'<0$, $\omega''\geq0$ and therefore Eq.~\ref{B1} do not admit any  zero in the upper half plane for a very usual permittivity like  Eq.~\ref{B3}. Actually, the case $\omega'=0$ should be handled separately. We find from Eq.~\ref{B4} that for such value $\varepsilon''=0$. This is acceptable in order to have a zero existence in agreement with Eq.~\ref{B2}. However, from Eq.~\ref{B2} we find also that if a zero exists along the axis  $\omega'=0$ then we should have as well $\varepsilon'=-\omega_\alpha^2/\omega''^2<0$. This is in contradiction with Eq.~\ref{B4}  which implies $\varepsilon'>0$. This completes the proof for the permittivity given by Eq.~\ref{B3}.\\
The question concerning the generality of the  proof is however still handling. Huttner and  Barnett mentioned the existence of such a proof in the Landau and Lifschitz text-book~\cite{Lifshitzbook} but \begin{figure}[h!t]
\begin{center}
\includegraphics[width=8.2cm]{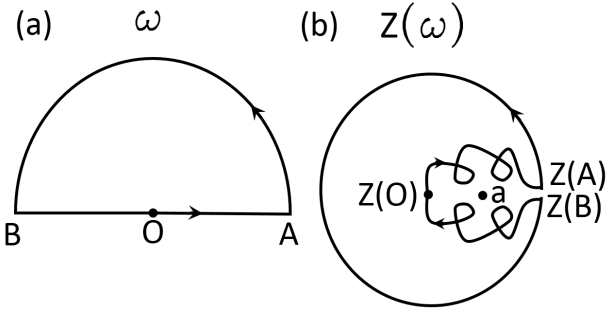}
\caption{Integration path deformation leading to the proof that Eq.~\ref{B1} has no solution in the upper frequency half-plane (see Refs.~\cite{Lifshitzbook,LandauStat}).} \label{figure1}
\end{center}
\end{figure}
it is relevant to detail the missing proof here. In order to get the complete result we will use a method used by Landau and Lifshitz (see Ref.~\cite{LandauStat} p. 380). In the complex $\omega$-plane we define $Z(\omega)=\tilde{\varepsilon}(\omega)\omega^2$. From the properties of $\tilde{\varepsilon}(\omega)$ we deduce $Z(-\omega^\ast)=Z(\omega)^\ast$. This implies that $Z$ is real along the imaginary axis and that $Z'(\omega')=Z'(-\omega')$ and  $Z''(\omega')=-Z''(-\omega')$ along the real axis. Furthermore due to causality we have $\tilde{\varepsilon}''(\omega')>0$ and therefore $Z''(\omega')>0$   if $\omega'>0$ along the real axis.   Now we consider the  closed contour integral (see Fig.~1)
\begin{eqnarray}
\oint_{C}\frac{ d\omega }{2i\pi}\frac{\frac{dZ(\omega)}{d\omega}}{Z(\omega)-a}\label{B5}
\end{eqnarray}
along $C$ made of the real axis and of the semi circle $C^+$ of infinite radius $R\rightarrow+\infty$ in the upper half plane. However, $a$ is a real number and since $Z$ is not real along the real axis  there is no pole along  $C$  unless $a$ is infinite or null. Therefore, since $Z$ is analytical in the upper half plane Eq.~\ref{B5} gives the numbers of zeros of  $Z(\omega)-a=0$ in this half space.\\
We then rewrite Eq.~\ref{B5}  as an integral in the complex $Z-plane$: \begin{eqnarray}
\oint_{C'}\frac{ dZ }{2i\pi}\frac{1}{Z-a}\label{B6}
\end{eqnarray}
where $C'$ is the image
of $C$ along the mapping $\omega\mapsto Z(\omega)$. In particular the origin $O$ is mapped on it self while the semi circle of radius $R$ is mapped onto the circle of radius $R^2$. The half real axis $OA$ corresponding to $\omega>0$  is mapped onto a complex curve located in the upper half plane of the complex $Z$-space (since $Z''>0$ along this half line). Similarly, the second half axis $OB
$ is mapped in the lower half plane. As shown on the figure if $0<a<+\infty$ then the contour integral omits the point $Z=a$ and there is no pole involved in the integral which therefore vanishes.   We thus deduce that  Eq.~\ref{B1} has no solution in the upper half-plane in the $\omega$ space which is the proof needed.
\section{polar expansion of a  causal Green function}
The calculation of $H_\alpha(\tau)$ defined by the Bromwich integral Eq.~\ref{95} for $\tau>0$:
\begin{eqnarray}
H_\alpha(\tau)=\int_{\gamma-i\infty}^{\gamma+i\infty}\frac{i dp}{2\pi}\frac{e^{p\tau}}{\omega_\alpha^2+(1+\bar{\chi}(p))p^2}\nonumber\\
=\int_{-\infty}^{+\infty}\frac{d\omega}{2\pi}\frac{e^{-i\omega\tau}}{\omega_\alpha^2-\tilde{\varepsilon}(\omega)(\omega+i\eta)^2}e^{+\eta\tau}\nonumber\\ 
\end{eqnarray}
can be handled after closing the contour integral in  the lower plane.
However, since $\frac{1}{\omega_\alpha^2-\tilde{\varepsilon}(\omega)\omega^2}$ is not analytical in such lower plane we must include the poles $\Omega_{\alpha,m}^{(\pm)}$ (all located in the lower plane see Appendix A), i.e. the residues, in the integral. We use the separation:
\begin{eqnarray}
\frac{1}{\omega_\alpha^2-\tilde{\varepsilon}(\omega)\omega^2}=\frac{1}{2\omega_\alpha}[\frac{1}{\omega_\alpha-\omega\sqrt{\tilde{\varepsilon}(\omega)}}+\frac{1}{\omega_\alpha+\omega\sqrt{\tilde{\varepsilon}(\omega)}}]\nonumber\\ \label{96}
\end{eqnarray} and express it as a function of  $\omega=\Omega_{\alpha,m}^{(\pm)}+\rho e^{i\varphi}$ near each poles (i.e, in the limit $\rho\rightarrow 0$). We get:
 \begin{eqnarray}
\frac{1}{\omega_\alpha\pm\omega\sqrt{\tilde{\varepsilon}(\omega)}}\thickapprox \pm\frac{1}{\rho e^{i\varphi}\frac{\partial(\omega\sqrt{\tilde{\varepsilon}(\omega)})}{\partial \omega}|_{\Omega_{\alpha,m}^{(\pm)}}}\nonumber\\ \label{97}
\end{eqnarray}
From the condition ${\Omega_{\alpha,m}^{(\pm)}}^\ast=-\Omega_{\alpha,m}^{(\mp)}$  and the equality $(\frac{\partial(\omega\sqrt{\tilde{\varepsilon}(\omega)})}{\partial \omega})^\ast=\frac{\partial(-\omega^\ast\sqrt{\tilde{\varepsilon}(-\omega^\ast)})}{\partial -\omega^\ast}=\frac{\partial(\omega\sqrt{\tilde{\varepsilon}(\omega)})}{\partial \omega}|_{-\omega^\ast}$ we then get for $\tau>0$ after integration in the lower plane Eq.~\ref{98} and $H_\alpha(\tau)=0$ for $\tau<0$ (after integration in the upper plane where no pole are present).
The value at $\tau=0$ deserves some careful analysis. Indeed, if $\tau=0$ the integration along the semicircle do not vanish exponentially with it radius $R$ and if we choose to integrate in the upper half plane  (where there is no pole)  we get $\int_{-\infty}^{+\infty}\frac{d\omega}{2\pi}\frac{1}{\omega_\alpha^2-\omega^2\tilde{\varepsilon}(\omega)}=-\int_0^\pi\frac{id\varphi R e^{i\varphi}}{2\pi}\frac{1}{\omega^2-R^2e^{2i\varphi}}=O(1/R)$ if $R\rightarrow+\infty$ since $\tilde{\varepsilon}(Re^{i\varphi})=1$ in this limit in the upper half plane.  Therefore we have indeed $H_\alpha(0)=0$ and the function is continuous at  $\tau=0$. Of course the null value for negative time $t'$ have no meaning since the Laplace transform is only interested in the evolution for positive time.\\
This leads to the sum rule:
\begin{eqnarray}
\sum_{m}\frac{1}{\omega_\alpha}\textrm{Imag}[\frac{1}{\frac{\partial(\omega\sqrt{\tilde{\varepsilon}(\omega)})}{\partial \omega}|_{\Omega_{\alpha,m}^{(-)}}}]=0.\label{101}
\end{eqnarray}This leads to the sum rule:
\begin{eqnarray}
\sum_{m}\frac{1}{\omega_\alpha}\textrm{Imag}[\frac{1}{\frac{\partial(\omega\sqrt{\tilde{\varepsilon}(\omega)})}{\partial \omega}|_{\Omega_{\alpha,m}^{(-)}}}]=0.\label{99}
\end{eqnarray}
\indent  the free term  $q_{\alpha,j}^{(0)}(t)$ is defined as
\begin{eqnarray}
q_{\alpha,j}^{(0)}(t)=\int_{\gamma-i\infty}^{\gamma+i\infty}\frac{i dp}{2\pi}\frac{(1+\bar{\chi}(p))[pq'_{\alpha,j}(0)+\dot{q}'_{\alpha,j}(0)]e^{p(t-t_0)}}{\omega_\alpha^2+(1+\bar{\chi}(p))p^2}\nonumber\\
=U_\alpha(t-t_0)\dot{q}_{\alpha,j}(t_0)+\dot{U}_\alpha(t-t_0)q_{\alpha,j}(t_0)\nonumber\\ \label{102}
\end{eqnarray} with  \begin{eqnarray}
U_\alpha(\tau)=\int_{\gamma-i\infty}^{\gamma+i\infty}\frac{i dp}{2\pi}\frac{(1+\bar{\chi}(p))e^{p\tau}}{\omega_\alpha^2+(1+\bar{\chi}(p))p^2}\nonumber\\
=\sum_{m}\frac{-\tilde{\varepsilon}(\Omega_{\alpha,m}^{(-)})}{2i\omega_\alpha}\frac{e^{-i\Omega_{\alpha,m}^{(-)}\tau}}{\frac{\partial(\omega\sqrt{\tilde{\varepsilon}(\omega)})}{\partial \omega}|_{\Omega_{\alpha,m}^{(-)}}} +cc.\label{103}
\end{eqnarray}  Like for $H_\alpha$ we get $U_\alpha(0)=0$. The last line of Eq.~\ref{102} is therefore justified  from the fact that the  Laplace transform  of  $\frac{d}{d\tau}U_\alpha(\tau)$ is $p\overline{U_\alpha}(p)-U_\alpha(0)=p\overline{U_\alpha}(p)$.  Now, the boundary condition at $t=t_0$ imposes
$\frac{d}{d\tau}U_\alpha(\tau)|_{\tau=0}=1$. Therefore, from Eq.~\ref{103} we deduce the second sum rule:
\begin{eqnarray}
\sum_{m}\frac{1}{\omega_\alpha}\textrm{Real}[\frac{\tilde{\varepsilon}(\Omega_{\alpha,m}^{(-)})\Omega_{\alpha,m}^{(-)}}{\frac{\partial(\omega\sqrt{\tilde{\varepsilon}(\omega)})}{\partial \omega}|_{\Omega_{\alpha,m}^{(-)}}}]=1.\label{104}
\end{eqnarray}
The value at $\tau=0$ is not defined univocally since $\frac{d}{d\tau}U_\alpha(\tau)$  defined through the Bromwich integral of $U_\alpha(\tau)$ is discontinuous. We point out that considering a direct integration at $\tau=0$ could lead to contradictions since the integration along  $C^\pm$ do not vanish. If we choose to integrate in the upper half plane  (where there is no pole)  we get 
\begin{eqnarray}\int_{-\infty}^{+\infty}\frac{d\omega}{2\pi}\frac{-i\omega\tilde{\varepsilon}(\omega)}{\omega_\alpha^2-\tilde{\varepsilon}(\omega)\omega^2}\nonumber\\=-\int_0^\pi\frac{d\varphi R^2 e^{2i\varphi}}{2\pi}\frac{1}{\omega^2-R^2e^{2i\varphi}}=1/2\label{105}
\end{eqnarray}
if $R\rightarrow+\infty$ since $\tilde{\varepsilon}(Re^{i\varphi})=1$ in this limit in the upper half plane. We would get $\frac{d}{d\tau}U_\alpha(\tau)|_{\tau=0}=1/2$ (a similar calculation could be done in the lower space including poles and residues and we would obtain once again $1-1/2=1/2$). Here we considered carefully the limit to prevent us from such a contradiction.\\
\section{The source field: a vectorial and scalar potentials discussion}
\indent   The source field can be  written as: $\mathbf{F}^{(s)}=\sum_{\alpha,j}q_{\alpha,j}^{(s)}\boldsymbol{\hat{\epsilon}}_{\alpha,j}\Phi_\alpha$. After some algebras we get: 
\begin{eqnarray}
\mathbf{F}^{(s)}(\mathbf{x},t)=\int_{\gamma-i\infty}^{\gamma+i\infty}\frac{i dp}{2\pi}e^{p(t-t_0)}\int d^3\mathbf{x'}G_\chi(|\mathbf{x}-\mathbf{x'}|,ip)\nonumber\\ \cdot\boldsymbol{\nabla'}\times\overline{\mathbf{P}'}^{(0)}(\mathbf{x'},p))\nonumber\\
=\int_{-\infty}^{+\infty}d\omega e^{-i\omega t}\int d^3\mathbf{x'}G_\chi(|\mathbf{x}-\mathbf{x'}|,\omega+i0^+) \nonumber\\ \cdot \boldsymbol{\nabla'}\times\widetilde{\underline{\mathbf{P}}}^{(0)}(\mathbf{x'},\omega)\nonumber\\ \label{103}
\end{eqnarray} where we introduced the Green function:\begin{eqnarray}
G_\chi(|\mathbf{x}-\mathbf{x'}|,ip)=c^2\sum_{\alpha}\frac{\Phi_\alpha(\mathbf{x})\Phi_\alpha^\ast(\mathbf{x'})}{\omega_\alpha^2+(1+\bar{\chi}(p))p^2}\nonumber\\=c^2\int \frac{d^3\mathbf{k}}{(2\pi)^3}\frac{e^{i\mathbf{k}\cdot(\mathbf{x}-\mathbf{x'})}}{c^2k^2+(1+\bar{\chi}(p))p^2}
\nonumber\\=\frac{e^{-p\sqrt{1+\bar{\chi}(p)}|\mathbf{x}-\mathbf{x'}|/c}}{4\pi |\mathbf{x}-\mathbf{x'}|}\label{104}
\end{eqnarray} computed by contour integration  in the complex plane and solution of $\boldsymbol{\nabla}^2G_\chi(|\mathbf{x}-\mathbf{x'}|,ip)-(1+\bar{\chi}(p))\frac{p^2}{c^2}G_\chi(|\mathbf{x}-\mathbf{x'}|,ip)=\delta^3(\mathbf{x}-\mathbf{x'})$. Of course,  along the real axis $\gamma\rightarrow0^+$ we get $G_\chi(|\mathbf{x}-\mathbf{x'}|,\omega+i0^+)=\frac{e^{i\omega\sqrt{\tilde{\varepsilon}(\omega)}|\mathbf{x}-\mathbf{x'}|/c}}{4\pi |\mathbf{x}-\mathbf{x'}|}$ which is the usual Green function for an homogenous medium.
We can also write this field without introducing  $\omega$ by using the Green propagator $\Delta_\chi(\tau,R)=\int_{\gamma-i\infty}^{\gamma+i\infty}\frac{i dp}{2\pi}e^{p\tau}G_\chi(R,ip)$ which leads to:
 \begin{eqnarray}
\mathbf{F}^{(s)}(\mathbf{x},t)=\int_0^{t-t_0}d\tau\int d^3\mathbf{x'}\Delta_\chi(\tau,|\mathbf{x}-\mathbf{x'}|)\nonumber\\ \cdot\boldsymbol{\nabla'}\times\mathbf{P}^{(0)}(\mathbf{x'},t-\tau)\nonumber\\
 \label{105}
\end{eqnarray}
We have also \begin{eqnarray}
\Delta_\chi(t-t',|\mathbf{x}-\mathbf{x'}|)=c^2\sum_{\alpha}H_\alpha(t-t')\Phi_\alpha^\ast(\mathbf{x'})\Phi_\alpha(\mathbf{x})\nonumber\\
=c^2\sum_{\alpha,m}\frac{-1}{2i\omega_\alpha}\frac{e^{-i\Omega_{\alpha,m}^{(-)}(t-t')}\Phi_\alpha^\ast(\mathbf{x'})\Phi_\alpha(\mathbf{x})}{\frac{\partial(\omega\sqrt{\tilde{\varepsilon}(\omega)})}{\partial \omega}|_{\Omega_{\alpha,m}^{(-)}}} +cc.\nonumber\\ 
\label{106}
\end{eqnarray} which represent the generalization of retarded propagator expansion for a lossy and dispersive medium. The role of causality is here crucial since the modes are always damped when the time is growing in the future direction as expected from pure thermodynamical considerations. This means in particular that $\Delta_\chi(t,|\mathbf{x}-\mathbf{x'}|)$ tends to vanish exponentially as $t$ goes to infinity. Importantly, in the vacuum limit ($\chi(\tau)\rightarrow 0$) we get naturally 
\begin{eqnarray}
\Delta_v(\tau,R)=c^2\sum_{\alpha}\frac{\sin{\omega_\alpha \tau}}{\omega_\alpha}\Phi_\alpha^\ast(\mathbf{x'})\Phi_\alpha(\mathbf{x})
\nonumber\\=\frac{\delta(\tau-R/c)}{4\pi R}\label{107}\end{eqnarray} 
and in the limit $t_0\rightarrow -\infty$ we obtain  the retarded potential
 \begin{eqnarray}
\mathbf{F}^{(s)}(\mathbf{x},t)=\int d^3\mathbf{x'}\frac{\boldsymbol{\nabla'}\times\mathbf{P}^{(0)}(\mathbf{x'},t-|\mathbf{x}-\mathbf{x'}|/c)}{4\pi|\mathbf{x}-\mathbf{x'}|} \label{108}
\end{eqnarray}
However, in the vacuum limit we have also:  $\mathbf{P}^{(0)}(\mathbf{x},t)\rightarrow 0$ therefore   $\mathbf{F}^{(s)}(\mathbf{x},t)$ actually vanishes and we get in this limit
$\mathbf{F}(\mathbf{x},t)=\mathbf{F}^{(0)}(\mathbf{x},t)$ as it should be. Now, from Eq.~\ref{105} and from the field definition we easily get the integral formulas  for $\mathbf{D}^{(s)}(\mathbf{x},t)$ and $\mathbf{B}^{(s)}(\mathbf{x},t)$:
\begin{eqnarray}
\mathbf{D}^{(s)}(\mathbf{x},t)=\boldsymbol{\nabla}\times\boldsymbol{\nabla}\times\int_0^{t-t_0}d\tau\int d^3\mathbf{x'}\Delta_\chi(\tau,|\mathbf{x}-\mathbf{x'}|)\nonumber\\ \cdot\mathbf{P}^{(0)}(\mathbf{x'},t-\tau)\nonumber\\
\mathbf{B}^{(s)}(\mathbf{x},t)=\boldsymbol{\nabla}\times\int_0^{t-t_0}d\tau\int d^3\mathbf{x'}\frac{1}{c}\Delta_\chi(\tau,|\mathbf{x}-\mathbf{x'}|)\nonumber\\ \cdot\partial_{t-\tau}\mathbf{P}^{(0)}(\mathbf{x'},t-\tau)\nonumber\\
+\boldsymbol{\nabla}\times\int d^3\mathbf{x'}\frac{1}{c}\Delta_\chi(t-t_0,|\mathbf{x}-\mathbf{x'}|)\cdot\mathbf{P}^{(0)}(\mathbf{x'},t_0)\nonumber\\
 \label{109}
\end{eqnarray} These equations have a clear interpretation in term of generalized Hertz potentials. In particular, taking the limit $t_0\rightarrow- \infty$ and using the properties of convolutions , together with the fact that $\Delta_\chi(t-t_0,|\mathbf{x}-\mathbf{x'}|)\rightarrow 0$,  we get 
 \begin{eqnarray}
\mathbf{D}^{(s)}(\mathbf{x},t)=\boldsymbol{\nabla}\times\boldsymbol{\nabla}\times\boldsymbol{\Xi}(\mathbf{x},t)\nonumber\\
\mathbf{B}^{(s)}(\mathbf{x},t)=\boldsymbol{\nabla}\times\frac{1}{c}\partial_t\boldsymbol{\Xi}(\mathbf{x},t)
 \label{110}
\end{eqnarray} with 
\begin{eqnarray}\boldsymbol{\Xi}(\mathbf{x},t)=\int_{-\infty}^{+\infty}dt'\int d^3\mathbf{x'}\Delta_\chi(t-t',|\mathbf{x}-\mathbf{x'}|)\mathbf{P}^{(0)}(\mathbf{x'},t'). \nonumber\\  \label{111} \end{eqnarray}
\section{Complement concerning the Green Dyadic tensor in a homogeneous medium}
By rewriting  $\mathbf{G}(\mathbf{x},\mathbf{x'},ip)$ in Eq.~\ref{cool} we get after some rearrangements: 
\begin{eqnarray}
 \mathbf{G}_{\chi}(\mathbf{x},\mathbf{x'},ip)=G_{\chi}(\mathbf{x},\mathbf{x'},ip)\mathbf{I}\nonumber\\-\frac{c^2}{p^2(1+\bar{\chi}(p))}\boldsymbol{\nabla}\otimes\boldsymbol{\nabla}G_{\chi}(\mathbf{x},\mathbf{x'},ip)\nonumber\\ 
=-\frac{c^2\boldsymbol{\nabla}\times\boldsymbol{\nabla}\times[G_{\chi}(\mathbf{x},\mathbf{x'},ip)\mathbf{I}]}{p^2(1+\bar{\chi}(p))}\label{137}
\end{eqnarray}
 which involves the scalar Green function defined in Eq.~\ref{104} and from $\mathbf{S}_{\chi}(\mathbf{x},\mathbf{x'},ip)=\boldsymbol{\nabla}\times\boldsymbol{\nabla}\times\mathbf{G}_{\chi}(\mathbf{x},\mathbf{x'},ip)$
\begin{eqnarray}
 \mathbf{S}_{\chi}(\mathbf{x},\mathbf{x'},ip) =\boldsymbol{\nabla}\times\boldsymbol{\nabla}\times[G_{\chi}(\mathbf{x},\mathbf{x'},ip)\mathbf{I}]\nonumber\\+
\mathbf{I}\delta^{3}(\mathbf{x}-\mathbf{x'})\label{138}
\end{eqnarray} These formulas must be taken carefully  since they are not actually valid at the source location, i.e., if $\mathbf{x}\rightarrow\mathbf{x'}$ due to the bad convergence of the series defining the dyadic Green function. After regularization we can obtain the result 
\begin{eqnarray}
 \mathbf{G}_{\chi}(\mathbf{x},\mathbf{x'},ip)=\textrm{P.V.}[G_{\chi}(\mathbf{x},\mathbf{x'},ip)\mathbf{I}\nonumber\\-\frac{c^2}{p^2(1+\bar{\chi}(p))}\boldsymbol{\nabla}\otimes\boldsymbol{\nabla}G_{\chi}(\mathbf{x},\mathbf{x'},ip)]\nonumber\\ +\frac{c^2}{p^2(1+\bar{\chi}(p))}\mathbf{L}\delta^{3}(\mathbf{x}-\mathbf{x'})\label{139}
\end{eqnarray}
and
\begin{eqnarray}
 \mathbf{S}_{\chi}(\mathbf{x},\mathbf{x'},ip) =(\mathbf{I}-\mathbf{L})\delta^{3}(\mathbf{x}-\mathbf{x'})\nonumber\\+\textrm{P.V.}[\boldsymbol{\nabla}\times\boldsymbol{\nabla}\times[G_{\chi}(\mathbf{x},\mathbf{x'},ip)\mathbf{I}]]\label{140}
\end{eqnarray}
with $\mathbf{L}$ a dyadic term depending on the way we define the principal value~\cite{Yaghjian1980,Novotny}: 
\begin{eqnarray}
 \mathbf{L}=\oint_{(\Sigma)} \frac{\textbf{n}\otimes\textbf{R}}{4\pi R^2}dS\label{141}
\end{eqnarray}
For a small exclusion spherical volume surrounding  the point $\mathbf{x'}$ we get $\mathbf{L}=\mathbf{I}/3$, i.e., the depolarization  field predicted by the Clausius-Mosotti formula~\cite{Jackson1999}.\\
\indent In the particular case $\chi=0$ we write $\mathbf{G}_{\chi}(\mathbf{x},\mathbf{x'},ip)=\mathbf{G}_{v}(\mathbf{x},\mathbf{x'},ip)$ and similarly for other Green functions. We also consider the time evolution which in vacuum relies on the propagators
 \begin{eqnarray}
\mathbf{Q}_v(\tau,\mathbf{x},\mathbf{x'})=\int_{\gamma-i\infty}^{\gamma+i\infty}\frac{i dp}{2\pi}e^{p\tau}\mathbf{S}_{v}(\mathbf{x},\mathbf{x'},ip)\nonumber\\
 \mathbf{U}_v(\tau,\mathbf{x},\mathbf{x'})=\int_{\gamma-i\infty}^{\gamma+i\infty}\frac{i dp}{2\pi}e^{p\tau}\mathbf{U}_{v}(\mathbf{x},\mathbf{x'},ip)\label{A10}
\end{eqnarray}
Explicit calculations  lead to:
\begin{eqnarray}
\mathbf{Q}_v(\tau,\mathbf{x},\mathbf{x'})=\sum_{\alpha,j}\omega_\alpha^2\frac{\sin{(\omega_\alpha\tau)}}{\omega_\alpha}\Phi_\alpha(\mathbf{x})\Phi_\alpha^\ast(\mathbf{x'})\boldsymbol{\hat{\epsilon}}_{\alpha,j}\otimes\boldsymbol{\hat{\epsilon}}_{\alpha,j}\nonumber\\ \label{A11}
\end{eqnarray} and similarly  for the transverse part of $\mathbf{U}_v(\tau,\mathbf{x},\mathbf{x'})$:
\begin{eqnarray}
\mathbf{U}_{v,\bot}(\tau,\mathbf{x},\mathbf{x'})=\sum_{\alpha,j}c^2\frac{\sin{(\omega_\alpha\tau)}}{\omega_\alpha}\Phi_\alpha(\mathbf{x})\Phi_\alpha^\ast(\mathbf{x'})\boldsymbol{\hat{\epsilon}}_{\alpha,j}\otimes\boldsymbol{\hat{\epsilon}}_{\alpha,j}\nonumber\\ \label{A12}
\end{eqnarray} while  for the longitudinal part we get:  
\begin{eqnarray}
\mathbf{U}_{v,||}(\tau,\mathbf{x},\mathbf{x'})=c^2\tau\sum_{\alpha}\Phi_\alpha(\mathbf{x})\Phi_\alpha^\ast(\mathbf{x'})\boldsymbol{\hat{k}}_{\alpha}\otimes\boldsymbol{\hat{k}}_{\alpha}
\label{A13}
\end{eqnarray} We deduce automatically the boundary conditions $\mathbf{U}_{v,\bot}(0,\mathbf{x},\mathbf{x'})=\mathbf{U}_{v,||}(0,\mathbf{x},\mathbf{x'})=\mathbf{Q}_v(0,\mathbf{x},\mathbf{x'})=0$. We also obtain 
\begin{eqnarray}
\partial_\tau\mathbf{U}_{v,\bot}(\tau,\mathbf{x},\mathbf{x'})=\sum_{\alpha,j}c^2\cos{(\omega_\alpha\tau)}\Phi_\alpha(\mathbf{x})\Phi_\alpha^\ast(\mathbf{x'})\boldsymbol{\hat{\epsilon}}_{\alpha,j}\otimes\boldsymbol{\hat{\epsilon}}_{\alpha,j}\nonumber\\ \label{A14}
\end{eqnarray} and 
\begin{eqnarray}
\partial_\tau\mathbf{U}_{v,||}(\tau,\mathbf{x},\mathbf{x'})=c^2\sum_{\alpha}\Phi_\alpha(\mathbf{x})\Phi_\alpha^\ast(\mathbf{x'})\boldsymbol{\hat{k}}_{\alpha}\otimes\boldsymbol{\hat{k}}_{\alpha}
\label{A15}
\end{eqnarray} from which we obtain the boundary condition: $\partial_\tau\mathbf{U}_{v}(\tau,\mathbf{x},\mathbf{x'})_{\tau=0}=c^2\textbf{I}\sum_{\alpha}\Phi_\alpha(\mathbf{x})\Phi_\alpha^\ast(\mathbf{x'})=c^2\textbf{I}\delta^3(\mathbf{x}-\mathbf{x}')$. We thus obtain: 
\begin{eqnarray}
\mathbf{Q}_v(\tau,\mathbf{x},\mathbf{x'})=\boldsymbol{\nabla}\times\boldsymbol{\nabla}\times\mathbf{U}_v(\tau,\mathbf{x},\mathbf{x'})\nonumber\\=-\frac{\partial_\tau^2\mathbf{U}_v(\tau,\mathbf{x},\mathbf{x'}) }{c^2}.\label{definiti}
\end{eqnarray}
Finally, from Eq.~\ref{139} we find explicitly for the time dependent $\mathbf{Q}_v(\tau,\mathbf{x},\mathbf{x'})$ field:  
\begin{eqnarray}
 \mathbf{Q}_v(\tau,\mathbf{x},\mathbf{x'}) =(\mathbf{I}-\mathbf{L})\delta^{3}(\mathbf{x}-\mathbf{x'})\delta(\tau)\nonumber\\+\textrm{P.V.}[\boldsymbol{\nabla}\times\boldsymbol{\nabla}\times[\Delta_v(\tau,\mathbf{x},\mathbf{x'})\mathbf{I}]].\label{A16}\end{eqnarray}
\section{Transverse polaritons in the Hopfield model: a consistency check}
From Eq.~\ref{B} we deduce in the Hopfield model
\begin{eqnarray}
(\omega_0^2-\Omega^2)\widetilde{\textbf{P}}(\Omega)=\omega_p^2\widetilde{\textbf{E}}(\Omega).
\end{eqnarray} 
This leads to the solution
\begin{eqnarray}
\widetilde{\textbf{P}}(\Omega)=\frac{\omega_p^2}{(\omega_0^2-(\Omega+i0^+)^2)}\widetilde{\textbf{E}}(\Omega)+\widetilde{\textbf{P}}(\Omega)^{(in)}.\label{reeu}
\end{eqnarray} 
and therefore to 
\begin{eqnarray}
\boldsymbol{\nabla}\times\boldsymbol{\nabla}\times\widetilde{\textbf{E}}(\Omega)-\frac{\Omega^2}{c^2}\widetilde{\varepsilon}(\Omega)\widetilde{\textbf{E}}(\Omega)=\frac{\Omega^2}{c^2}\widetilde{\textbf{P}}(\Omega)^{(in)}\label{A}
\end{eqnarray} 
where the permittivity is given by  the lossless Lorentz-Drude formula $\widetilde{\varepsilon}(\Omega)=1+\frac{\omega_p^2}{(\omega_0^2-(\Omega+i0^+)^2)}$. 
For the transverse fields we  expand the different fields as 
\begin{eqnarray}
\widetilde{\textbf{E}}_\bot(\mathbf{x},\Omega)=\sum_{\alpha,j}\widetilde{E}_{\alpha,j}(\Omega)\boldsymbol{\hat{\epsilon}}_{\alpha,j}\Phi_\alpha(\mathbf{x}) \nonumber\\
\widetilde{\textbf{P}}_\bot(\mathbf{x},\Omega)=\sum_{\alpha,j}\widetilde{P}_{\alpha,j}(\Omega)\boldsymbol{\hat{\epsilon}}_{\alpha,j}\Phi_\alpha(\mathbf{x}) 
\end{eqnarray}
with $\widetilde{E}_{\alpha,j}(\Omega)^\ast=\eta_j\widetilde{E}_{-\alpha,j}(-\Omega)$ and  $\widetilde{P}_{\alpha,j}(\Omega)^\ast=\eta_j\widetilde{P}_{-\alpha,j}(-\Omega)$. Now, we write $\widetilde{\textbf{P}}_{\bot}^{(in)}(\Omega)=\boldsymbol{\gamma}\delta(\Omega-\omega_0)+\boldsymbol{\gamma}^\ast\delta(\Omega+\omega_0)$ with $\boldsymbol{\gamma}(\mathbf{x})$ a transverse vector field. We thus have 
$\widetilde{P}_{\alpha,j}(\Omega)^{(in)}=\gamma_{\alpha,j}\delta(\Omega-\omega_0)+\eta_j\gamma_{-\alpha,j}^\ast\delta(\Omega+\omega_0)$ with $\gamma_{\alpha,j}=\int d^3\mathbf{x}\boldsymbol{\gamma}(\mathbf{x})\cdot \boldsymbol{\hat{\epsilon}}_{\alpha,j}\Phi_\alpha(\mathbf{x})^\ast$.
We thus  get the equation
\begin{eqnarray}
(\omega_\alpha^2-\Omega^2\widetilde{\varepsilon}(\Omega))\widetilde{E}_{\alpha,j}(\Omega)=\Omega^2\widetilde{P}_{\alpha,j}(\Omega)^{(in)}
\end{eqnarray} 
Near the singularities $\omega_0$ we get : 
\begin{eqnarray}
[\omega_\alpha^2-\omega_0^2-i\frac{\pi\omega_p^2\omega_0}{2}\delta(\Omega-\omega_0)]\widetilde{E}_{\alpha,j}(\Omega)=\omega_0^2\gamma_{\alpha,j}\delta(\Omega-\omega_0)
\end{eqnarray} 
and therefore supposing the regularity (as for the longitudinal case) we have $-i\frac{\pi\omega_p^2}{2\omega_0}\widetilde{E}_{\alpha,j}(\omega_0)=\gamma_{\alpha,j}$.
The secular equation $[\omega_\alpha^2-\Omega^2\widetilde{\varepsilon}(\Omega)]\widetilde{E}_{\alpha,j,\pm}(\Omega)=0$ associated with the transverse modes can be easily solved 
and this indeed leads to \begin{eqnarray}
\widetilde{\textbf{E}}_\bot(\mathbf{x},\Omega)=\sum_{\alpha,j,\pm}\widetilde{E}_{\alpha,j,\pm}(\Omega)\boldsymbol{\hat{\epsilon}}_{\alpha,j}\Phi_\alpha(\mathbf{x}) \label{defi}
\end{eqnarray} with $\widetilde{E}_{\alpha,j,\pm}(\mathbf{x},\Omega)=\phi_{\alpha,j,\pm}\delta(\Omega-\Omega_{\alpha,\pm})+\eta_j\phi_{-\alpha,j,\pm}^\ast\delta(\Omega+\Omega_{\alpha,\pm})$ with $\phi_{\alpha,j,\pm}$ an amplitude coefficient for the  transverse polariton mode. Importantly we again get from the regularity condition $\widetilde{E}_{\alpha,j}(\omega_0)=0$ and thus $\gamma_{\alpha,j}=0$ like for the longitudinal mode. This implies $\widetilde{\textbf{P}}(\Omega)^{(in)}=0$ and from Eq.~\ref{reeu} we get $\widetilde{\textbf{P}}_\bot(\Omega)=\frac{\omega_p^2}{(\omega_0^2-(\Omega+i0^+)^2)}\widetilde{\textbf{E}}_\bot(\Omega)$. At the end of the day we obtain the following fields amplitudes 
\begin{eqnarray}
E_{\alpha,j,\pm}(t)=\phi_{\alpha,j,\pm}e^{-i\Omega_{\alpha,\pm}t}+\eta_j\phi_{-\alpha,j,\pm}e^{-i\Omega_{\alpha,\pm}t}\nonumber\\
B_{\alpha,j,\pm}(t)=\frac{\omega_\alpha}{\Omega_{\alpha,\pm}}(\eta_j\phi_{\alpha,j,\pm}e^{-i\Omega_{\alpha,\pm}t}-\phi_{-\alpha,j,\pm}e^{-i\Omega_{\alpha,\pm}t})\nonumber\\\end{eqnarray}  and
\begin{eqnarray}
\frac{P_{\alpha,j,\pm}(t)}{\omega_p^2}=\frac{(\phi_{\alpha,j,\pm}e^{-i\Omega_{\alpha,\pm}t}+\eta_j\phi_{-\alpha,j,\pm}e^{-i\Omega_{\alpha,\pm}t})}{\omega_0^2-\Omega_{\alpha,\pm}^2}\nonumber\\
\frac{i\dot{P}_{\alpha,j,\pm}(t)}{\omega_p^2\Omega_{\alpha,\pm}}=\frac{(\phi_{\alpha,j,\pm}e^{-i\Omega_{\alpha,\pm}t}-\eta_j\phi_{-\alpha,j,\pm}e^{-i\Omega_{\alpha,\pm}t})}{\omega_0^2-\Omega_{\alpha,\pm}^2}\nonumber\\ \label{hopfieldmatrix}
\end{eqnarray} (using definitions similar to Eq.~\ref{defi}). These define the Hopfield transformation between the old variables  $E_{\alpha,j,\pm}(t)$, $B_{\alpha,j,\pm}(t)$, $P_{\alpha,j,\pm}(t)$, $\dot{P}_{\alpha,j,\pm}(t)$ and the new polaritonic variables representing the normal coordinates of the problem $\phi_{\alpha,j,\pm}$ and $\phi_{-\alpha,j,\pm}$. Up to a normalization this is equivalent to the work by Hopfield.

\section{Milonni's model and rising lowering polariton operators }
We start with the amplitude  for the transverse polariton field
\begin{eqnarray}
E_{\alpha,j,m}^{(+)}(t)=\int_{\delta \omega_{\alpha,m}} d\omega\frac{\omega^2}{\omega_\alpha^2-\tilde{\varepsilon}(\omega)\omega^2}\sqrt{\frac{\hbar\sigma_\omega}{\pi\omega}}f^{(0)}_{\omega,\alpha,j}(t). 
\end{eqnarray} where $\delta \omega_{\alpha,m}$ is a frequency window centered on the polariton pulsation $\omega_c\simeq\textrm{Re}[\Omega_{\alpha,m}]:=\Omega'_{\alpha,m}$.
This field is equivalently written as
 \begin{eqnarray}
E_{\alpha,j,m}^{(+)}(t)=\int_{0}^{+\infty} \frac{d\omega F_{\alpha,m}(\omega)\omega^2}{\omega_\alpha^2-\tilde{\varepsilon}(\omega)\omega^2}\sqrt{\frac{\hbar\sigma_\omega}{\pi\omega}}f^{(0)}_{\omega,\alpha,j}(t). 
\end{eqnarray} where $F_{\alpha,m}(\omega)$ is a window function such as $F_{\alpha,m}(\omega)\simeq 1$ if   $\omega$ belongs to the interval $\delta \omega_{\alpha,m}$  and $F_{\alpha,m}(\omega)\simeq 0$ otherwise. From the commuting properties of  $f^{(0)}_{\omega,\alpha,j}(t)$ (see Eqs.~\ref{33} and \ref{commuting}) we deduce the commutator:
\begin{eqnarray}
[E_{\alpha,j,m}^{(+)}(t),E_{\beta,l,n}^{(-)}(t)]=\frac{\hbar}{\pi}\delta_{\alpha,\beta}\delta_{j,l}\nonumber\\ \cdot\int_0^{+\infty} \frac{d\omega\omega^4\tilde{\varepsilon}''(\omega)}{|\omega_\alpha^2-\tilde{\varepsilon}(\omega)\omega^2|^2} F_{\alpha,m}(\omega)F_{\beta,n}(\omega)
\end{eqnarray} with $F_{\alpha,m}(\omega)F_{\beta,n}(\omega)\simeq \delta_{n,m}F_{\alpha,m}(\omega)$. Now consider the integral:
  \begin{eqnarray}
I=\int_0^{+\infty} d\omega\frac{\omega^4\tilde{\varepsilon}''(\omega)}{|\omega_\alpha^2-\tilde{\varepsilon}(\omega)\omega^2|^2}F_{\alpha,m}(\omega)\nonumber\\
=\int_0^{+\infty} d\omega\frac{\omega^4\tilde{\varepsilon}''(\omega)}{(\omega_\alpha^2-\tilde{\varepsilon}'(\omega)\omega^2)^2+\omega^4\tilde{\varepsilon}''(\omega)^2}F_{\alpha,m}(\omega).
\end{eqnarray}
 Since for weak losses the  integrand is extremely  peaked  on the value $\Omega_{\alpha,m}$ we can  write $I$ as
 \begin{eqnarray}
I\simeq \int_{-\infty}^{+\infty} d\omega\frac{\Omega_{\alpha,m}^4\tilde{\varepsilon}''(\Omega_{\alpha,m})}{(\omega_\alpha^2-\tilde{\varepsilon}'(\omega)\omega^2)^2+\Omega_{\alpha,m}^4\tilde{\varepsilon}''(\Omega_{\alpha,m})^2}.
\end{eqnarray}
We use the approximation $\omega_\alpha^2-\tilde{\varepsilon}'(\omega)\omega^2\simeq \omega_\alpha^2-\tilde{\varepsilon}'(\Omega_{\alpha,m})\Omega_{\alpha,m}^2-(\omega-\Omega_{\alpha,m})\frac{d(\tilde{\varepsilon}'(\omega)\omega^2)}{d\omega}|_{\Omega_{\alpha,m}}=-(\omega-\Omega_{\alpha,m})\frac{d(\tilde{\varepsilon}'(\omega)\omega^2)}{d\omega}|_{\Omega_{\alpha,m}}$  which is valid near the pole where the conditions $\omega_\alpha^2\simeq\tilde{\varepsilon}'(\Omega_{\alpha,m})\Omega_{\alpha,m}^2$ approximately holds for transverse polaritons. We thus get 
 \begin{eqnarray}
I=\frac{1}{\tilde{\varepsilon}''(\Omega_{\alpha,m})} \int_{-\infty}^{+\infty} du\frac{\gamma^2}{u^2+\gamma^2}=\frac{\pi\Omega_{\alpha,m}^2}{\frac{d(\tilde{\varepsilon}'(\omega)\omega^2)}{d\omega}|_{\Omega_{\alpha,m}}}\nonumber\\=\frac{\pi\Omega_{\alpha,m}}{2}\frac{d\Omega_{\alpha,m}^2}{d\omega_\alpha^2}.
\end{eqnarray}
with $\gamma^2=\left(\frac{\Omega_{\alpha,m}^2\tilde{\varepsilon}''(\Omega_{\alpha,m})}{\frac{d(\tilde{\varepsilon}'(\omega)\omega^2)}{d\omega}|_{\Omega_{\alpha,m}}}\right)^2$ and $u=\omega--\Omega_{\alpha,m}$.
and where we used the integral $\int_{-\infty}^{+\infty} du\frac{\gamma^2}{u^2+\gamma^2}=\pi\gamma$ which is easily calculated in the complex plane. From this we finally obtain the commutator  of Eq.~\ref{window}. We point out that the result does not explicitly depends on the extension of the frequency windows $\delta \omega_{\alpha,m}$ if losses and dispersion are weak enough to have $\gamma\ll \delta \omega_{\alpha,m}$.\\
For the longitudinal polariton field we have  a similar calculation. Starting with the definition:
    \begin{eqnarray}
E_{\alpha,m}^{(+)}(t)=\int_{0}^{+\infty} d\omega F_{\alpha,m}(\omega)\frac{-1}{\tilde{\varepsilon}(\omega)}\sqrt{\frac{\hbar\sigma_\omega}{\pi\omega}}f^{(0)}_{\omega,\alpha,||}(t). 
\end{eqnarray} 
we can calculate the commutator $[E_{\alpha,m,||}^{(+)}(t),E_{\beta,n,||}^{(-)}(t)]$. We have
\begin{eqnarray}
[E_{\alpha,m,||}^{(+)}(t),E_{\beta,n,||}^{(-)}(t)]=\frac{\hbar}{\pi}\delta_{\alpha,\beta}\nonumber\\ \cdot \int_0^{+\infty} d\omega\frac{\tilde{\varepsilon}''(\omega)}{|\tilde{\varepsilon}(\omega)|^2} F_{\alpha,m}(\omega)F_{\beta,n}(\omega)
\end{eqnarray} with again  $F_{\alpha,m}(\omega)F_{\beta,n}(\omega)\simeq \delta_{n,m}F_{\alpha,m}(\omega)$.
We have to evaluate the integral 
\begin{eqnarray}
I=\int_0^{+\infty} d\omega\frac{\tilde{\varepsilon}''(\omega)}{|\tilde{\varepsilon}(\omega)|^2}F_{\alpha,m}(\omega)\nonumber\\
=\int_0^{+\infty} d\omega\frac{\tilde{\varepsilon}''(\omega)}{(\tilde{\varepsilon}'(\omega))^2+\tilde{\varepsilon}''(\omega)^2}F_{\alpha,m}(\omega).
\end{eqnarray}
which like for the transverse modes in the limit of weak losses and dispersion leads after straightforward calculations to 
\begin{eqnarray}
I\simeq\frac{\pi}{|\frac{d\tilde{\varepsilon}'(\omega)}{d\omega}|_{\Omega'_{\alpha,m}}|}.
\end{eqnarray}From this we deduce the commutator given in Eq.~\ref{windowlong}.


\begin{thebibliography}{99}
\bibitem{footnote}
We have the symmetries for $\mathbf{k}_\alpha=-\mathbf{k}_{-\alpha}$: $\Phi_{-\alpha}(\mathbf{x})=\Phi_\alpha^\ast(\mathbf{x})$, $\boldsymbol{\hat{\epsilon}}_{-\alpha,j}=\eta_{j}\boldsymbol{\hat{\epsilon}}_{\alpha,j}$ with $\eta_1=-1$ and $\eta_2=+1$. The field being real valued, i.e., $\mathbf{F}(\mathbf{x},t)=\mathbf{F}^\ast(\mathbf{x},t)$ we deduce $q_{-\alpha,j}(t)=\eta_j q_{\alpha,j}^\ast(t)$.
\bibitem{footnotebis}
More precisely we have near the singularity $\omega_0$ the condition $\boldsymbol{\alpha}\delta(\Omega-\omega_0)=\frac{i\pi\omega_p^2}{2\omega_0}\delta(\Omega-\omega_0)\widetilde{\textbf{P}}_{||}(\omega_0)+\widetilde{\textbf{P}}_{||}(\Omega)$. If we suppose $\widetilde{\textbf{P}}_{||}(\omega_0)$ to be regular we get the result discussed in the text.  If $\widetilde{\textbf{P}}_{||}(\omega_0)$ is singular then $\frac{i\pi\omega_p^2}{2\omega_0}\delta(\Omega-\omega_0)\widetilde{\textbf{P}}_{||}(\omega_0)$ would be even more singular and this would contradict the equation. Therefore regularity at $\pm\omega_0$ must be imposed.
\end{thebibliography}
\end{document}